\pdfoutput=1

\documentclass[12pt,reqno]{article}
\usepackage{jheppub}
\usepackage{amsmath,amssymb,amsfonts}

\usepackage{bbold}

\usepackage{float}

\usepackage[usenames,dvipsnames]{xcolor}
\usepackage{epsfig}
\usepackage{epstopdf}
\usepackage{dcolumn}
\usepackage{tikz}
\usetikzlibrary{shapes.geometric, arrows}
\usepackage{upgreek}
\usepackage{setspace}
\usepackage{enumitem}
\usepackage{array,multirow,bigdelim}

\def\be{\begin{equation}}
\def\ee{\end{equation}}
\def\ba{\begin{eqnarray}}
\def\ea{\end{eqnarray}}

\def\volpsl{{\rm Vol}(PSL(2,\mathbb{C}))}

\def\a{\alpha}
\def\b{\beta}

\def\b#1{\overline{#1}}

\def\CP1{\mathbb{CP}^1}
\def\SL2C{\mathrm{SL}(2,\mathbb{C})}

\def\Z2{\mathbb{Z}_2}

\def\su2{{SU(2)}}

\def\a{{\alpha}}

\def\[{\left[}
\def\]{\right]}

\def\L{\Lambda}

\def\s{\sigma}
\def\a{\alpha}
\def\b{\beta}

\def\({\left(}
\def\){\right)}
\def\[{\left[}
\def\]{\right]}

\def\<{\langle}
\def\>{\rangle}

\def\i2{\frac{i}{2}}

\def\2F1{\,_2{\rm F}_1}

\newcommand{\beq}{\begin{equation}}
\newcommand{\eeq}{\end{equation}}
\newcommand{\beqq}{\begin{equation*}}
\newcommand{\eeqq}{\end{equation*}}
\newcommand\beqa{\begin{eqnarray}}
\newcommand\eeqa{\end{eqnarray}}
\newcommand\beqaa{\begin{eqnarray*}}
\newcommand\eeqaa{\end{eqnarray*}}
\newcommand\bea{\begin{array}}
\newcommand\eea{\end{array}}

\title{$\Lambda$ Scattering Equations}

\author{Humberto Gomez}
\affiliation{Instituto de Fisica -- Universidade de S\~ao Paulo,\\
Caixa Postal  66318, 05315-970 S\~ao Paulo, SP, Brazil\\
Facultad de Ciencias Basicas,  Universidad Santiago de Cali,\\
Calle 5 $N^\circ$  62-00 Barrio Pampalinda, Cali, Valle, Colombia.}

\emailAdd{humgomzu@gmail.com}

\abstract{The CHY representation of scattering amplitudes is based on integrals over the moduli space of a punctured sphere. We replace the punctured sphere by a double-cover version. The resulting scattering equations depend on a parameter $\Lambda$ controlling the opening of a branch cut. The new representation of scattering amplitudes possesses an enhanced redundancy which can be used to fix, modulo branches, the location of four punctures while promoting $\Lambda$ to a variable. Via residue theorems we show how CHY formulas break up into sums of products of smaller (off-shell) ones times a propagator. This leads to a powerful way of evaluating CHY integrals of generic rational functions, which we call the $\L$ algorithm.
}

\begin{document}
{\setstretch{1}
\maketitle
}

\onehalfspacing
\section{Introduction}

The complete tree-level S-matrix of a large variety of field theories of massless particles are now known (or conjectured) to have a description in terms of contour integrals over ${\cal M}_{0,n}$, the moduli space of $n$-punctured Riemann sphere
\cite{Cachazo:2013hca,Cachazo:2013iea,Cachazo:2014xea,Mason:2013sva,Dolan:2013isa,Berkovits:2013xba,Adamo:2013tsa,Gomez:2013wza,Dolan:2014ega,
Geyer:2014fka,Cachazo:2014nsa,Ohmori:2015,MasonC}. Some of these theories are Yang-Mills, Einstein gravity, Dirac-Born-Infeld, and the $U(N)$ non-linear sigma model \cite{Cachazo:2014xea,Ohmori:2015}. The new formulas for the scattering of $n$ particles are given as a sum over multidimensional residues \cite{harris} on ${\cal M}_{0,n}$.

In addition, many algorithms have been created in order to compute these kind of contour integrals \cite{Kalousios:2013eca,Kalousios:2015fya,humbertoF,jacobrules,jacobSF,Cardonaone,Cardonatwo,fengalgorithm,LamECHY,Weinzierl:2014vwa,mafra}, as well as much progress have been done at loop level \cite{Adamo:2015,Adamo:2013tsa,Casali:2014hfa,Masonloop1,jacobrulesloop,
Qcutone,Qcuttwo,fengplanarloop,Masonloop2,ellisloop1,ellisloop2}.

Denoting the position of $n$ punctures on a sphere by $\{ z_1,z_2,\ldots z_n\}$ and using $PSL(2,\mathbb{C})$ to fix three of them, say $z_1,z_2,z_3$, there is a rational map from ${\mathbb C}^{n-3}\to {\mathbb C}^{n-3}$ which is a function of the space of kinematic invariants for the scattering of $n$ massless particles ($k_a^2=0$), $s_{ab} = k_a\cdot k_b$,
\be
E_a(z) =\sum_{b=1,b\neq a}^n\frac{s_{ab}}{z_a-z_b} \,,\quad {\rm for} \quad a\in \{1,2,\ldots ,n\},
\ee
with $a\in \{4,5,\ldots ,n\}$.

Using this map scattering amplitudes are defined as the sum over the residues of
\be\label{form1}
{\cal M}_n = \int d^{n-3}z\frac{|1,2,3|^2\,\,H(z)}{E_4(z)E_5(z)\cdots E_n(z)}.
\ee
over all the zeroes of the map $\{E_4,E_5,\ldots ,E_n\}$. Here $|1,2,3|=(z_1-z_2)(z_2-z_3)(z_3-z_1)$ and  $H(z)$ is a rational function that depends on the theory under consideration. The equations defining the zeroes, $E_a=0$, are known as the scattering equations \cite{Fairlie:1972,Roberts:1972,Fairlie:2008dg,Gross:1987ar,Witten:2004cp,Caputa:2011zk,Caputa:2012pi,Makeenko:2011dm,Cachazo:2012da}. In section \ref{preliminaries} we will give a short review about this ideas.

The representation \eqref{form1} of scattering amplitudes makes many properties manifest. Some of them are gauge invariance, soft limits, BCJ relations and the existence of KLT formulas \cite{SWsoft,freddysoft,chileG,soft1,soft2,soft3,Kawai:1985xq,Bern:2008qj,Bern:1998sv,humbertoF,freddyklt1,freddyklt2}. The drawback is that integrals of the form \eqref{form1} require the solution of polynomial equations whose degree increases with the number of particles.

In this paper we reformulate the formula for scattering amplitudes in terms of a double cover of the punctured sphere. More precisely, we consider a sphere as a curve in ${\mathbb CP}^2$ defined by
\be\label{initialcurve}
y^2 = \sigma^2-\Lambda^2,
\ee
where $\L$ is a non zero constant parameter. 
Clearly, the curve is invariant under a simultaneous scaling of the coordinates $(y,\sigma,\Lambda)$. The new formulation is schematically given by
\be\label{form2}
{\cal M}_n ={1\over \volpsl} \int \prod_{a=1}^n \left(\frac{(y_a\,d y_a)}{\prod_{a=1}^n (y_a^2-\sigma_a^2 + \Lambda^2)}\right)
\prod_{b=1}^n\frac{d\s_b\,\,\,H(\sigma,y,\Lambda)}{ E_b(\sigma,y,\Lambda )}.
\ee
Integration over both residues of the curve implements the sums over choices of branches.

In the double cover description each puncture is specified by a pair of complex numbers $(\sigma_a,y_a)$. The value of $y_a$ indicates the branch where the puncture is located. The new form of the components of the map $E_a$ is
\be
E_a(\sigma,y,\Lambda ) = \sum_{b=1,b\neq a}^n\frac{1}{2}\left(\frac{y_b}{y_a}+1\right)\frac{s_{ab}}{\sigma_a-\sigma_b}.
\ee
This form is very easy to motivate and it is done in section \ref{sthreedc}.

The differential form being integrated in the double cover version of the formula for ${\cal M}_n$ is invariant under the global rescaling inherited from ${\mathbb CP}^2$. This ${\mathbb C}^*$ group can be promoted to a full redundancy of the description introducing the scale measure
\begin{equation}\label{mlambda}
\frac{1}{{\rm vol}({\mathbb C}^*)}\frac{d\L}{\L},
\end{equation}
%
where the $\L$ factor is proportional to the square root of the  discriminant  of the quadratic  curve in \eqref{initialcurve},  $\Delta=4\L^2$.
The ${\mathbb C}^*$ action is non-trivial on the puncture locations, this means that one can combine the new $\mathbb{C}^*$ action with the $PSL(2,\mathbb{C})$ group of the sphere and use it to fix the $\sigma$ coordinate of four punctures. Doing so leaves $\Lambda$ as an integration variable to be fixed by the scattering equations. This is done in section \ref{new gauge fixing}.

In section \ref{residuetheorem} we show that the global residue theorem can be used to replace one of the components of the map, say $E_n$, by $\Lambda$. As it turns out the residue theorem only picks up poles at $\Lambda=0$ and at $\Lambda=\infty$ and both are identical. At $\Lambda=0$ the branch cut connecting the two branches of the double cover closes and the integrals separates into sectors. Each sector is determined by the distribution of the punctures between the two branches. The amplitude then becomes schematically
\be
{\cal M}_n = \sum_{U\cup D}{\cal M}_U^{\rm off-shell} \frac{1}{P^2_U} {\cal M}_D^{\rm off-shell}.
\ee
where the sum is over possible distributions of punctures and ${\cal M}^{\rm off-shell}$ refers to amplitudes where one particle, corresponding to the puncture created by the closing of the branch cut is off-shell.

We apply this procedure to more general integrals over the moduli space which appear as parts of physical amplitudes in their CHY representation. In these more general cases, when the integrand has at most double poles on the boundary of the moduli space ${\cal M}_{n,0}$ then the propagator is a standard Feynman propagator. An example of an integral with at most double poles is
\be
\int d^n z\frac{1}{E_1(z)E_2(z)\cdots E_n(z)}\frac{1}{(123\cdots n)^2}
\ee
where
\be\label{PTdef}
(123\cdots n) := (z_1-z_2)(z_2-z_3)\cdots (z_n-z_1).
\ee
This integral is known to give the sum over all Feynman diagrams computing a partial amplitude in a cubic scalar theory in the bi-adjoint representation of $U(N)\times U(M)$. Iterating the procedure gives rise to a novel set of diagrams where the buliding blocks are four-particle amplitudes and propagators.

When the integrand has higher order poles on the moduli space one finds generalized propagators which are made from higher powers of kinematic invariant. One example, explicitly compute in section \ref{mainsection}, is a six-particle integral
\be
\int d^6 z\frac{1}{E_1(z)E_2(z)\cdots E_6(z)}\frac{1}{(1234)^2(56)^2}
\ee
with $(56)=(z_5-z_6)(z_6-z_5)$ consistent with the definition \eqref{PTdef}. This integral has poles of the form $1/s_{56}^3$.

A very familiar way of understanding this process is by analogy with the Stukelberg procedure for taking massless limits of massive vector bosons \cite{procafield}. The mass parameter is played by the kinematic invariant controlling the factorization limit while the Stukelberg field is played by the $\Lambda$ parameter.
All this process is shown in section \ref{mainsection}, where we formulate a new algorithm and in section \ref{examples}  we  give three non trivial examples.

In section \ref{jacobvslambda} we compare our method with the rules given in \cite{jacobrules} by Baadsgaard {\bf et al}.  We also generalize the new algorithm to non trivial numerators and we give a simple example.

Finally, we end in section \ref{discussion}  with discussions.

\section{Preliminaries}\label{preliminaries}

In this section we review the basic CHY construction \cite{Cachazo:2013hca,Cachazo:2013iea,Cachazo:2014xea} and show some examples that motivate the double-cover construction.

\subsection{CHY Construction}

Consider the scattering of $n$ massless particles. The scattering data is determined in terms of a set of $n$ momentum vectors $\{ k_1^\mu,k_2^\mu,\ldots ,k_n^\mu \}$ and $n$ wave functions $\{ \epsilon_1^\mu,\epsilon_2^\mu,\ldots ,\epsilon_n^\mu \}$. Here we take the wave functions to be polarization vectors as higher spin wave functions, e.g. for gravitons, can be constructed using tensor products. In a slightly different terminology from the original CHY construction, one introduces $n$ rational functions of the puncture locations, $z_a$, defined by \cite{Gross:1987ar,Cachazo:2013hca}
\be
E_a = \sum_{b=1,b\neq a}^n \frac{s_{ab}}{z_a-z_b}.
\ee
It is easy to show that three linear combinations vanish
\be
\sum_{a=1}^n z_a^m E_a = 0 \quad {\rm for} \quad m \in \{0,1,2\}.
\ee

Recalling that different configurations of punctures on a ${\mathbb CP}^1$ are to be identified if they differ by a $PSL(2,\mathbb{C})$ transformation. This means that the location of three punctures can be fixed. It is possible to show that for any rational function $H(z)$ which transforms as
\be\label{int_trans}
H(z)\to H(z)\prod_{a=1}^n(\gamma z_a+\delta )^4 ,\quad {\rm when} \quad z_a \to \frac{\alpha z_a+\beta}{\gamma z_a+\delta} \quad{\rm and}\quad\a\delta-\b\gamma=1,
\ee
the contour integral \cite{Cachazo:2013hca}
\be
\int \prod_{a=1,a\neq \{i,j,k\}}^n dz_a \frac{|ijk|_z |pqr|_z}{\prod_{c=1,c\neq \{p,q,r\}}^n E_c(z)}H(z)
\ee
that computes one of the local residues at  a zero of the map ${\mathbb C}^{n-3}\to {\mathbb C}^{n-3}$ is independent of the choice of fixed punctures $\{z_i,z_j,z_k\}$ and of equations eliminated $\{E_p,E_q,E_r\}$ to construct the map. In this formula $|ijk|$ stands for the Vandermonde determinant of $z_i,z_j,z_k$.

One way to see that this is the case is to realize that the generators of $PSL(2,{\mathbb C})$ are
\be
L_1 = \sum_{a=1}^n\partial_{z_a}, \quad L_0 = \sum_{a=1}^n z_a \partial_{z_a}, \quad L_{-1} = \sum_{a=1}^n z^2_a \partial_{z_a}.
\ee
Treating the $PSL(2,{\mathbb C})$ as a redundance of the integral and using a gauge fixing procedure one can check that the Fadeev-Popov determinant is indeed
\be\label{zgfixing}
|ijk|_z = \left|
  \begin{array}{ccc}
    1 & ~z_i & ~z_i^2 \\
    1 & ~z_j & ~z_j^2 \\
    1 & ~z_k & ~z_k^2 \\
  \end{array}
\right|.
\ee

\subsection{Examples}

The CHY representation of many theories are known (or are conjectured). In this subsection we review some of them in order to motivate the constructions in this paper.

Let us start with Einstein gravity \cite{Cachazo:2013hca,Cachazo:2013iea}. The integrand $H$ is computed as the reduced determinant of a matrix
a $2n\times 2n$ antisymmetric matrix
\be\label{Psi}
\Psi = \left(
         \begin{array}{cc}
           A &  -C^{\rm T} \\
           C & B \\
         \end{array}
       \right),
\ee
where $A$, $B$ and $C$ are $n\times n$ matrices. The first two matrices have components
\be
A_{ab} = \begin{cases} \displaystyle \frac{k_{a}\cdot k_b}{z_{a}-z_{b}} & a\neq b,\\
\displaystyle \quad ~~ 0 & a=b,\end{cases} \qquad B_{ab} = \begin{cases} \displaystyle \frac{\epsilon_a\cdot\epsilon_b}{z_{a}-z_{b}} & a\neq b,\\
\displaystyle \quad ~~ 0 & a=b,\end{cases}
\label{ABmatrix}
\ee
while the third is given by
\be
C_{ab} = \begin{cases} \displaystyle \frac{\epsilon_a \cdot k_b}{z_{a}-z_{b}} &\quad a\neq b,\\
\displaystyle -\sum_{c=1;c\neq a}^n\frac{\epsilon_a \cdot k_c}{z_{a}-z_{c}} &\quad a=b.\end{cases}
\ee
This matrix depends on the momenta $k^\mu_a$ and on polarization vectors $\epsilon_a^\mu$.

The diagonal components of the $C$ matrix can be written in a manifestly $PSL(2,{\mathbb C})$ covariant way by choosing a momentum vector, say $k_n$ if $a\neq n$, and eliminating it using momentum conservation
\be
C_{aa} = -\sum_{b=1,b\neq a}^{n-1}\epsilon_a \cdot k_c\frac{(z_{n}-z_{c})}{(z_{a}-z_{c})(z_{a}-z_{n})}.
\ee
The integrand is given by
\be
H^{\rm gravity}(z) = {\rm det}'\Psi = \frac{1}{(z_i-z_j)^2}\det\Psi_{ij}^{ij},
\ee
where $\Psi^{ij}_{ij}$ is the $(n-2)\times (n-2)$ matrix obtained from $\Psi$ by removing the rows $(i,j)$ and the columns $(i,j)$.

The second example is that of the scattering of gluons in a $U(N)$ Yang-Mills theory \cite{Cachazo:2013hca,Cachazo:2013iea}. The coefficient of the trace ${\rm Tr}(T^{a_1}T^{a_2}\cdots T^{a_n})$ is computed by the integrand
\be
H^{\rm YM}(z) = \frac{1}{(123\cdots n)}{\rm Pf}'\Psi   ,
\ee
where ${\rm Pf}'\Psi = (z_i-z_j)^{-1}{\rm Pf}\Psi^{ij}_{ij}$ and $(123\cdots n) = (z_1-z_2)(z_2-z_3)\cdots (z_n-z_1)$.

The third example is that of a scalar theory in the bi-adjoint representation of $U(N)\times U(\tilde N)$ \cite{Cachazo:2013iea}. The coefficient of the trace ${\rm Tr}(T^{a_1}T^{a_2}\cdots T^{a_n}){\rm Tr}({\tilde T}^{a_{w(1)}}{\tilde T}^{a_{w(2)}}\cdots {\tilde T}^{a_{w(n)}})$ with $w$ some permutation of labels, is given by
\be\label{scalarH}
H^{\rm scalar}(z) = \frac{1}{(123\cdots n)}\times \frac{1}{(w(1)w(2)w(3)\cdots w(n))}.
\ee

The last two examples are also purely scalar theories but with derivative interactions \cite{Cachazo:2014nsa,Cachazo:2014xea}.

The fourth example is a special Galileon theory (sGal) that possesses more non-linearly realized symmetries than a generic Galileon. Amplitudes in this theory are computed using
\be
H^{\rm sGal}(z) = \left({\rm det}' A\right)^2  ,
\ee
where ${\rm det}' A = (z_i-z_j)^{-2}{\rm det}A^{ij}_{ij}$.

The fifth and final example is the $U(N)$ non-linear sigma model. The term proportional to the trace ${\rm Tr}(T^{a_1}T^{a_2}\cdots T^{a_n})$ is computed by
\be
H^{\rm NLSM}(z) = \frac{1}{(123\cdots n)}{\rm det}'A   ,
\ee

In order to illustrate the kind of integrals we are interested in performing let us consider ${\rm det}' A$ for four particles,
\be
{\rm det}' A_4 = \frac{1}{(z_1-z_2)^2}\left|
                                                  \begin{array}{cc}
                                                    0 & \frac{s_{34}}{z_3-z_4} \\
                                                    \frac{s_{34}}{z_4-z_3} & 0 \\
                                                  \end{array}
                                                \right|.
\ee
This means that the integrands of the Galileon and NLSM are
\be
H^{\rm sGal}_4 = s_{34}^4\times \frac{1}{(z_1-z_2)^4(z_3-z_4)^4}, \quad H^{\rm NLSM}_4 = s_{34}^2\times \frac{1}{(1234)}\frac{1}{(z_1-z_2)^2(z_3-z_4)^2}.
\ee

\subsection{Singularities on ${\cal M}_{0,n}$}

The examples given above make it clear that a variety of integrands $H(z)$ can appear. One way to classify them is by the kind of singularities they have as different boundaries on the moduli space of a punctured sphere are approached. The largest codimension singularities are when two punctures approach each other. Consider for example the integrands for four particles \cite{Cachazo:2013iea,Cachazo:2014nsa,Cachazo:2014xea}
\be
H^{\rm \phi^3}_4 \!\!\sim \frac{1}{(1234)^2},~~ H^{\rm NLSM}_4 \!\!\sim \frac{1}{(1234)}\frac{1}{(z_1-z_2)^2(z_3-z_4)^2}, ~~H^{\rm sGal}_4 \!\!\sim \frac{1}{(z_1-z_2)^4(z_3-z_4)^4}\nonumber.
\ee
Clearly, the first integrand has double poles as any two consecutive punctures approach each other $z_a\to z_{a+1}$ and no other poles. The second integrand has a triple poles when $z_1\to z_2$ and when $z_3\to z_4$ and simple poles when $z_2\to z_3$ and when $z_4\to z_1$. Finally, the last integrand only has fourth order poles when $z_1\to z_2$ and when $z_3\to z_4$. It is easy to show that the order of the pole is related to the order of the propagator associated to the coincident punctures. If the integrand as a $(m+1)^{\rm th}$ order pole when $z_a\to z_b$ then the integral has a pole of the form $1/s_{ab}^m$.

In the rest of this paper we develop a double cover formulation which is tailored for exploiting the behavior of integrands near boundaries of the moduli space. This method not only becomes a powerful tool in the evaluation of integrals but it also makes physical properties manifest such as crossing and factorization.

\section{Double-Cover Formulation}\label{sthreedc}

We consider a sphere as a curve in ${\mathbb CP}^2$ defined by \cite{harris}
\be\label{ac}
y^2 = \sigma^2-\Lambda^2.
\ee
We call this curve $\Sigma$ and it can be interpreted as two sheets joined by a branch cut. We take $\sigma$ as the coordinate on a sheet and $y$ as the variable determining the branch. $\Lambda$ is taken to be a constant parameter that controls the opening of the branch cut joining the branch points $\sigma=-\Lambda$ and $\sigma=\Lambda$.

The location of $n$-punctures on $\Sigma$ is given by $n$ pairs $\{(\sigma_a,y_a)\}$. We would like to find formulation of the maps $E_a$ defining the scattering equations for this curve. Clearly, $E_a$ must have a simple pole when puncture $a$ coincides with puncture $b$. On $\Sigma$, it is not enough to have $\sigma_a \to \sigma_b$ but we also need $y_a\to y_b$, i.e., they must be on the same branch. When $\sigma_a\to \sigma_b$ one can have either $y_a\to y_b$ or $y_a\to -y_b$. So we need a projector, $P_{b}^{(a)}$, that gives one in the former and zero in the latter. One choice is
\be
P_{b}^{(a)} = \frac{1}{2}\left( \frac{y_b}{y_a}+1 \right).
\ee
This turns out to be the correct choice and one has
\be
E_a(\sigma,y ) = \sum_{b=1,b\neq a}^n\frac{1}{2}\left(\frac{y_b}{y_a}+1\right)\frac{s_{ab}}{\sigma_a-\sigma_b}.
\ee
One important condition the equations have to satisfy is that they must be covariant under the exchange of $\sigma$ and $y$ (with $\Lambda\to i\Lambda$) which is a symmetry of the curve $\Sigma$. It is easy to check that on the support of $y_b^2 = \sigma_b^2-\Lambda^2$, the function $y_a E_a$ is invariant.

Having found the new version of the maps $E_a$ which give rise to the scattering equations, the next step is to translate the rational function $H(z)$ which defines the theory under consideration. All such functions can be decomposed as linear combinations of functions of the form \cite{humbertoF}
\be\label{general_H}
H(z ) = \frac{1}{(\a(1)\a(2)\cdots \a(n))(\gamma(1)\gamma(2)\cdots \gamma(n))}f(r_{ijkl}),
\ee
where $(\a(1)\a(2)\cdots \a(n))$ and $(\gamma(1)\gamma(2)\cdots \gamma(n))$ are  Parke-Taylor factors with ordering $\a$ and $\gamma$  (see \eqref{PTdef} for the Parke-Taylor factor definition \cite{Parke:1986gb}). $f$ is a rational function of $r_{ijkl}$ which are general cross ratios, i.e.,
\be
r_{ijkl} \equiv \frac{z_{ij}\, z_{kl}}{z_{il}\, z_{jk}},
\ee
where we have introduced a convenient shorthand notation
\be
z_{ab} \equiv z_a - z_b \quad \quad (\sigma_{ab} \equiv \sigma_a-\sigma_b).
\ee 
In order to map $H(z)$ to $H(\s,y)$, we define any combinations of factors of the form
\be
(z_{a_1}-z_{a_2})(z_{a_2}-z_{a_3})\cdots (z_{a_{m-1}}-z_{a_m})(z_{a_{m}}-z_{a_1})
\ee 
as a chain $(a_1~a_2~\cdots a_{m-1}~a_m)$ of length $m$. Chains are taken to have lengths $2\leq m\leq n$. A chain of length $2$ is given by
\be
(a_1~a_2) = (z_{a_1}-z_{a_2})(z_{a_2}-z_{a_1}).
\ee

It is straightforward to check
\begin{equation}
r_{ijkl}\equiv \frac{z_{ij}\, z_{kl}}{z_{il}\, z_{jk}}=\frac{z_{ij} z_{jl} z_{lk} z_{ki}  }{z_{kj} z_{jl} z_{li} z_{ik} }=\frac{(ijlk)}{(ikjl)}  .
\end{equation}

Now we propose to use the following replacement into the chain so as to construct the integrand $H(\s,y)$, 
\be\label{replace}
\frac{1}{z_{ab}} ~\mapsto ~\tau_{a:b} \equiv{1\over 2} \left(\frac{y_a+y_b+\sigma_{ab}}{y_a}\right)\frac{1}{\sigma_{ab}}.
\ee
Note that while the left hand side is antisymmetric in the  $a$ and $b$ labels the right hand side is not and hence the notation $\tau_{a:b}$. This fact becomes irrelevant when the substitution is made into chains and hence the importance of the appearance of them in the integrands. So, we  complete  the map $H(z)\rightarrow H(\s,y)$ by
\begin{equation}\label{general_Hsy}
 H(\s,y)=(\tau_{\a(1):\a(2)}\cdots  \tau_{\a(n):\a(1)} ) (\tau_{\gamma(1):\gamma(2)}\cdots  \tau_{\gamma(n):\gamma(1)} ) f \left( \frac{\tau_{i:k} \tau_{k:j} \tau_{j:l} \tau_{l:i}  } {\tau_{i:j} \tau_{j:l} \tau_{l:k} \tau_{k:i} } \right).
\end{equation}
In addition one can check, in a simple way, the chain property
\begin{align}\label{cproperty}
(a_1\,a_2\cdots a_{m-1}\,a_m)&= (-1)^m (a_m\, a_{m-1} \cdots a_2 a_1 ),\\
(\tau _{a_1:a_2} \cdots \tau_{a_{m-1}:a_{m}} \,\tau_{a_m:a_1} ) &=(-1)^m ( \tau_{a_{m}:a_{m-1}} \, \tau_{a_{m-1}:a_{m-2}}  \cdots
\tau _{a_2:a_1} \,\tau_{a_1:a_m} ),
\end{align}
and the inverse map works in  the same way, $\tau_{a:b}\rightarrow \frac{1}{z_{ab}}$.

Moreover, it is straightforward to check the scattering equations can be written as
\begin{equation}
E_a(\s,y)=\sum_{b\neq a}^n\,s_{ab}\, \tau_{a:b}=\frac{1}{y_a}\sum_{b\neq a}^n\,s_{ab}\,\tilde \tau_{a:b}= \frac{1}{y_a}\tilde E_a(\s,y),
\end{equation}
where we have denoted $\tilde\tau_{a:b}$ and $\tilde E_a$ as
\begin{equation}\label{tautilde}
\tilde\tau_{a:b}=\frac{y_a+y_b+\s_{ab}}{2\,\s_{ab}}, \qquad \tilde E_a=\sum_{b\neq a}^n\,s_{ab}\,\tilde \tau_{a:b}.
\end{equation}

It is not obvious how chains appear in integrands that are computed using the Pfaffian or the determinant of the matrices $\Psi$ of $A$.

\subsection{Redundancies}\label{R_part1}

Next we move to the discussion of the redundancies and how to gauge fix them. This subsection is only the first part of the discussion in which we show how to perform the standard gauge fixings. In the second part, presented in section \ref{new gauge fixing}, we perform a different gauge fixing which allow us to use residue theorems to break up contour integrals into integrals with smaller number of punctures. 

%
%
We consider the following integral
\be\label{intres}
{\cal I}=\frac{1}{\volpsl}\int \prod_{a=1}^n\,\frac{d\s_a \,(y_a\, d y_a)}{(y^2_a-\sigma_a^2+\Lambda^2)}\times\frac{H(\s,y)}{\prod_{b=1}^n E_b(\s,y)}
\ee
%
%
where $\L$ is a non-zero constant parameter and  $H(\sigma, y)$ is a general rational function as in \eqref{general_Hsy}. 
%
%
%
The factor $\volpsl$ in the integral is there only as a reminder that the integral has a redundancy that has to be gauge fixed. The $PSL(2,\mathbb{C})$ action is generated by the vectors (on the support of the algebraic curve $y_a^2=\s_a^2-\L^2,\,a=1,\cdots n$) 
\be\label{redun}
L_{\pm 1} = \sum_{a=1}^n \frac{1}{\Lambda}y_a(\sigma_a \mp y_a)\partial_a, \qquad L_0 = \sum_{a=1}^n y_a\partial_a,
\ee
%
%
%
where $\partial_a \equiv \partial /\partial \sigma_a$ and  they satisfy the algebra
\begin{equation}
[L_{\pm 1},L_0]=\pm\,L_{\pm 1},\qquad [L_1,L_{-1}]=2L_0.
\end{equation}

The covariance of the $E_a$ maps under these transformations imply that there are three linear combinations among them\footnote{We would like to thank B. Feng for letting us know this typo.}
\be
\sum_{a=1}^n \,\, y_a\, E_a = 0, ~~
\sum_{a=1}^n \,\sigma_a\, y_a\, E_a = 0,~~
\sum_{a=1}^n \,\, y_a^2\, E_a = 0.
\ee 

In order to define local residues in \eqref{intres}, one must remove three of the elements of the map $(\sigma_1,\sigma_2,\ldots ,\sigma_n)\mapsto (E_1,E_2,\ldots ,E_n)$ from $\mathbb{C}^n\mapsto \mathbb{C}^{n}$. This is welcome as one can use the $PSL(2,\mathbb{C})$ group to fix the location of three $\sigma_a$ variables. Using the standard Fadeev-Popov procedure one has 
\be\label{partGF}
{\cal I}=\frac{1}{2^3}\int_{\Gamma} \prod_{a\neq i,j,k}d\sigma_a \prod_{b=1}^n \frac{(y_b\,dy_b)}{(y^2_b-\sigma_b^2+\Lambda^2)} \times \frac{|i,j,k| |p,q,r|}{\prod_{d\neq p,q,r} E_d}H(\sigma,y),
\ee
where the Fadeev-Popov determinants are given by
\be
|p,q,r| = \frac{1}{\Lambda^2} \left|
                            \begin{array}{ccc}
                              y_p & \,\, y_p(\sigma_p+y_p)\, \,&   y_p(\sigma_p-y_p)  \\
                              y_q & \,\, y_q(\sigma_q+y_q) \,\, & y_q(\sigma_q-y_q) \\
                              y_r  & \,\,y_r(\sigma_r+y_r)\,\,& y_r(\sigma_r-y_r) \\
                            \end{array}
                          \right|=
                           \frac{2\,y_p y_q y_r}{\Lambda^2}\left|
                            \begin{array}{ccc}
                              1 & \,\, y_p \,\,& \s_p \\
                              1 &\,\, y_q \,\,& \s_q \\
                              1 &\,\, y_r \,\,& \s_r \\
                            \end{array}
                          \right|,
\ee 
likewise for $|i,j,k|$ and $\Gamma$ is the integration cycle defined  by the solutions of the $2n-3$ equations
\begin{align}\label{cycle_one}
& y_b^2-\s_b^2+\L^2=0, ~ ~  b=1,\ldots n, \\
&  E_d=0, ~~ {\rm with} ~ d= 1,\ldots n ~{\rm and}~ d\neq p,q,r.  
\end{align}
The $2^3$ factor  appears when the $PSL(2,\mathbb{C})$ symmetry is fixed and the $(\mathbb{Z}_2)^3 $ symmetry ($\s_i\rightarrow-\s_i, \s_j\rightarrow-\s_j,\s_k\rightarrow-\s_k$) is broken.

Note that the values of $\sigma_i,\sigma_j$ and $\sigma_k$ have been fixed but  their branches do not, i.e.  $y_i,y_j$ and $y_k$ can still take any of the solutions to $y_b^2-\sigma_b^2+\Lambda^2=0$.

\subsubsection{Promoting $\L$ to variable}

In the previous prescription, \eqref{partGF},  $\L$ is a constant parameter. In this section we show how to introduce $\L$ as a variable.

It is straightforward to check that  the  integral in \eqref{intres}  is invariant by the scale transformation 
\begin{equation}\label{scaletransformation}
(\s_a,y_a,\L)\rightarrow \rho (\s_a,y_a,\L), \,\rho\in \mathbb{C}^* ~ {\rm and}~ a=1,\dots,n, 
\end{equation}
Note that  the $PSL(2,\mathbb{C})$ measure
\begin{equation}\label{mgltwo}
\frac{d\s_i d\s_j d\s_k}{|i,j,k| },
\end{equation}
is also invariant by the scale transformation in \eqref{scaletransformation}.

In order to promote the $\L$  parameter  to a variable we introduce the scale invariant measure $\frac{d\L}{\L}$. Thus,  the  new measure 
\begin{equation}
\frac{d\L}{\L}\frac{d\s_i d\s_j d\s_k}{|i,j,k| },
\end{equation}
is also scale and  $PSL(2,\mathbb{C})$ invariant, i.e $GL(2,\mathbb{C})$  invariant. Clearly, the generators of this $GL(2,\mathbb{C})$ symmetry are given by the elements  $\{L_0,L_{-1},L_1\}$ and the scale generator
\be\label{redun_D}
D = \sum_{a=1}^n \sigma_a\partial_a + \Lambda\partial_\Lambda.
\ee
Its algebra is given by
\begin{equation}
[L_{\pm 1},L_0]=\pm\,L_{\pm 1},\quad [L_1,L_{-1}]=2L_0,\quad [D,L_m]=0,\,\,\,m\in\{-1,0,1\},
\end{equation}
on the support of the algebraic curve $y_a^2 = \s_a^2-\L^2$. 

Now,  note that the denominator in  \eqref{mgltwo} can be written as the following  determinant
\be
\L\,|i,j,k| = \frac{1}{\Lambda^2} \left|
                            \begin{array}{cccc}
                              y_i\,\, & \,\, y_i(\sigma_i+y_i)\, \,& \,\,  y_i(\sigma_i-y_i)\,\,&\,\,\s_i  \\
                              y_j \,\,& \,\, y_j(\sigma_j+y_j) \,\, &\,\, y_j(\sigma_j-y_j) \,\,&\,\,\s_j\\
                              y_k  \,\,& \,\,y_k(\sigma_k+y_k)\,\,&\,\, y_k(\sigma_k-y_k) \,\,&\,\,\s_k\\
                              0 \,\, & \,\,0\,\,&\,\, 0 \,\,&\,\, \L \\
                            \end{array}
                          \right| \equiv \Delta_{\rm FP}(ijk;\L).
\ee 
This determinant  is just the  Fadeev-Popov determinant for the gauge fixing of the three  punctures $(\s_i,\s_j,\s_k)$  and the branch cut variable $\L$.  

Finally, we can rewrite the \eqref{intres} prescription  as
\be\label{general_prescription}
{\cal I}=\frac{1}{{\,\rm Vol}(GL(2,\mathbb{C}))}\int \frac{d\Lambda}{\Lambda}\int d^n\sigma\prod_{b=1}^n\,\frac{(y_b\, d y_b)}{(y^2_b-\sigma_b^2+\Lambda^2)} \times \frac{H(\sigma,y)}{\prod_{d=1}^n E_d}  .
\ee
Fixing the $\{E_p,E_q,E_r\}$ scattering equations , the $(\s_i,\s_j,\s_k)$ punctures  and the $\L$ branch cut variable one obtains 
\be\label{fullGF}
{\cal I}={1\over 2^3 }\int_{\Gamma} \prod_{a\neq i,j,k}d\sigma_a \prod_{b=1}^n \frac{(y_b\, dy_b)}{(y^2_b-\sigma_b^2+\Lambda^2)}\times \frac{\Delta_{\rm FP}(ijk;\L)\,\, |p,q,r|}{\L}\times\frac{H(\sigma,y)}{\prod_{d\neq p,q,r} E_d},
\ee
which is the same expression as in \eqref{partGF}.

\subsection{Equivalence with the CHY Construction}\label{ECHY}

The idea of this section is to show how the \eqref{intres}  prescription is in fact equivalent to the original CHY approach.

Let us define a map from the double-cover version of the sphere into a single cover of $\mathbb{CP}^1$. This  should take us back to the original CHY construction. Such a map is very well known and it is given by
\be\label{mapcpone}
\sigma_a = \frac{\Lambda}{2}\left(z_a+\frac{1}{z_a}\right),
\ee
where $\L\neq\{0,\infty\}$ is a constant and $z_a$ are the coordinates on $\mathbb{C}P^1$ (CHY coordinates).  The first observation is that if all the punctures are located on the same branch, say the upper sheet, i.e.  $y_a=+\sqrt{\s_a^2-\L^2}$, then
\be
\tau_{a:b}=\frac{1}{2}\left(\frac{y_a+y_b+\sigma_{ab}}{y_a}\right)\frac{1}{\sigma_{ab}} = \left(\frac{2}{\L} \right)\,\,\frac{z_a^2}{ (z_a^2-1)}\times \frac{1}{z_{ab}}.
\ee
In this expression it is easy to see that the lack of antisymmetry in the labels translate into an overall factor in the $z_a$ variables. Also it is  simple to show
\be
d\sigma_a = \left(\frac{\L}{2}\right)\,\,\frac{ (z^2_a-1)}{z_a^2}\,\, dz_a,
\ee
which means that 
\be
\frac{1}{2}\left(\frac{y_a+y_b+\sigma_{ab}}{y_a}\right)\frac{d\sigma_a}{\sigma_{ab}}  = \frac{dz_a}{z_{ab}}.
\ee
This is indeed the natural differential form on $\Sigma$ with simple poles at $(\sigma_a,y_a) = (\sigma_b,y_b)$ and at $\sigma_a=\infty$ with residues $1$ and $-1$ respectively.

Therefore, it is straightforward to conclude 
\begin{align}
H(\s,y) &= \left(\frac{2}{\L}\right)^{2n}\left(\prod_{a=1}^n\frac{z_a^2}{z_a^2-1}\right)^2 H(z)\\
E_a(\s,y)&=\sum_{i\neq a}^n s_{ai}\,\tau_{a:i}=\left(\frac{2}{\L}\right)\left(\frac{z_a^2}{z_a^2-1}\right)\sum_{i\neq a}^n \frac{s_{ai}}{z_{ai}}=\left(\frac{2}{\L}\right)\left(\frac{z_a^2}{z_a^2-1}\right)\,E_a(z),
\end{align}
where $H(z)$ is as in \eqref{general_H}.

Carrying out the integration over the $y_a$ variables on the contour given by the solutions $y_a=+\sqrt{\s_a^2-\L^2}$ and performing the map \eqref{mapcpone}, then \eqref{intres} becomes 
\begin{eqnarray}\label{CHYequiv}
&&\left.\frac{1}{{\,\rm Vol}(PSL(2,\mathbb{C}))}\int \prod_{a=1}^n\frac{(y_a\, d y_a) }{(y^2_a-\sigma_a^2+\Lambda^2)} \left(d^n\s\,\frac{H(\sigma,y)}{\prod_{b=1}^n E_b(\s,y)}\right) \right|_{y_a=+\sqrt{\s_a^2-\L^2}}\\
&&=\left(\frac{1}{2^n}\right)\frac{1}{\volpsl}\int\prod_{a=1}^n dz_a  \frac{H(z)}{\prod_{b=1}^n E_b(z)},\nonumber
\end{eqnarray}
where the $\frac{1}{2^n}$ factor comes from the integral
\begin{equation}
\left.\int \prod_{a=1}^n\frac{(y_a\, d y_a) }{(y^2_a-\sigma_a^2+\Lambda^2)}\right|_{y_a=+\sqrt{\s_a^2-\L^2}}=\frac{1}{2^n}.
\end{equation}

Computing the integral over all possible configurations,  this means the $2^n$ way of choosing $(y_1=\pm\sqrt{\s_1^2-\L^2},...,y_n=\pm\sqrt{\s_n^2-\L^2})$, and performing the map \eqref{mapcpone}, one obtains
\begin{eqnarray}
&&\frac{1}{{\,\rm Vol}(PSL(2,\mathbb{C}))}\int \prod_{a=1}^n\frac{(y_a\, d y_a) }{(y^2_a-\sigma_a^2+\Lambda^2)} \left(d^n\s\,\frac{H(\sigma,y)}{\prod_{b=1}^n E_b(\s,y)}\right)\\
&&=\frac{1}{\volpsl}\int\prod_{a=1}^n dz_a  \frac{H(z)}{\prod_{b=1}^n E_b(z)}.\nonumber
\end{eqnarray}
This result agrees with the original CHY formula.

\section{New Gauge Fixing}\label{new gauge fixing}

In this section we find that by using the full $GL(2)$ redundancy one can gauge fix the location of  four  punctures, modulo branches. Thus, promoting $\Lambda$ to a variable to be fixed by the scattering equations, one has the possibility of using a global  residue theorem \cite{harris}  that leads to a new diagrammatic expansion of general amplitudes. Moreover, the residue theorem allows the analytic evaluation of integrals with rational functions whose answers have non-local poles and thus are hard to obtain by other means. 

\subsection{New Gauge Fixing}

Let us start by reviewing the generations of the $GL(2,\mathbb{C})$ redundancy, as we did in \eqref{redun} in section \ref{sthreedc},
\be\label{redun2}
L_{\pm 1} = \sum_{a=1}^n \frac{1}{\Lambda}y_a(\sigma_a \mp y_a)\partial_a, \qquad  L_0 = \sum_{a=1}^n y_a\partial_a, \qquad D = \sum_{a=1}^n \sigma_a\partial_a + \Lambda\partial_\Lambda,
\ee
where $y_a^2=\s_a^2-\L^2$.
Since all four vectors act on $\sigma$'s one can use them to fix four of the punctures' $\sigma$. For simplicity of notation let us assume that they are $\sigma_1,\sigma_2,\sigma_3$ and $\sigma_4$. The Fadeev-Popov jacobian, $\Delta_{\rm FP}$, is now
\be
\Delta_{\rm FP}(1234) = \frac{1}{\Lambda^2} {\rm det}\left(
                            \begin{array}{cccc}
                              y_1\,\, &\,\,y_1(\sigma_1+y_1)\,\,  & \,\, y_1(\sigma_1-y_1)\,\,  &\,\, \s_1 \\
                             y_2\,\, &\,\,y_2(\sigma_2+y_2)\,\,  & \,\, y_2(\sigma_2-y_2)\,\,  &\,\, \s_2 \\  
                               y_3\,\, &\,\,y_3(\sigma_3+y_3)\,\,  & \,\, y_3(\sigma_3-y_3)\,\,  &\,\, \s_3 \\
                                y_4\,\, &\,\,y_4(\sigma_4+y_4)\,\,  & \,\, y_4(\sigma_4-y_4)\,\,  &\,\, \s_4 \\
                            \end{array}
                          \right). 
\ee
In addition to this one still has to remove three elements from the map $\{ E_1,E_2,\ldots ,E_n\}$. This procedure is not affected by the new gauge choice and the formula used in \eqref{partGF} is still valid. Putting all together and removing, without loss of generality, the scattering equations $E_1,\,E_2$ and $E_3$  we arrive at the new formula
\be\label{newGF}
{\cal I}={1\over 2^3}\int_{\Gamma} \frac{d\Lambda}{ \Lambda} \prod_{a=5}^n d\sigma_a \prod_{b=1}^n\frac{(y_b\, dy_b)}{(y^2_b-\sigma_b^2+\Lambda^2)} \times\frac{\,\Delta_{\rm FP}(1234)\,|1,2,3|\;   }{\prod_{d=4}^n E_d}H(\sigma,y),
\ee
where $\Gamma$ is the integration cycle defines as in \eqref{cycle_one}, given by the equations 
\begin{align}\label{cycle_two_Gamma}
& y_b^2-\s_b^2+\L^2=0, ~ ~  b=1,\ldots n, \\
&  E_d=0, ~~ {\rm with} ~ d= 4,\ldots n,\nonumber 
\end{align}
for the $2n-3$ variables $(\L,\s_5,\ldots,\s_n,y_1,y_2,\ldots ,y_n)$. 

It is interesting to note that the opening of the branch cut connecting the two branches (sheets)  becomes  a function of the kinematic invariants $k_a\cdot k_b$. This means that as we move in the space of kinematic invariants the branch cut also moves. This is what makes factorization and crossing natural properties to address using this formulation.

\section{Residue Theorem and Diagrammatic Expansion}\label{residuetheorem}

The equations obtained at the end of the previous section are polynomial equations of increasing degree as the number of particles increases. In fact, the equations \eqref{cycle_two_Gamma} lead to higher degree polynomials than the original CHY scattering equations. This seems to be an obstacle. However, using a residue theorem we will effectively replace one of the $E_a=0$ equations by the equation $\Lambda=0$. This might come as a surprise as closing the cut is intuitively related to a factorization limit. Instead, what we will see is that once the cut closes a new puncture appears that represents an off-shell particle. The sum over solutions to the equations $y_b^2 = \sigma_b^2-\Lambda^2$ give rise to $y_b = \pm \sigma_b$ and determine the branch location of the $b^{\rm th}$-puncture. For a given distribution of particles, say a subset $U$ ($L$) is on the upper (lower) branch, the equation $E_a$ that was eliminated gives rise to the propagator $1/P_U^2$ where $P_U$ is the sum over the momenta of all external particles on the upper branch. In this way, the integral  given in \eqref{newGF} 
\be\label{general_integral}
{\cal I}\,=\,{1\over 2^3}\int_\Gamma \frac{d\Lambda}{ \Lambda}  \prod_{a=5}^n d\sigma_a \prod_{b=1}^n \frac{(y_b\,dy_b)}{(y^2_b-\sigma_b^2+\Lambda^2)}\times \frac{ \Delta_{\rm FP}(1234)\,\, |1,2,3| }{\prod_{d=4}^n E_d(\s,y)}H(\sigma,y)
\ee
becomes a sum of products of contour integrals with smaller number of particles. By iterating the process we find a diagrammatic  description. The most important outcome is that at each step in the iteration process the degree of the scattering equation is lowered and analytic evaluations become possible.

\subsection{Residue Theorem}

Following the similar ideas as in section \ref{ECHY} les us consider the general integral
\be\label{Gintegral}
{\cal I}\,=\frac{1}{{\rm Vol} (GL(2,\mathbb{C}))}\int_\gamma d\Lambda\left(\prod_{b=1}^n \frac{\,dy_b}{(y^2_b-\sigma_b^2+\Lambda^2)} \right) \left(  \prod_{a=1}^n \frac{ d\sigma_a}{ \tilde E_a(\s,y)} \right)\frac{H(\s,y) \prod_{c=1}^n y_c^2}{\L},
\ee
where $\gamma$ is the contour defined by the equations
\begin{equation}
y_a^2-\s_a^2+\L^2=0,~~ \tilde E_a(\s,y)=\sum_{b\neq a}\frac{k_a\cdot k_b}{\s_{ab}}(y_b+y_a) =0,~~{\rm for}~~ a\in\{1,2,\ldots,n\}, 
\end{equation}
and $H(\s,y)\prod_{c=1}^n y_c^2/\L$ is the integrand. Clearly there are  more integration variables than contours,  nevertheless,  when the  $GL(2,\mathbb{C})$ symmetry is fixed then the number of integration variables becomes equal to the contour cycles. 

Note that \eqref{Gintegral} depends over the $\L$ variable by the expression
\begin{equation}
\int \frac{ d\Lambda}{\prod_{b=1}^n (y^2_b-\sigma_b^2+\Lambda^2)} \times \frac{1}{\L},
\end{equation}
where $\prod_{b=1}^n (y^2_b-\sigma_b^2+\Lambda^2)$ defines the integration cycle.  Naively, using the global residue theorem, it is straightforward to see  that the previous expression can be written as 
\begin{equation}
\int \frac{ d\Lambda}{\prod_{b=1}^n (y^2_b-\sigma_b^2+\Lambda^2)} \times \frac{1}{\L} = -\int \frac{ d\Lambda}{\L\,\,\prod_{b\neq l}^n (y^2_b-\sigma_b^2+\Lambda^2)} \times \frac{1}{y^2_l-\sigma_l^2+\Lambda^2}, 
\end{equation}
where $\L\,\, \prod_{b\neq l}^n (y^2_b-\sigma_b^2+\Lambda^2)$ defines the new contour and $1/(y^2_l-\sigma_l^2+\Lambda^2)$ becomes part of the integrand.  

Nevertheless, in order to apply the global residue theorem one must also verify  if the point at infinity is a pole.  One says that the \eqref{Gintegral} integral has a pole at infinity if and only if \cite{harris}
\begin{equation}\label{infty_condition}
{\rm deg}(g)> (d_1+\cdots + d_{2n})-((2n+1)+1), 
\end{equation}
where $g(\s,y,\L)$ is the integrand
\begin{equation}
g(\s,y,\L)= \frac{H(\s,y) \prod_{c=1}^n y_c^2}{\L}\,\,\,\Rightarrow\,\,\, {\rm deg}(g) = -1,
\end{equation}
$d_1+\cdots + d_{2n}$ is sum over all  degrees of  the polynomials that define the integration contour, i.e.
\begin{equation}
d_1+\cdots + d_{2n}={\rm deg}\left(\prod_{b=1}^n (y^2_b-\sigma_b^2+\Lambda^2)\,\tilde E_b\right)=2n
\end{equation}
and $2n+1$ is the number of integration variables, i.e. $(\L,\s_1,\ldots,\s_n,y_1,\ldots,y_n)$.  Clearly  \eqref{Gintegral}  has a pole at infinity  and it  must be integrated when the global residue theorem is performed.  

Since the integrand in  \eqref{Gintegral} is not well defined when $\L=0$, then this implies that  the \eqref{Gintegral} integrand is given on the $\L\neq 0$ chart . Thus,   so as to explore the pole at infinity we consider the following  transformation
\begin{equation}
\L\, \rightarrow  \, \L^\prime=\frac{1}{\L},\quad  y_i \, \rightarrow  \,  y_i^\prime=\frac{y_i}{\L^2},\quad  \s_i \, \rightarrow  \,   \s_i^\prime=\frac{\s_i}{\L^2}.
\end{equation}
Under this transformation  \eqref{Gintegral}  becomes invariant, i.e.
\begin{equation}
{\cal I}=\frac{1}{{\rm Vol} (GL(2,\mathbb{C}))}\int_{\gamma^{\prime}} d\Lambda^\prime\left(\prod_{b=1}^n \frac{\,dy^\prime_b}{(y^{\prime\, 2}_b-\sigma_b^{\prime\,2}+\Lambda^{\prime\,2})} \right) \left(  \prod_{a=1}^n \frac{ d\sigma_a^\prime}{ \tilde E_a(\s^\prime,y^\prime)} \right)\frac{H(\s^\prime,y^\prime) \prod_{c=1}^n y_c^{\prime\,2}}{\L^\prime},
\end{equation}
where $\gamma^\prime$ is the contour defined by the equations
\begin{equation}
y_a^{\prime\, 2}-\s_a^{\prime\, 2}+\L^{\prime \,2}=0,~~ \tilde E_a(\s^\prime,y^\prime)=\sum_{b\neq a}^n \,\frac{k_{a}\cdot k_b }{\s^\prime_{ab}} (y_a^\prime+y_b^\prime)=   0, 
\end{equation}
and $H(\s^\prime,y^\prime)$ is defined  with the $\tau_{a:b}^\prime$'s forms
\begin{equation}
\tau_{ab}^\prime = \frac{1}{2\,\, y_a^\prime}\left(\frac{y_a^\prime+y_b^\prime+\s_{ab}^\prime}{\s_{ab}^\prime}\right).
\end{equation}
Note that the minus sign $d\L/\L\rightarrow -d\L^\prime/\L^\prime$ is used to  reorient the $\L$ contour.  Finally, we can now integrate around the point $\L^\prime=0$ which is the pole at infinity, therefore performing the global residue theorem the \eqref{Gintegral} integral could be read as
\begin{equation}\label{Gintegral2}
{\cal I}=\frac{-2}{{\rm Vol} (GL(2,\mathbb{C}))}\int_{\tilde\Gamma} \frac{d\Lambda}{\L}\left(\frac{\prod_{i=1}^n dy_i}{\prod_{b\neq l}^n(y^{2}_b-\sigma_b^{2}+\Lambda^{2})} \right) \left(  \prod_{a=1}^n \frac{ d\sigma_a}{ \tilde E_a(\s,y)} \right)\frac{H(\s,y) \prod_{c=1}^n y_c^{2}}{(y^{2}_l-\sigma_l^{2}+\Lambda^{2})},
\end{equation}
where the new $\tilde\Gamma$ contour is now defined by the $2n$ equations
\begin{equation}
\L=0,~~\, y_b^2-\s_b^2+\L^2=0,~~{\rm for}~~ b\neq l,~~ \tilde E_a(\s,y)=0,~~{\rm for}~~ a\in\{1,2,\ldots,n\}.
\end{equation}

\subsection{Recovering The Curve}
Note that in the result obtained  in \eqref{Gintegral2} the  $y_l^2-\s_l^2+\L^2=0$ constraint is lost, i.e. we are not anymore on the support of the curve $y_l^2=\s_l^2-\L^2$.   Since our aim is to be on a sphere  then this constraint must be recovered.  In order to get back the  $y_l^2-\s_l^2+\L^2=0$ equation we perform the residue theorem but now using the $y_l$ variable. 

Before applying the residue theorem  it is useful to remember the following, first, in full view,  the integration contour is defined by polynomials over the $y_i$'s variables 
$$
\L\,\prod_{b\neq l}^n(y^{2}_b-\sigma_b^{2}+\Lambda^{2}))\prod_{a=1}^n\tilde E_a(\s,y)=\L \,\prod_{b\neq l}^n(y^{2}_b-\sigma_b^{2}+\Lambda^{2}))\prod_{a=1}^n \left[ \sum_{i\neq a}^n s_{ab}  \left(   \frac{y_a+y_i}{\s_{ai}}          \right)    \right],
$$
and secondly,  the integrand 
$$
{H(\s,y) \prod_{c=1}^n y_c^{2}\over (y^{2}_l-\sigma_l^{2}+\Lambda^{2})},
$$
has just one singularity over the $y_i$'s variables given by $(y_l^2-\s_l^2+\L^2)$. With this in mind, we are ready to use the residue theorem over $y_l$.  The integral by the $y_l$ variable is read as 
\begin{equation}
\int_{\tilde \Gamma} \frac{d y_l}{\prod_{a=1}^n \tilde E_a}\times {H(\s,y) \prod_{c=1}^n y_c^{2}\over (y^{2}_l-\sigma_l^{2}+\Lambda^{2})},
\end{equation}
so, performing the residue theorem one obtains\footnote{There is no contribution from the point at infinity.} 
\begin{equation}
\int_{\tilde \Gamma} \frac{d y_l}{\prod_{a=1}^n \tilde E_a}\times {H(\s,y) \prod_{c=1}^n y_c^{2}\over (y^{2}_l-\sigma_l^{2}+\Lambda^{2})}=
-\int_\Gamma \frac{d y_l}{(y^{2}_l-\sigma_l^{2}+\Lambda^{2})\,\prod_{a\neq l}^n \tilde E_a}\times {H(\s,y) \prod_{c=1}^n y_c^{2}\over \tilde E_l},
\end{equation}
where $\Gamma$ is the  new integration contour  defined by the $2n$ equations
\begin{equation}
\L=0,~~\, \tilde E_b(\s,y)=0,~~{\rm for}~~ b\neq l,~~\, y_a^2-\s_a^2+\L^2 =0,~~{\rm for}~~ a\in\{1,2,\ldots,n\}.
\end{equation}
Finally, the \eqref{Gintegral} integral is written as 
\begin{equation}
{\cal I}=\frac{2}{{\rm Vol} (GL(2,\mathbb{C}))}\int_{\Gamma} \frac{d\Lambda}{\L}\left(\prod_{b=1}^n \frac{dy_b}{(y^{2}_b-\sigma_b^{2}+\Lambda^{2})} \right) \left(   \frac{\prod_{a=1}^n d\sigma_a}{\prod_{i\neq l}^n  \tilde E_i(\s,y)} \right)\frac{H(\s,y) \prod_{c=1}^n y_c^{2}}{\tilde E_l}\nonumber.\\
\end{equation}
Fixing the $(\s_m,\s_n,\s_p,\s_q)$ puctures and the $(E_m,E_n,E_q)$ scattering equations the above integral becomes

\begin{equation}\label{Gintegral3}
{\cal I}=\frac{1}{2^2}\int_{\gamma} \left(\frac{d\Lambda}{\L}\right)\left(\prod_{b=1}^n \frac{(y_b\,\, dy_b)}{(y^{2}_b-\sigma_b^{2}+\Lambda^{2})} \right) \left(  \prod_{a\neq m,n,p,q} \frac{ d\sigma_a}{  E_a(\s,y)} \right)\frac{|m,n,q|\Delta_{\rm FP}(mnq,p)H(\s,y) }{ E_p},
\end{equation}
where the contour $\gamma$ is defined by the equations
\begin{equation}
\L=0,~~\, \tilde E_b(\s,y)=0,~~{\rm for}~~ b\neq \{m,n,p,q\},~~\, y_a^2-\s_a^2+\L^2 =0,~~{\rm for}~~ a=1,2,\ldots,n.
\end{equation}
We call this integral the {\bf $\L-$prescription}.

Note that we have chosen the same labels for the  punctures and scattering equations, this will be useful when we will formulate the $\L-$algorithm in section \ref{Lalgorithm}.

The $\L-$prescription, \eqref{Gintegral3},   recovered the support on the curves $y_a^2=\s_a^2-\L^2$, moreover, it must  be computed around the cycles  $\L\rightarrow 0$ and $\L\rightarrow \infty$, 
which are exactly the same, as one can see  in {\bf Fig. 5.1}
\begin{center}
\includegraphics[scale=0.4]{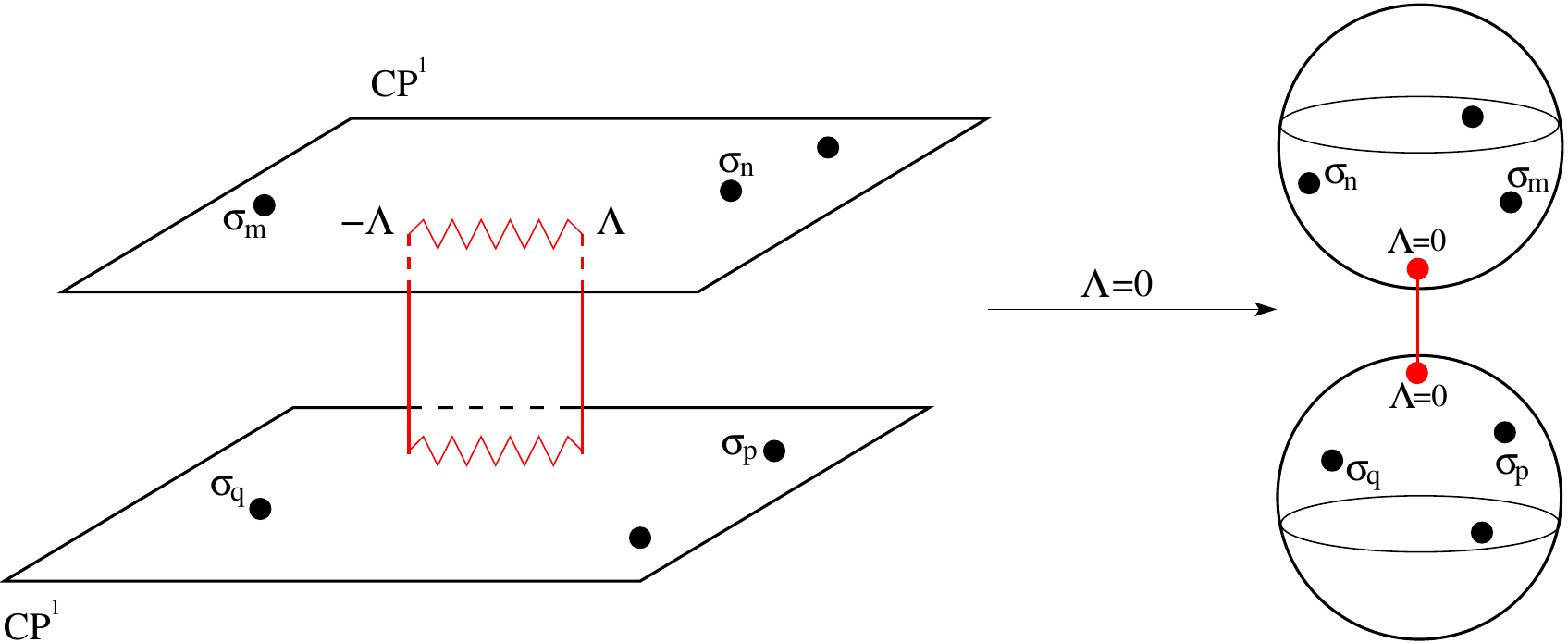}\,,
\begin{center}
(a)
\end{center}
\end{center}
\begin{center}
\includegraphics[scale=0.4]{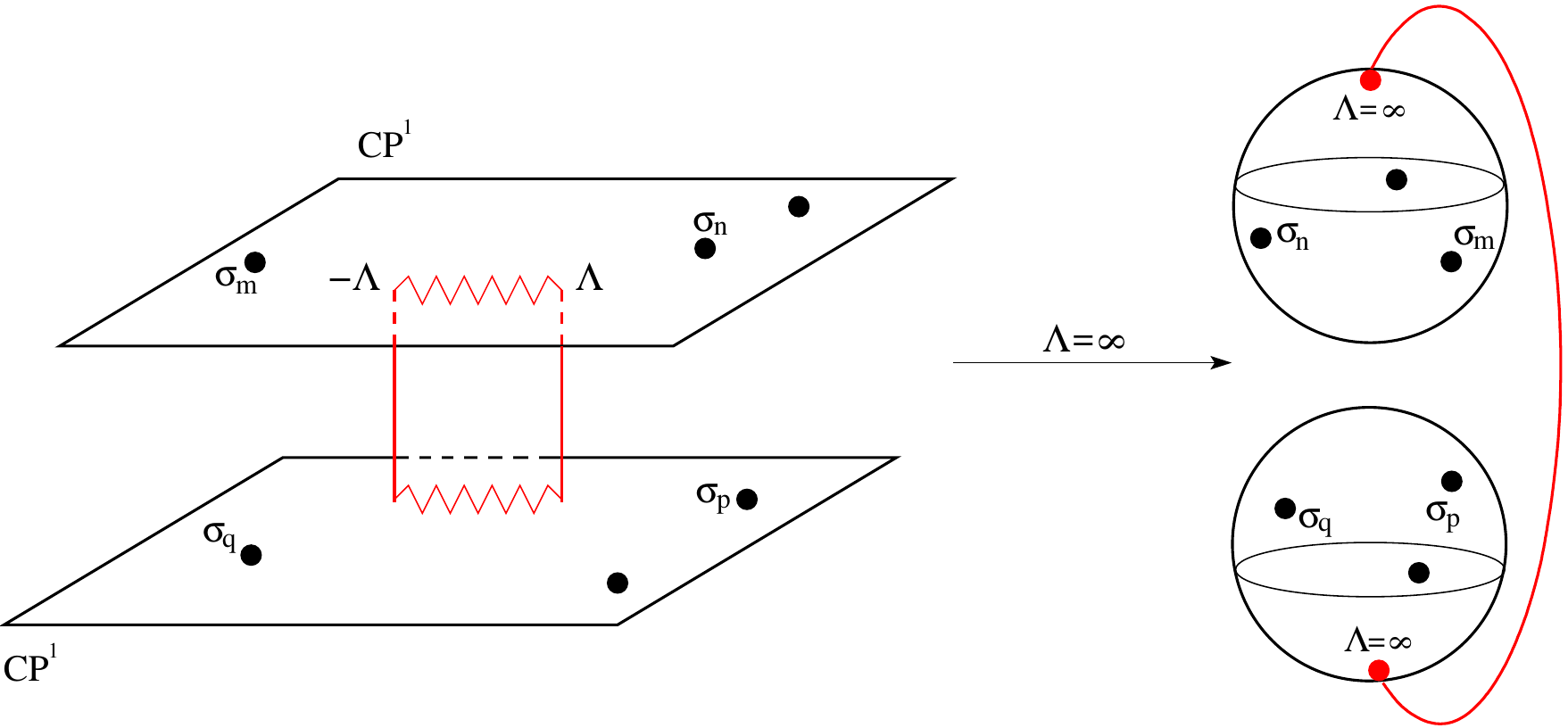}\,,
\begin{center}
(b)\\
{\bf Fig. 5.1} (a) Limit $\L\rightarrow 0$.  (b) Limit $\L\rightarrow \infty$.
\end{center}
\end{center}
In addition, one of the $(n-3)$ scattering equations becomes free, i.e. it now is part of the integrand, in \eqref{Gintegral3} it is $E_p$. 

This new point of view gives us  a new kind of diagrammatic representation,  {\bf Fig. 5.1}.  In the next section we will learn to use this new prescription  and we will propose a new algorithm (the $\L-$algorithm).

\section{ $\L$-Diagrams and  A New Algorithm}\label{mainsection}

Here we present a new algorithm,  which is a consequence of the new prescription  given in \eqref{Gintegral3}.

Before formulating the algorithm we introduce some notations. Let us remember the $s_{a_1\ldots a_n}$ Mandelstam variables are defined as\footnote{We have introduced the $(1/2)$ factor for 
convenience}
\begin{equation}
s_{a_1\ldots a_n}:={1\over 2}(k_{a_1}+\cdots + k_{a_n})^2.
\end{equation}
Nevertheless, it will be  useful for us  to use  the variables 
\begin{equation}
k_{a_1\ldots a_n}:=\sum_{a_i<a_j}^n k_{a_i}\cdot k_{a_j},
\end{equation}
Clearly, when the particles are massless, i.e. $k_i^2=0$, then $s_{a_1\ldots a_n}= k_{a_1\ldots a_n}$.

In the next two  section, \ref{colour} and \ref{ltheorem},  we give all tools  to formulate our new algorithm in section \ref{Lalgorithm}.  While we develope the sections \ref{colour} and \ref{ltheorem},  we apply all these tools on a simple and particular example and at the end we obtain the  result  for the \eqref{Gintegral3} integral.

\subsection{More Notations and a Simple Example}\label{colour}

In the same way as in \cite{humbertoF}, any $H(\s,y)$ integrand over ${\cal M}_{0,n}$ can be written as a linear combinations of integrands with  no zeros , i.e. integrands with just $2n-\tau_{a:b}$ factors. We call this kind of integrands as\footnote{The $D$ letter means that there are only $\sigma_a$ factors into denominator.} $H^D(\s)$. Each $H^D(\s)$ integrand has  associated  a 4-regular graph\footnote{A $G$ graph  is defined by the two finite sets, $V$ and $E$.  $V$ is the vertex set and $E$ is the edge set.}  (bijective map), which we denoted by $G=(V_G,E_G)$ \cite{humbertoF,graph1,graph2}.  The vertex set of $G$ is given by the $n$-labels (punctures) 
$$
V_G=\{1,2,\ldots,n\}
$$ 
and the edge set is given by the elements $\tau_{a:b}\,\leftrightarrow\,\overline{\,a\,b\,}\,,\,$ i.e.
$$
E_G = \{\,\,\overline{\,a\,b\,}\,\,/\,\,\tau_{a:b}\,\, \,\,\text{is a factor into the} ~~ H^D(\s)~~ \text{ integrand.}\,\,\}.
$$
Since $\tau_{a:b}$ always appears into a chain, for instance, let us remember the smallest chain is given by
\begin{equation}
\tau_{a:b}\tau_{b:a},
\end{equation}
 then the graph is not a directed graph,  as well as in \cite{humbertoF}.

For example, let us consider the integrand
\be\label{fexample}
H^D_4(1,2,3,4)=[1234]\times [1234]\,,
\ee
where the $[\,\cdot\,]$ bracket is defined as
\be
[1234]=(\tau_{1:2}\tau_{2:3}\tau_{3:4}\tau_{4:1})\,\,.
\ee 
This integrand is represented by the  $G$ graph
\begin{center}
\includegraphics[scale=0.35]{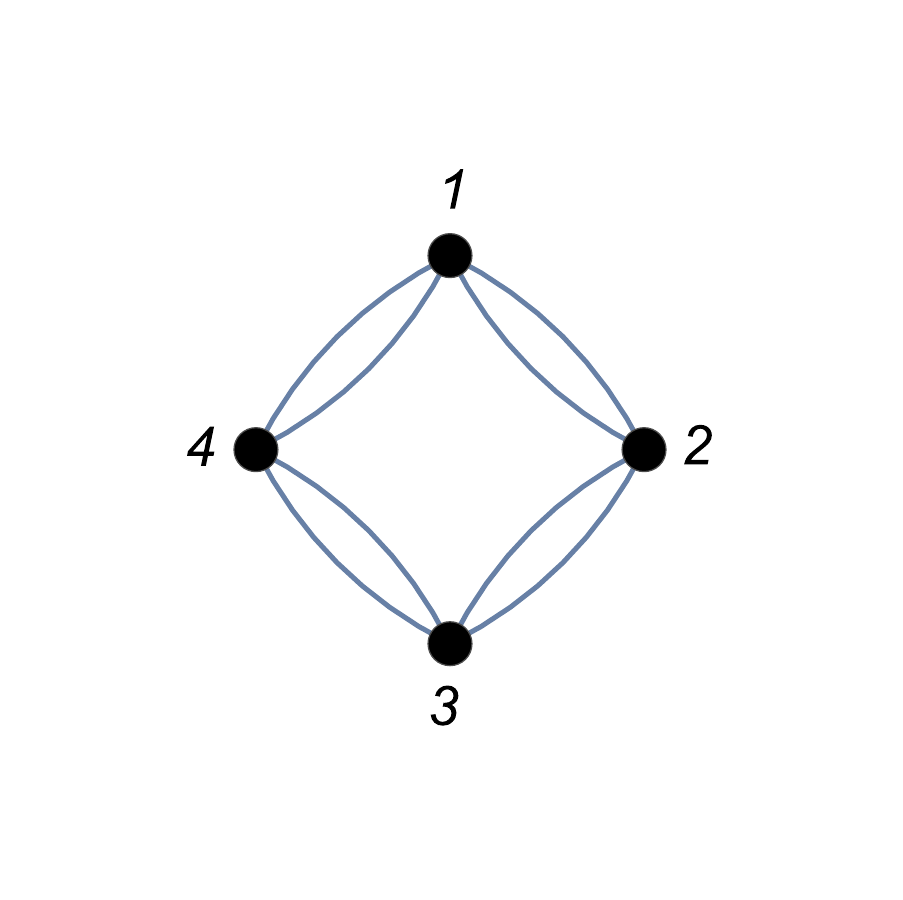}\,.
\begin{center}
({\bf Fig.6.1})\,{\small {\rm \,$G$ graph associated  to the integrand in \eqref{fexample}.}}
\end{center}
\end{center}

This is useful to clarify that the $G$ graph must be draw such that the number of intersection among the edges is as small as possible.

Note that the $G$ graph does not have any information of the $GL(2,\mathbb{C})$ symmetry and the $\L$ parameter or branch cut. In order to introduce this information on the graph we  coloured the vertex set in the following way

\begin{center}
\includegraphics[scale=0.35]{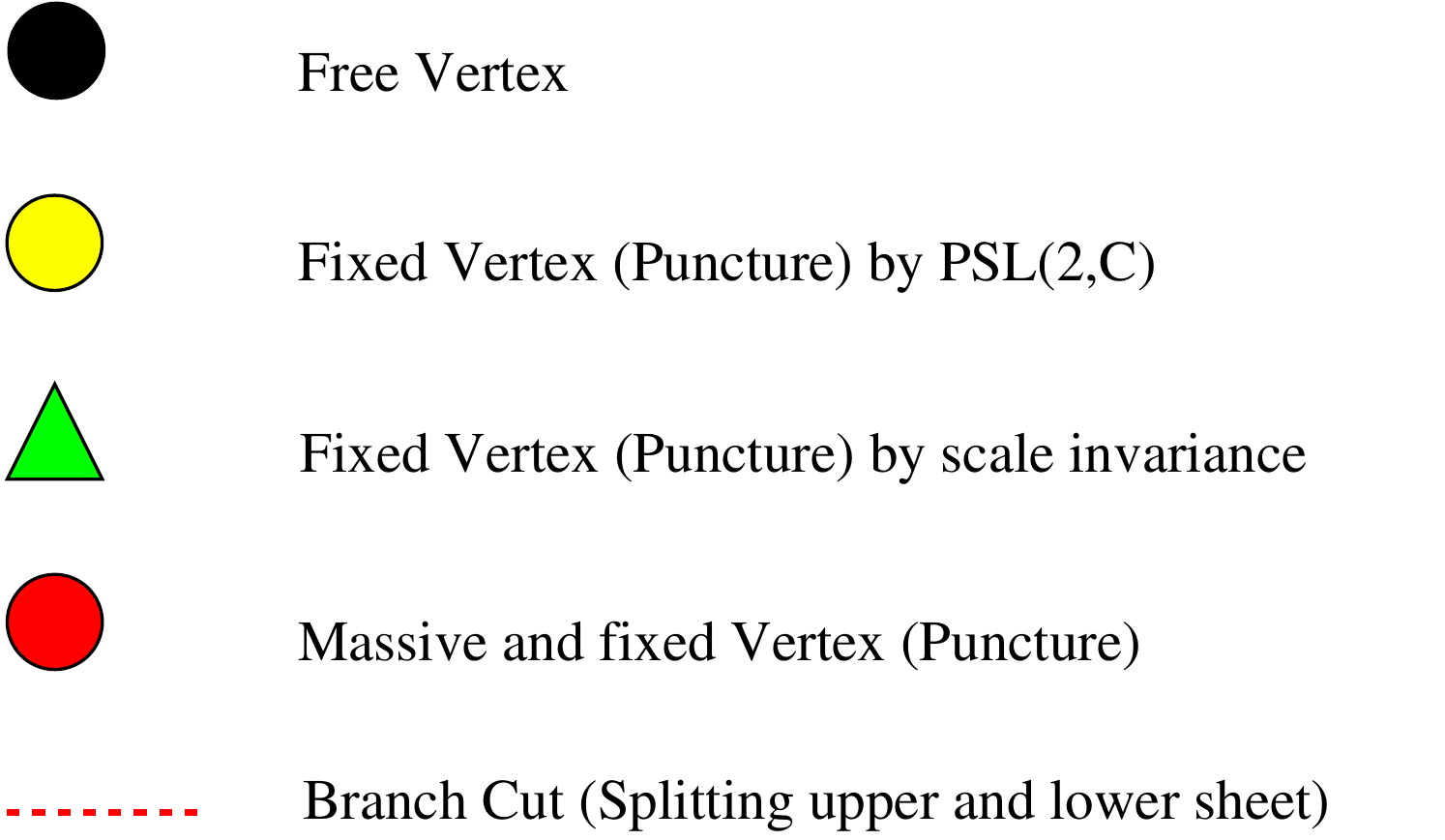}\,.
\begin{center}
({\bf Fig.6.2})\,{\small {\rm \,Coloured Vertices. }}
\end{center}
\end{center}
The $G$ graph can now contain the whole information of the integrand, i.e, it now represents the total integrand  ${\cal I}=|ijk|\Delta_{FP}(ijk,d) H(\s)$.

For example, using the $PSL(2,\mathbb{C})$ symmetry to fix the $(\s_1,\s_2,\s_3)$ punctures  and  the scale symmetry to fix the $\s_4$ puncture, the graph in  {\bf Fig.6.1}  becomes
\begin{center}
\includegraphics[scale=0.4]{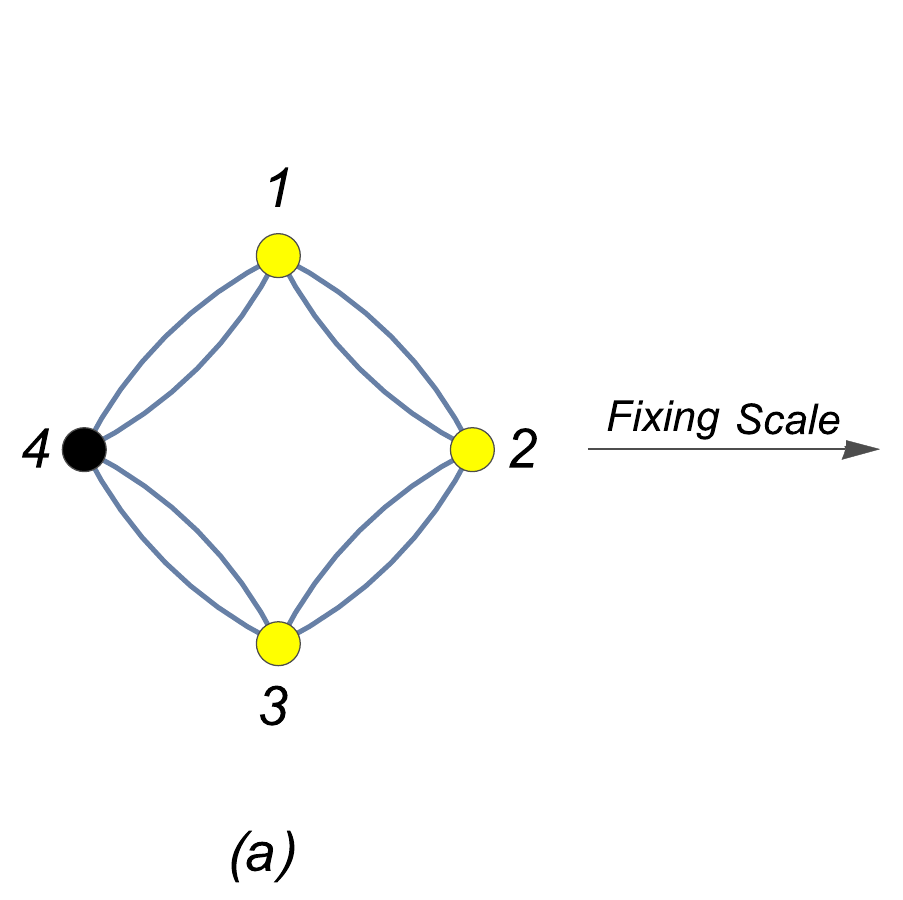}
\includegraphics[scale=0.4]{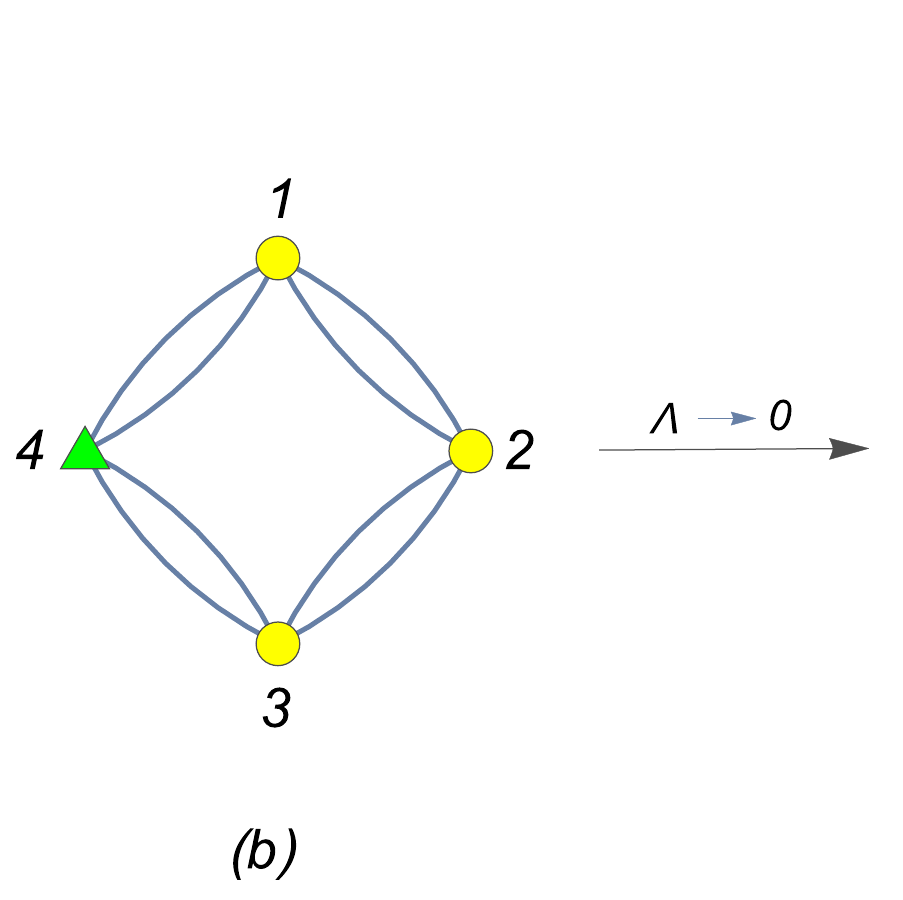}
\includegraphics[scale=0.4]{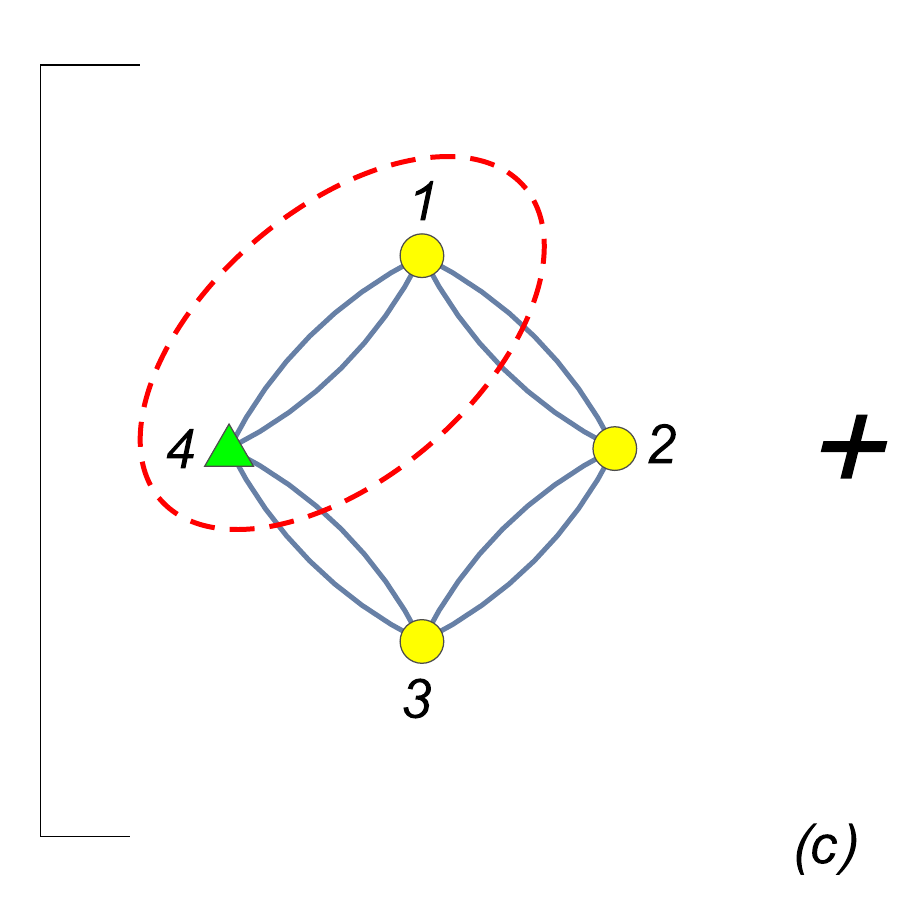}
\includegraphics[scale=0.4]{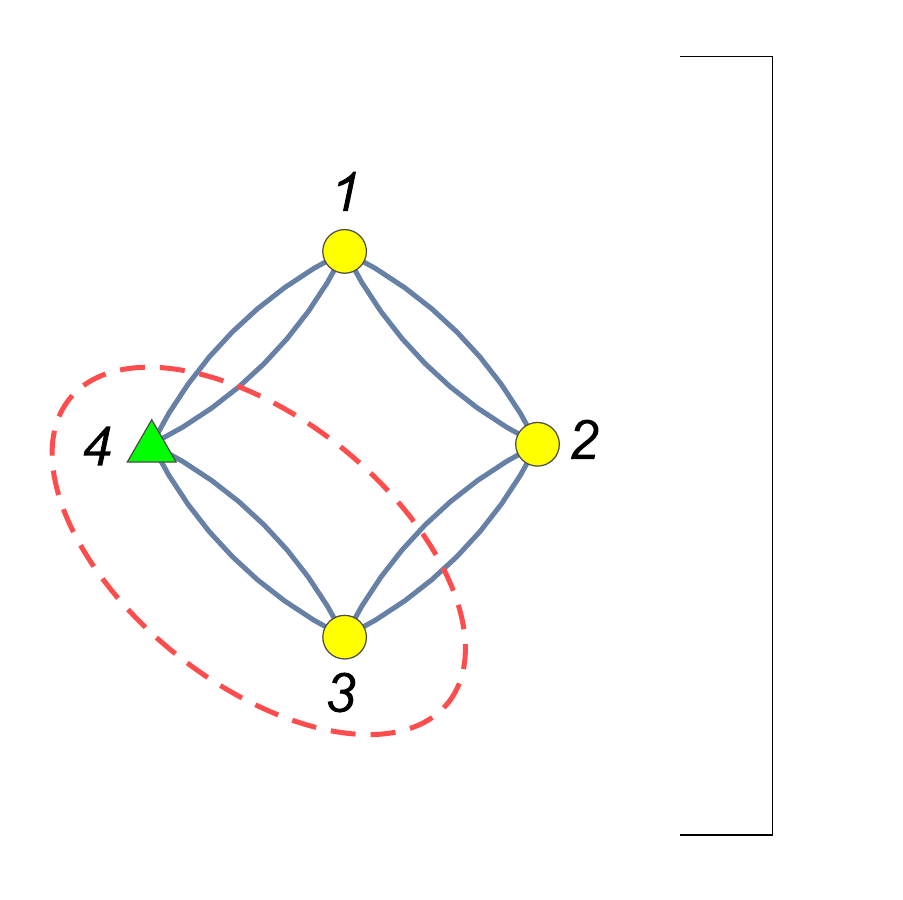},
\begin{center}
({\bf Fig.6.3})\,{\small {\rm \,(a) Fixing $(\s_1,\s_2,\s_3)$ from the $PSL(2,\mathbb{C})$ symmetry. (b) Fixing $\s_4$ from the scale symmetry. (c) Possibles contribution after performing the $ \Lambda$ integral.}}
\end{center}
\end{center}
where  {\bf Fig.6.3(c)} shows the whole possibles non-zero contributions or configurations, up to $\mathbb{Z}_2$ symmetry $y_a\,\rightarrow\, -y_a$, after performing the $\L$ integral  around  $\Lambda=0$. It is explained in detail in the next section.  

\subsubsection{Configurations and $\L-$Theorem}\label{ltheorem}

Although we previously have already used the word ``{\bf  configuration}", in this section we give a formal definition.  So,  the first thing we do in this section is to define what is  a configuration 
\begin{itemize}
\item {\bf Configuration}:   A configuration, which we denoted by $C$, is the integration over the $(y_1,\ldots y_n)$ variables around  one of the $2^n$ solutions of the equations
\begin{equation}
y_a^2-\s_a^2+\L^2=0,~~~{\rm for}~a=1\ldots n.
\end{equation}
\end{itemize}

This definition means that a $C$ configuration is the choosing of the $2^n$ possibilities given by $(y_1=\pm\sqrt{\s_1^2-\L^2},...,y_n=\pm\sqrt{\s_n^2-\L^2})$, i.e.  a configuration fixed the punctures on the upper or lower sheet.

Now, with this in mind we are ready to come back to our example and  note that besides of two configurations given in {\bf Fig.6.3(c)},  there are more possibles configurations  (up to $\mathbb{Z}_2$ symmetry)  such as
\begin{center}
\includegraphics[scale=0.4]{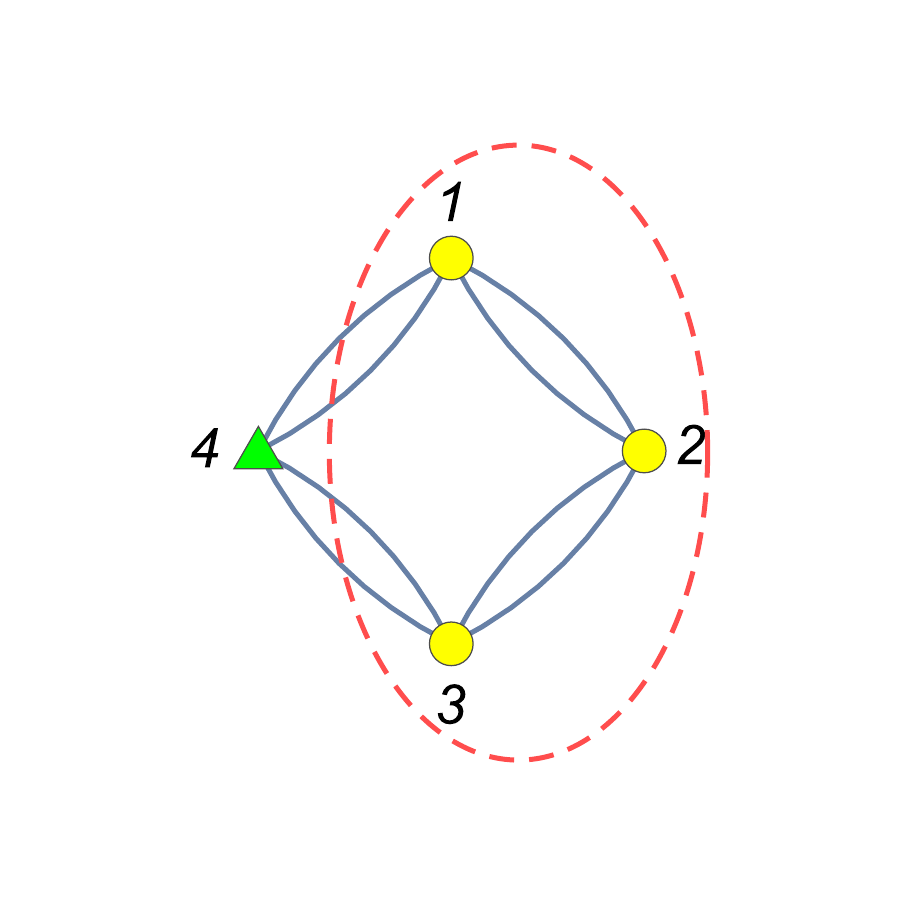}
\includegraphics[scale=0.4]{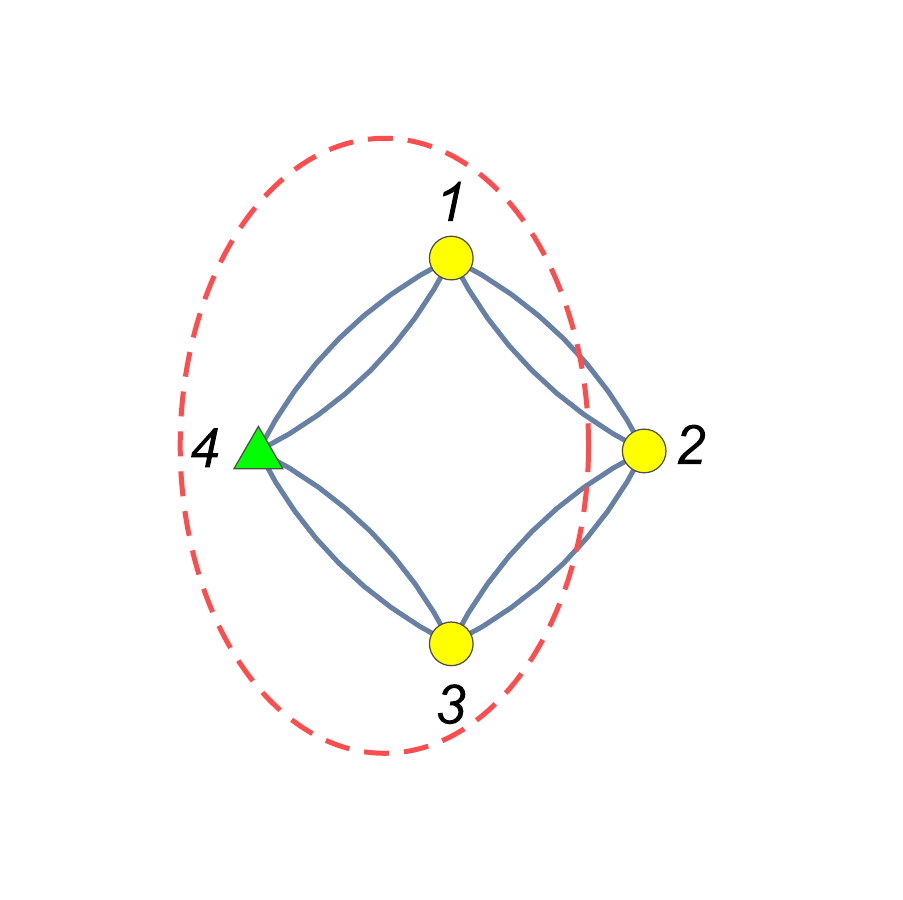}
\includegraphics[scale=0.4]{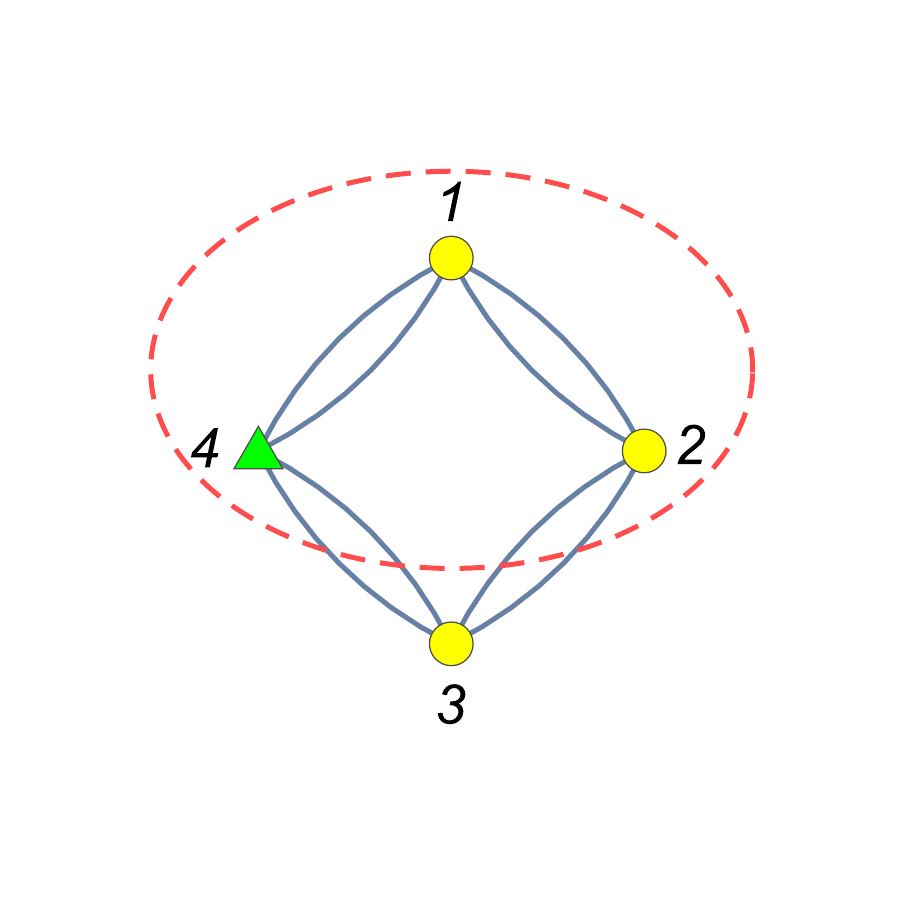}
\includegraphics[scale=0.4]{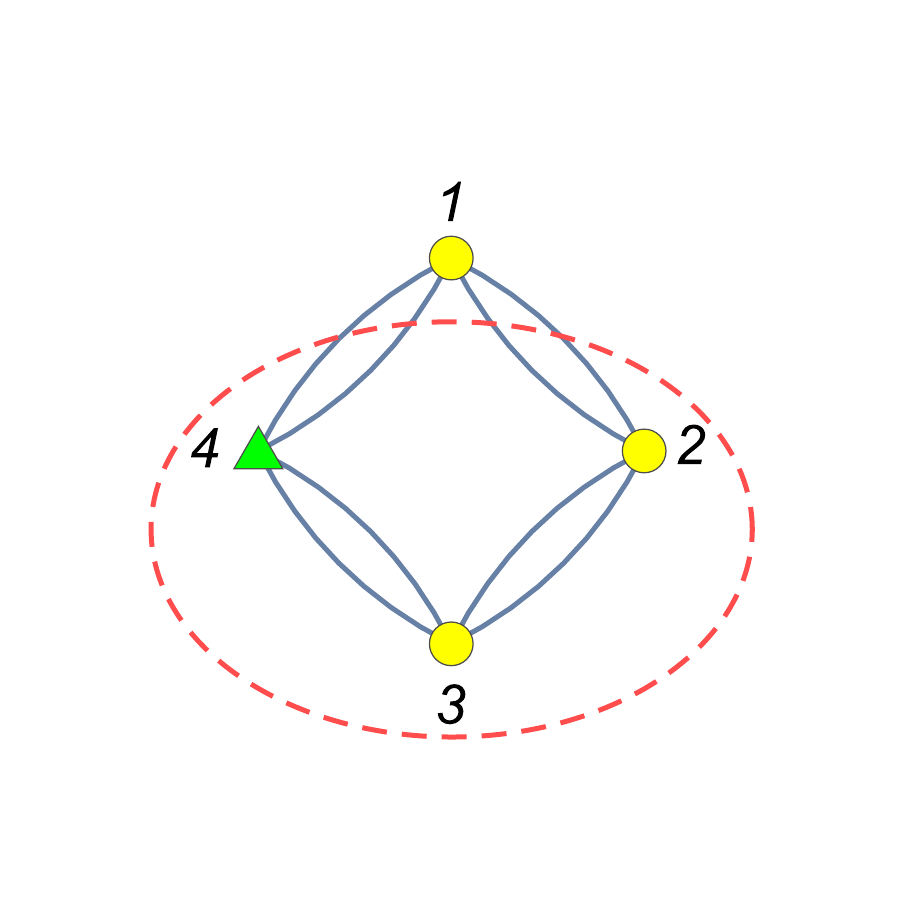}
\includegraphics[scale=0.4]{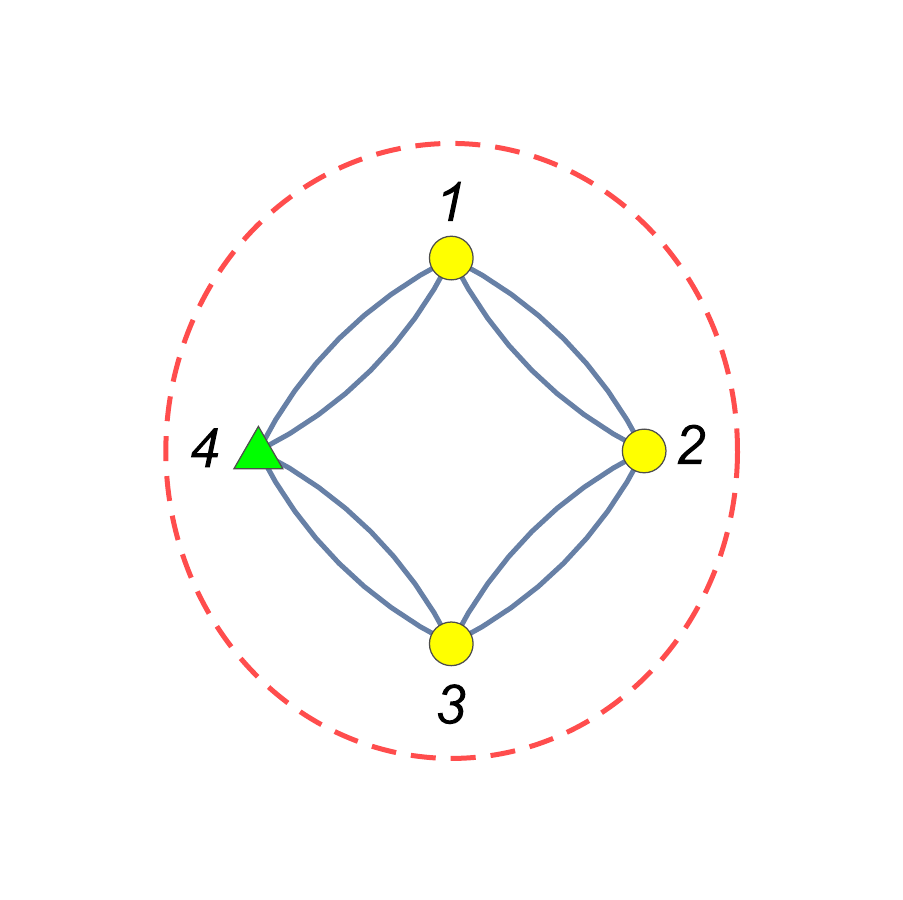},
\end{center}
where the red line enclose the punctures on the same branch cut, i.e. the red line is the  branch cut, which is controled by the $\L$ integration variable.

However, these five configurations vanish trivially because the $PSL(2,\mathbb{C})$ symmetry is breaking on upper and lower sheet when $\L\rightarrow 0$. This computation is straightforward. 
 
So as  to classify the different kind of configurations we introduce the following terminology 
\begin{itemize}
\item {\bf Allowable Configuration}:  Let $C$ be a configuration. We say $C$ is an {\it allowable configuration} if the number of fixed punctures on the upper and lower sheet is two. This implies that  in the  $\L\rightarrow 0$ limit the $PSL(2,\mathbb{C})$ symmetry is well defined (gauged) on each sheet. 
\end{itemize}
Clearly, for the diagram in {\bf Fig.6.3} there is one more {\it allowable  configuration}  given by 
\begin{center}
\includegraphics[scale=0.4]{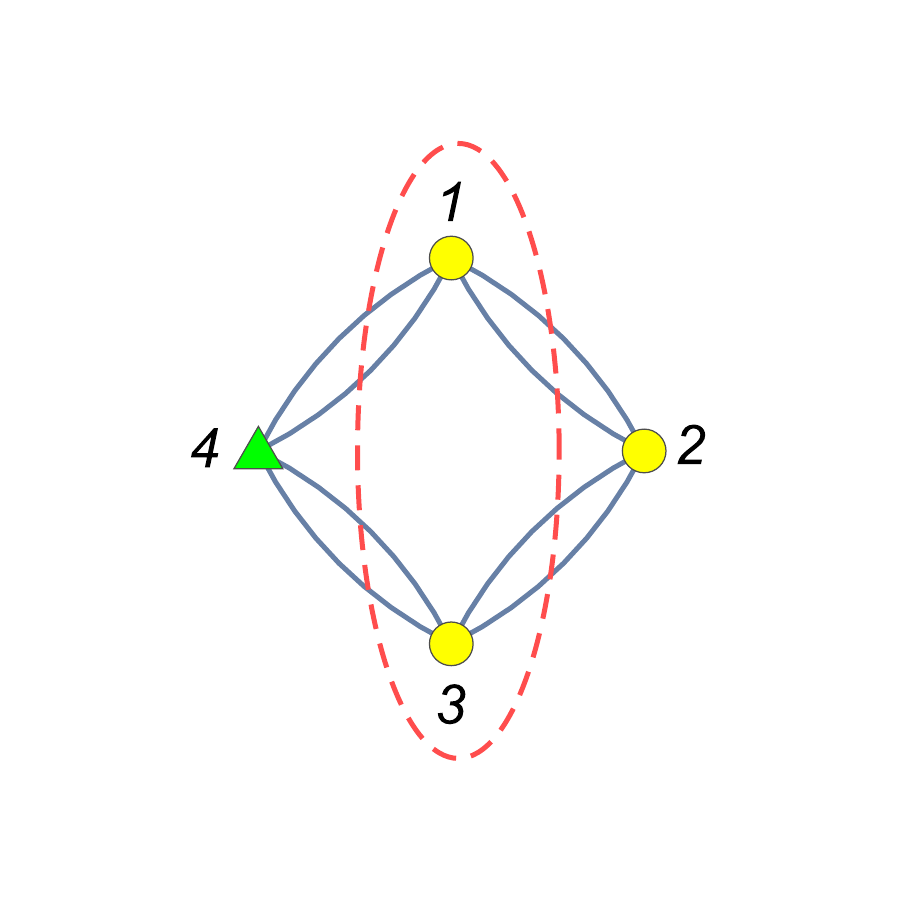}\,,
\begin{center}
({\bf Fig.6.4})\,{\small {\rm Allowable configuration which vanishes. \,}}
\end{center}
\end{center}
but this one also vanishes. 

The vanishing of this last  configuration is a consequence of the following theorem
\\

\begin{tabular}{| l |}
 \cline{1-1}  
 $\L-${\bf Theorem}\qquad\qquad\qquad\\
Let $C$ be an allowable configuration, then the integrand ${\cal I}=|ijk|\Delta_{FP}(ijk,d) H^D(\s)$ \\
on the $C$  configuration   has the $\Lambda-$behavior \\
\\
\hspace{5cm}$\mathcal {I}\,\Big|_{\L\rightarrow 0}^C\, \sim\, \L^{L-4}\,+\,{\cal O}(\L^{L-3})$\\
\\
around $\L= 0$, where  $L$ is the number of edges which are intersected by the red line.\\
\cline{1-1}
\end{tabular}
\\
\\

This theorem is proved in appendix \ref{appendix}.

So far, we have defined what is a configuration, an allowable configuration and we have formulated the $\L-$theorem. Now, with the intention to set down the $\L-$algorithm  it is  useful to  define a new kind of configuration 
\begin{itemize}
\item {\bf Singular Configuration}:  Let $C$ be a configuration. We say $C$ is an {\it  singular configuration} if $C$ is an  allowable configuration and the integrand, ${\cal I}=|ijk|\Delta_{FP}(ijk,d) H(\s)\sim \L^{-s},\,\, s>0$ around $\L=0$.
\end{itemize}

Following with our example, we note that expanding  the \\
${\cal I}=|123|\Delta_{FP}(123,4) H^D_4(1,2,3,4)$ total integrand and the $E_4$ scattering equation (S.E) around  $\Lambda=0$,  the two configurations in {\bf Fig.6.3(c)}  become
\begin{center}
\includegraphics[scale=0.45]{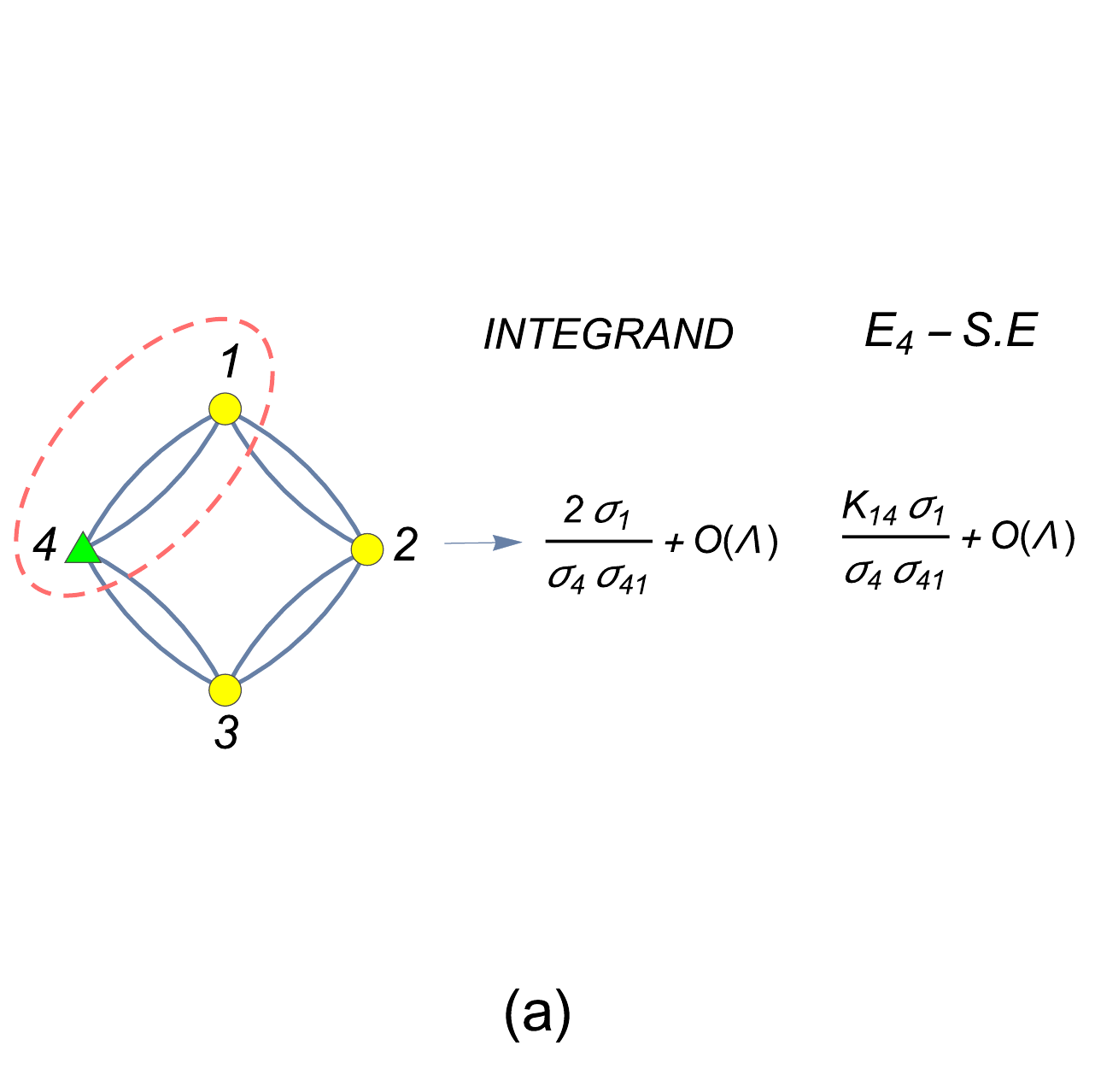}\qquad \qquad\
\includegraphics[scale=0.45]{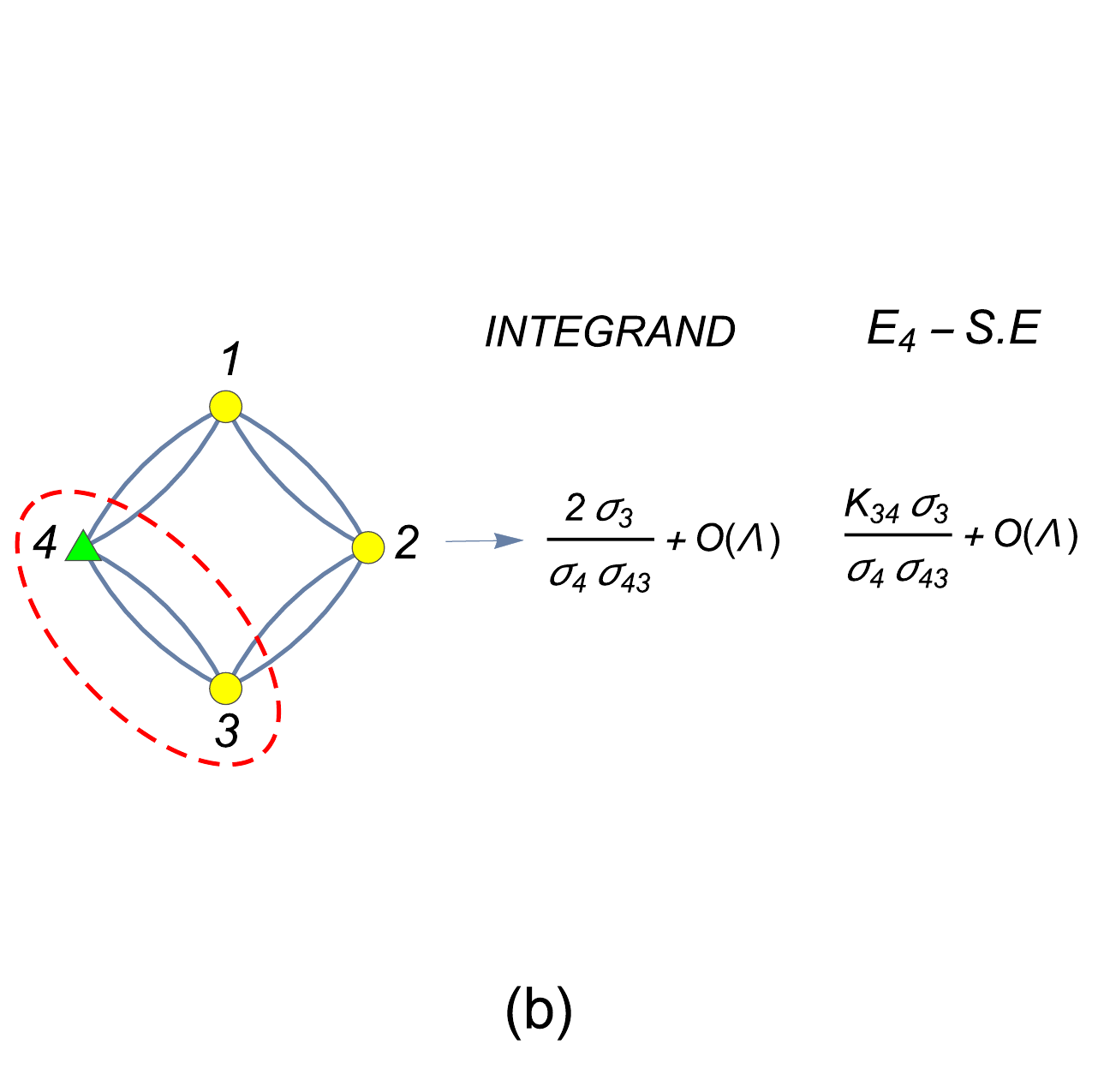}\,\, .
\begin{center}
({\bf Fig.6.5})\,{\small {\rm Computing the non-zero allowable configurations. \,}}
\end{center}
\end{center}
Thus, the integration over  $\L$ is straightforward and  the final result is 
\begin{center}
\includegraphics[scale=0.4]{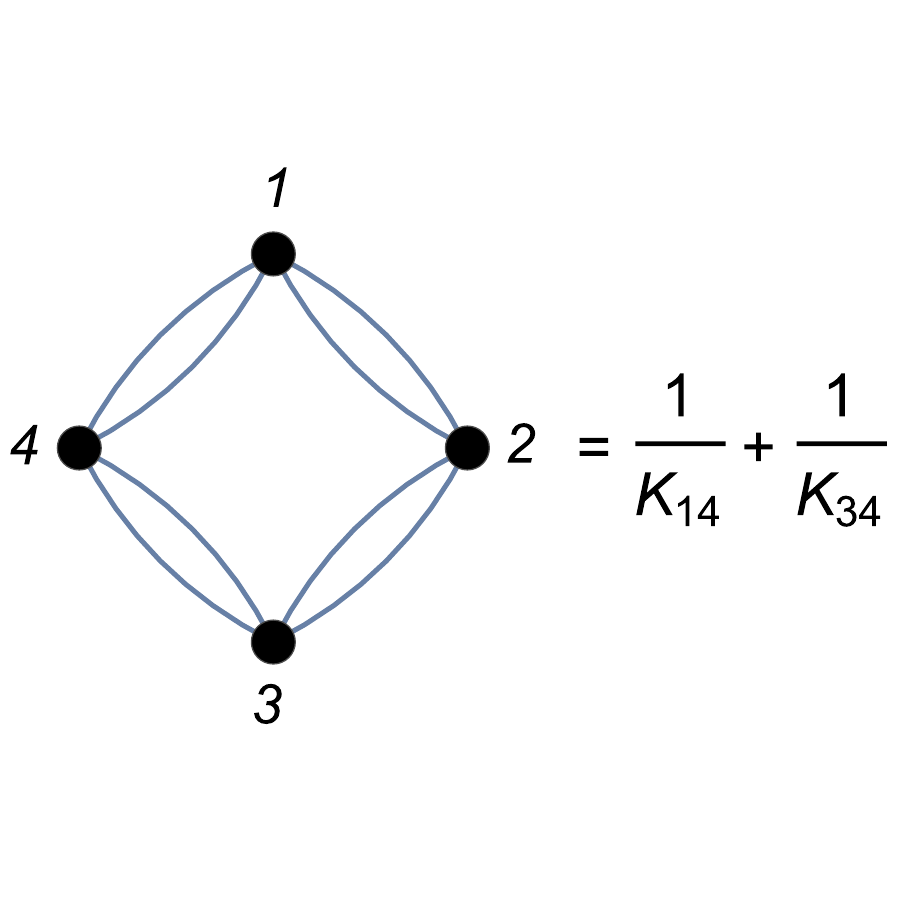}\,\,,
\begin{center}
({\bf Fig.6.6})\,{\small {\rm Final solution for the integrand in \eqref{fexample}. \,}}
\end{center}
\end{center}
which is the right answer. 

We call this method the $\L-$algorithm. In the next section we explain carefully this algorithm.  

\subsection{The $\L-$Algorithm}\label{Lalgorithm}

In this section we introduce formally the $\L-$algorithm, which is given up to $\mathbb{Z}_2$ symmetry,, $y_a\,\,\rightarrow\,\, -y_a$. 

We describe step by step the method.\\
{\bf $\L-$Algorithm  Steps}

\begin{itemize}
\item {\bf (1)}  To draw the graph to be computed.   Let us remember that the graph must be drawn such that the intersection number of the edges is the minimum.

This drawing must have three yellow vertices ($PSL(2,\mathbb{C})$ gauge  fixing) and one green vertex (scale symmetry fixing)
\begin{center}
\includegraphics[scale=0.38]{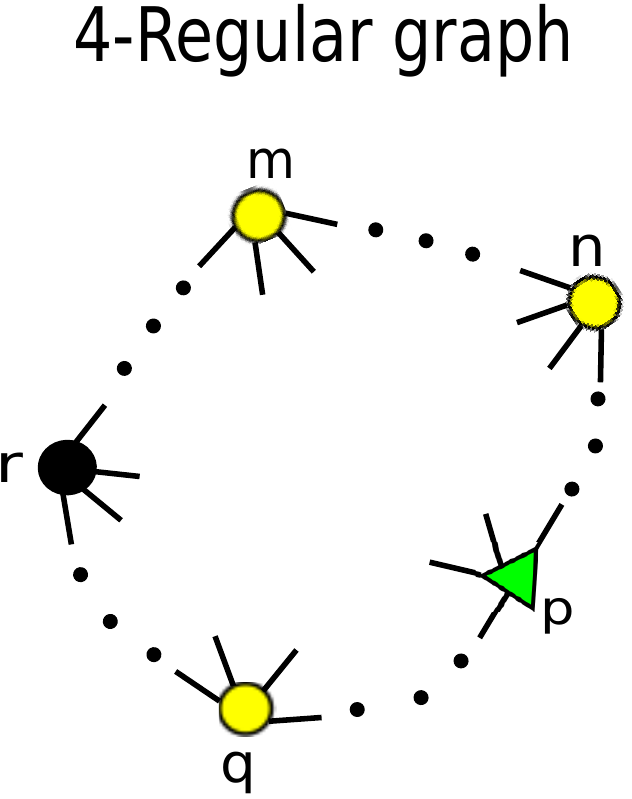}.
\begin{center}
(Figure (a))\,{\small {\rm 4-Regular graph. $PSL(2,\mathbb{C})$ (Yellow)  and scale symmetry (Green) gauge  fixing.\,}}
\end{center}
\end{center}

\item {\bf (2)} To find all non-zero allowable configurations
\begin{center}
\includegraphics[scale=0.38]{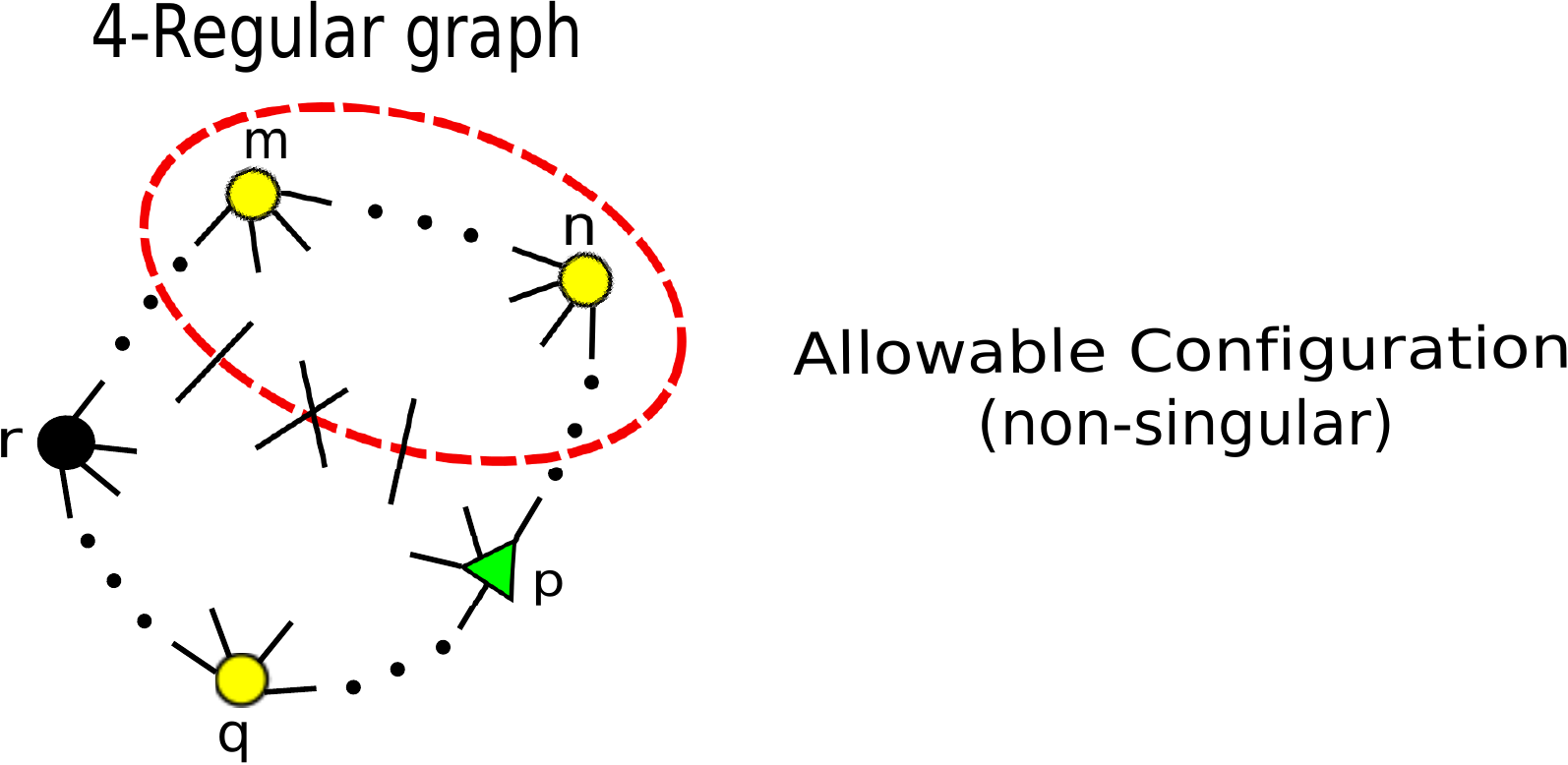}.
\begin{center}
(Figure (b))\,{\small {\rm One non-zero allowable configuration. \,}}
\end{center}
\end{center}
The gauge fixing from the  step (1) must be chosen  such that there are not singular configurations. This fact becomes clearer in section \ref{examples}.

If it is not possible to choose a gauge fixing such that it avoids singular configurations then the $\L-$algorithm can not be applied directly.

From the  $\L-$theorem, it is clear that  the red line in all non zero configurations intersects only 4 black lines, i.e. just 4 black lines go through the branch cut.

\item {\bf (3)} To compute the $\L$ integral around the cycle $|\L|=\epsilon$ (on all configurations found in the previous step).

\begin{itemize}

\item {\bf (i)}  After computing the $\L$ integral (on one particular configuration) the sphere is splitting into two spheres, the upper-sheet and the lower-sheet. This splitting is identified  by the red line. As a consequence  two new (massive) punctures arise, one on the upper and the other one on the lower-sheet. These punctures are fixed on each sheet at the point $\s_0=0$  and they are denoted by the red color. This process is shown in the following figure
\\
\begin{center}
\includegraphics[scale=0.38]{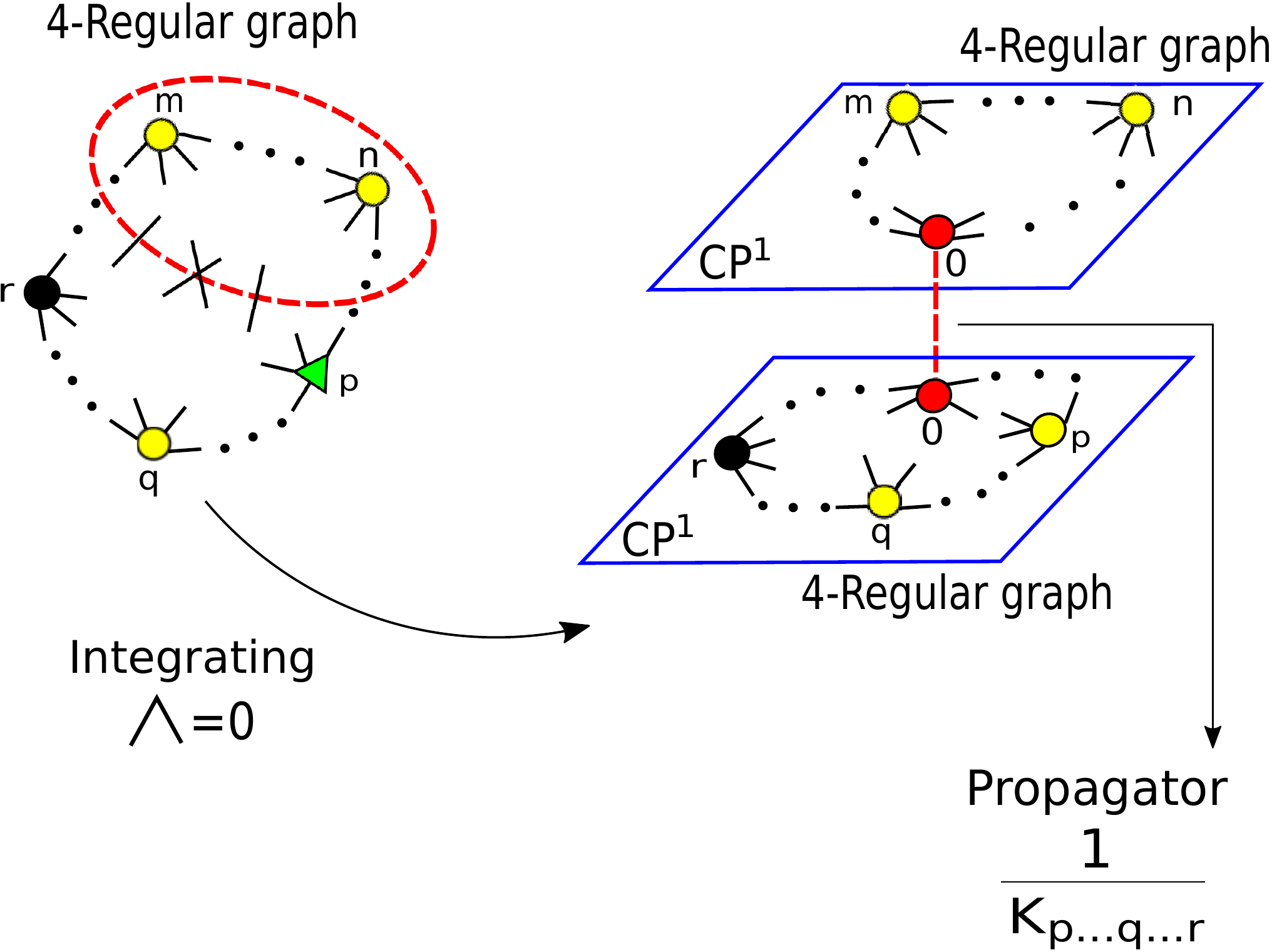}\,\,\, .
\begin{center}
(Figure (c))\,{\small {\rm Computating the $\L$ integral on one particular configuration. \,}}
\end{center}
\end{center}
The particles inside of the red line,  including now the new red massive puncture at $\s_0=0$ on the upper-sheet, shape a new 4-regular graph on the upper-sheet  (subdiagram) and the particles  outside of the red line, including the new red massive puncture at $\s_0=0$ on the lower-sheet,  shape the other new 4-regular graph  (subdiagram), such as it is shown in  figure (c).

The momentum of the red massive puncture on the upper-sheet is the sum over all momenta of the particles outside of the red line, i.e. 
\begin{equation}
k_0^{\rm upper}=k_p + \cdots + k_q +\cdots + k_r +\cdots,
\end{equation}
and the momentum of the red massive puncture on the lower-sheet  is the sum over all momenta of the particles inside of the red line, i.e. 
\begin{equation}
k_0^{\rm lower}=k_m + \cdots + k_n +\cdots \,.
\end{equation}

\item {\bf (ii)} The scattering equation associates to the puncture in the green triangle, in figure (c) it is $E_p$,  becomes.
$$
E_p=\sum_{a\neq {\rm upper \,\,sheet }}\frac{k_p\cdot k_a}{\s_{pa}}+\frac{k_p\cdot k_0^{\rm lower}}{\s_p}+{\cal O}(\L).
$$
Using the scattering equations (at $\L=0$) located on the same sheet as the green puncture, in figure(c) it is the lower sheet, i.e $E_r, \ldots $, it is straightforward to prove that
$$
E_p = -\frac{(\s_q-\s_0)(\s_0-\s_q)}{(\s_0-\s_p)(\s_p-\s_q)(\s_q-\s_0)}k_{p\ldots q\ldots r \ldots}+{\cal O}(\L)\,\, ,
$$
where $\s_0=0$. The  $(\s_0-\s_p)(\s_p-\s_q)(\s_q-\s_0)$ factor becomes  one of the two Faddeev-Popov determinants on the lower brach and the numerator, $(\s_q-\s_0)(\s_0-\s_q)$, cancels out  with the 
$|m,n,q|\,\Delta_{\rm FP}(mnq,p)$ Faddeev-Popov expansion given in appendix \ref{appendix}. Therefore, one can say that the $E_p$ scattering amplitude becomes the  propagator
$$
\frac{1}{E_p}   \quad  \rightarrow \quad\,\,\frac{1}{k_{p...q...r...}}\,\,\, .
$$

Note that although in our example (figure (c)) $k_{p...q...r...}=k_{m...n...}$, in general this is not true. Since the $\L-$algorithm is a iterative process then new massive particles arise  (red punctures) and the equality $k_{p...q...r...}=k_{m...n...}$ can be  broken.

Finally, the two new subdiagrams   are given  in the original CHY approach, where $(\s_0,\s_m,\s_n)$ are the gauged punctures on the upper-sheet and  $(\s_0,\s_p,\s_q)$ are the gauged punctures on the lower-sheet.  

\end{itemize}

\item {\bf (4)} To come back to the step (1). 

It is useful to remember that a 4-regular graph with 3 vertices is just 1 
\begin{center}
\includegraphics[scale=0.38]{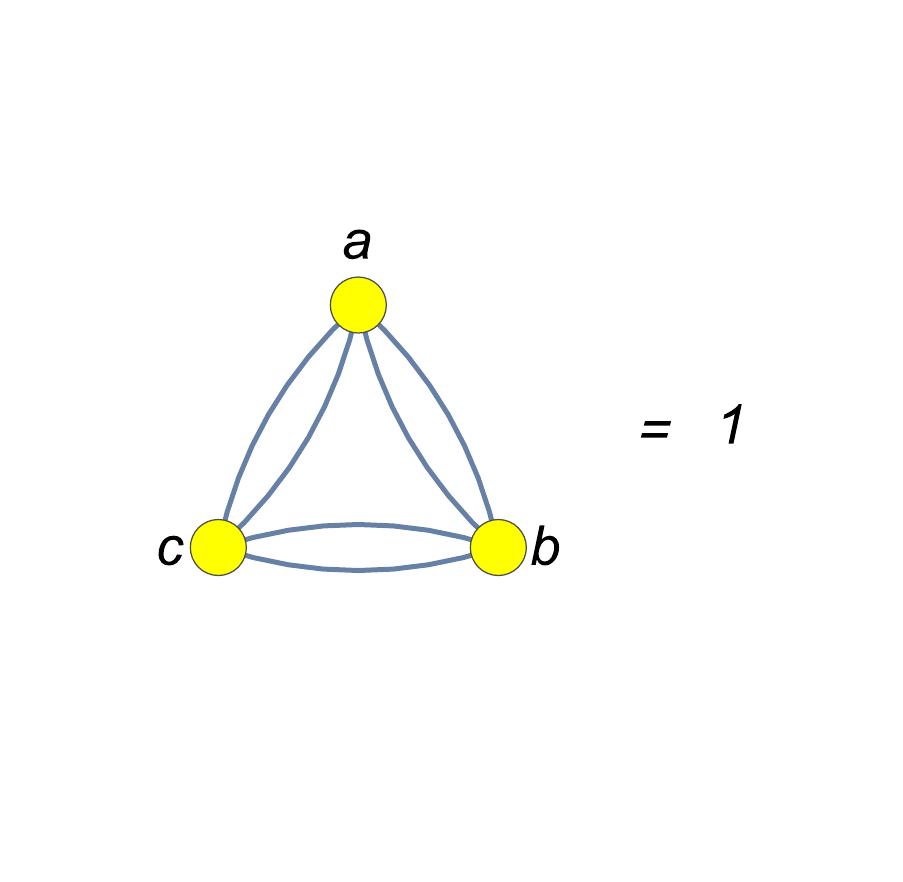}
\begin{center}
(Figure (d))\,{\small {\rm 3-point 4-regular graph\,}}
\end{center}
\end{center}

\end{itemize}

\subsection{Building Blocks}\label{bblocks}

Since that the $\L$-algorithm is an iterative process  then it is useful to construct  fundamental graphs or irreducible graphs  (building blocks).

Our building blocks  are given by the following diagrams of 4 and 5 vertices 
\begin{center}
\includegraphics[scale=0.32]{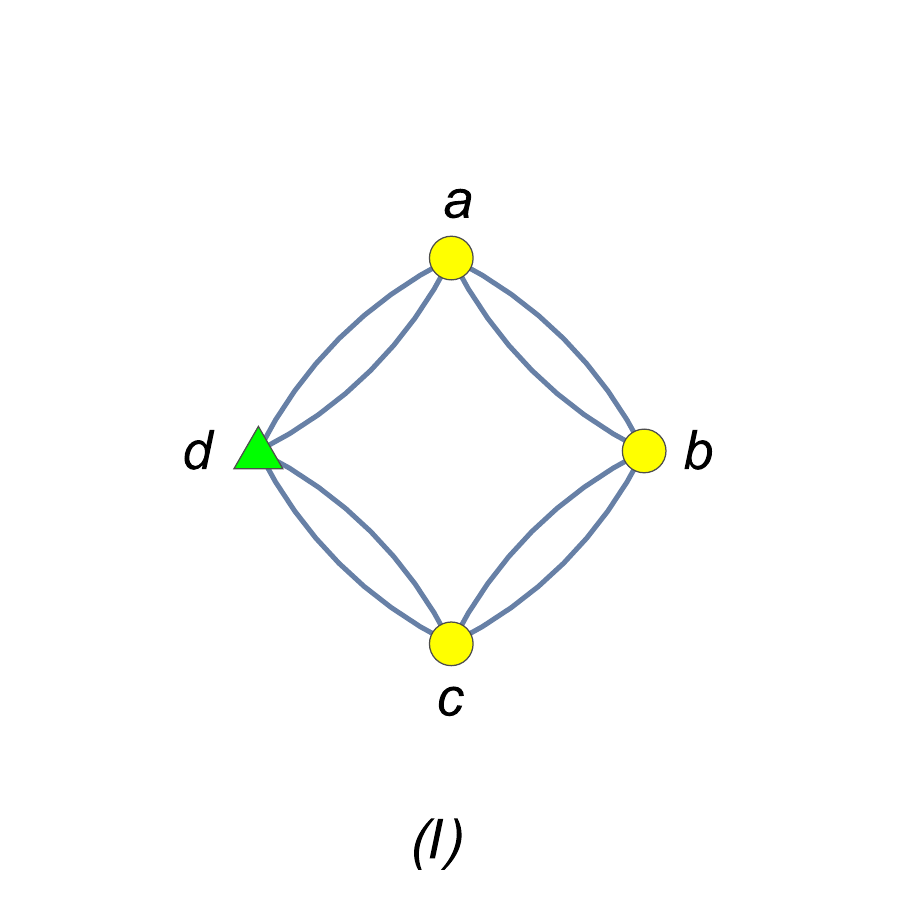}
\includegraphics[scale=0.32]{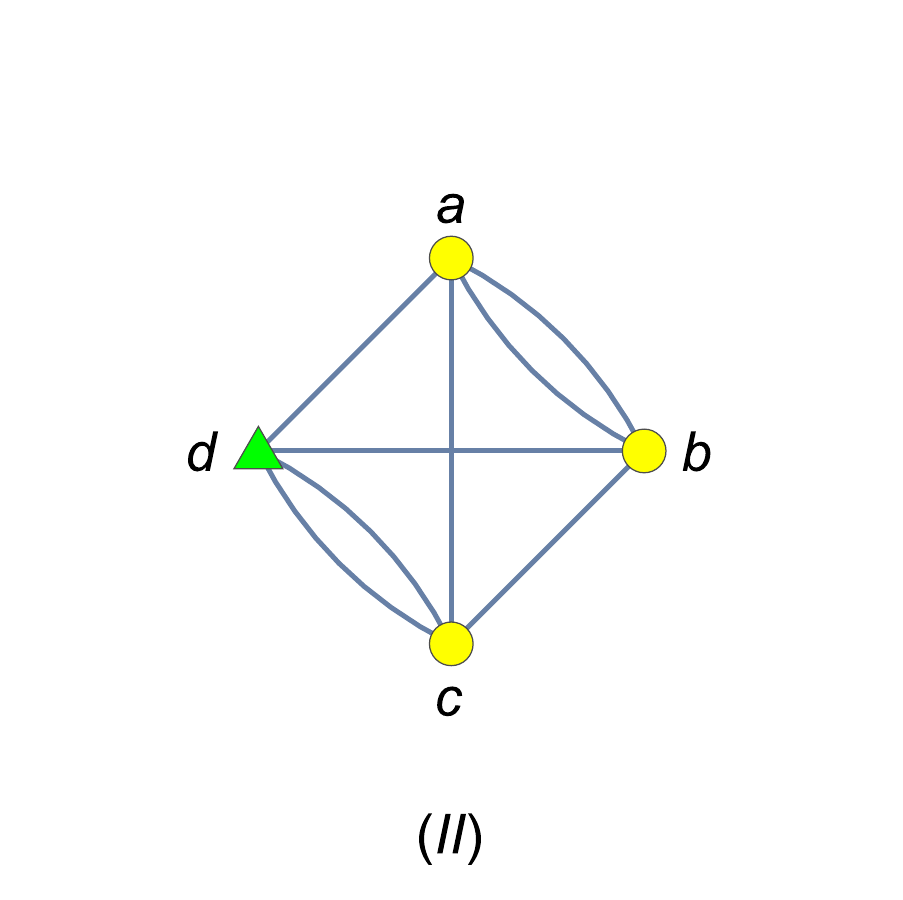}
\includegraphics[scale=0.32]{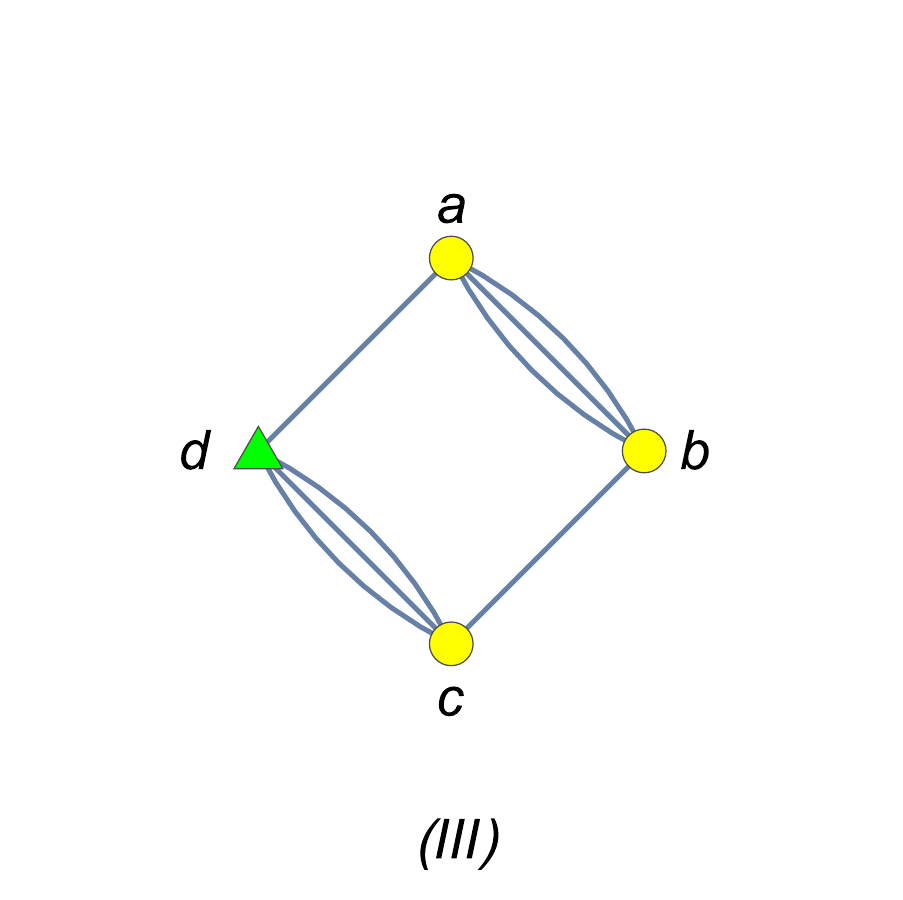}
\includegraphics[scale=0.32]{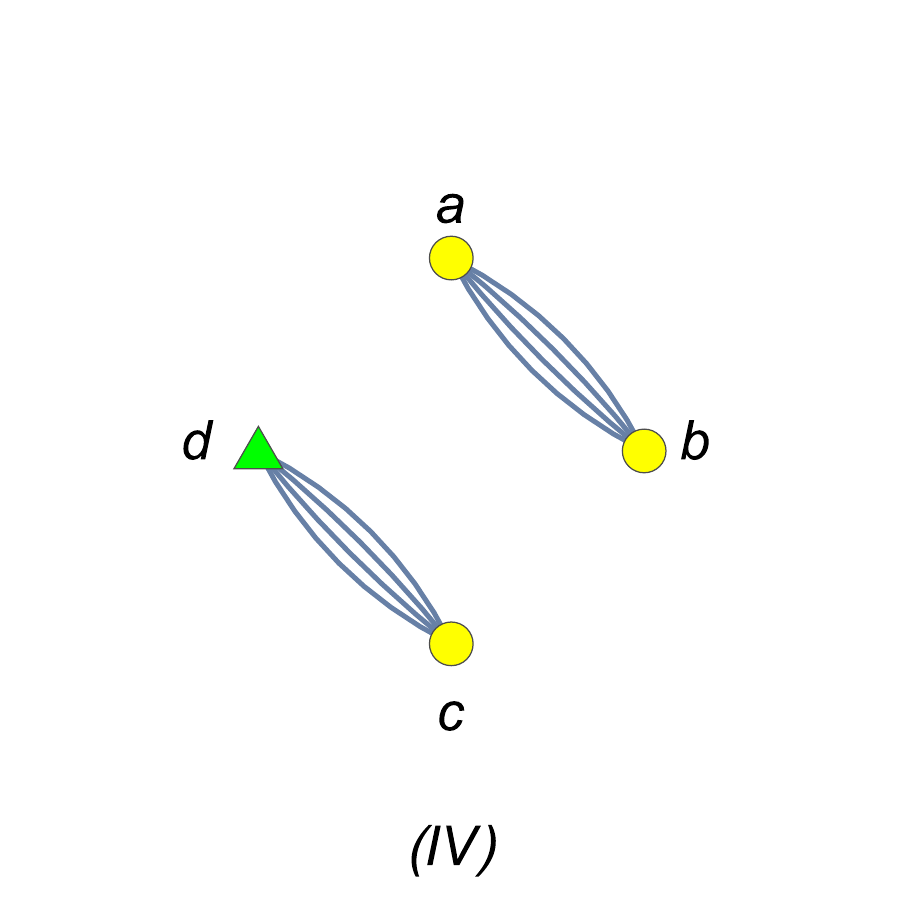}
\includegraphics[scale=0.32]{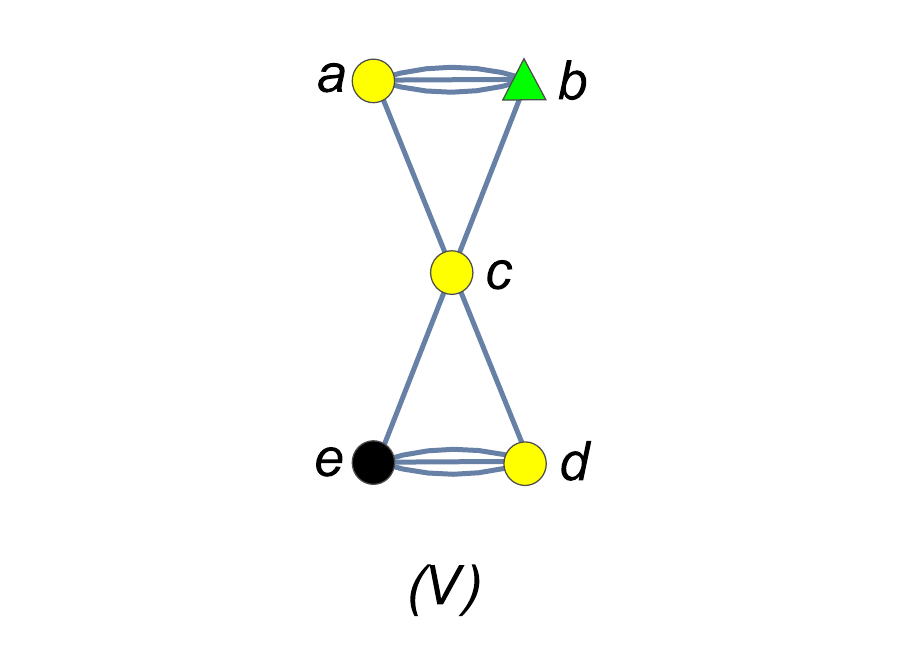}
\begin{center}
({\bf Fig.6.7})\,{\small {\rm Building Blocks. \,}}
\end{center}
\end{center}
The  $(I)$ graph, which was computed previously, and $(II)$ graph are trivials \cite{Cachazo:2013iea}
\begin{center}
\includegraphics[scale=0.4]{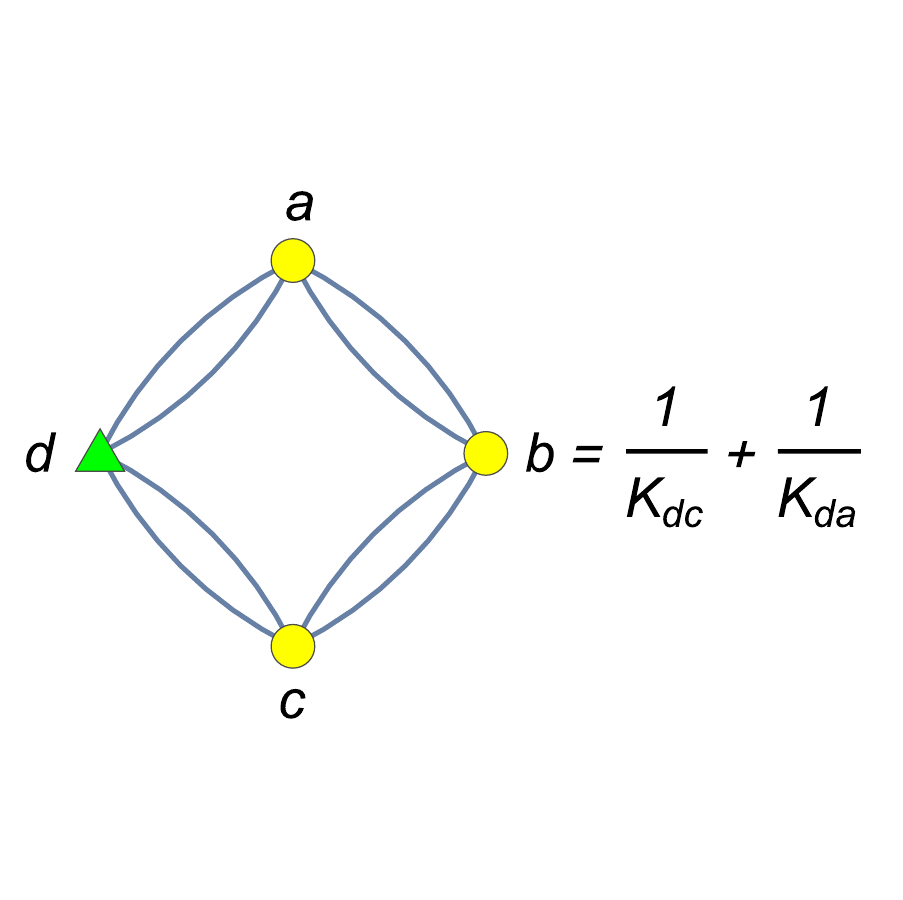}\qquad ,\qquad
\includegraphics[scale=0.4]{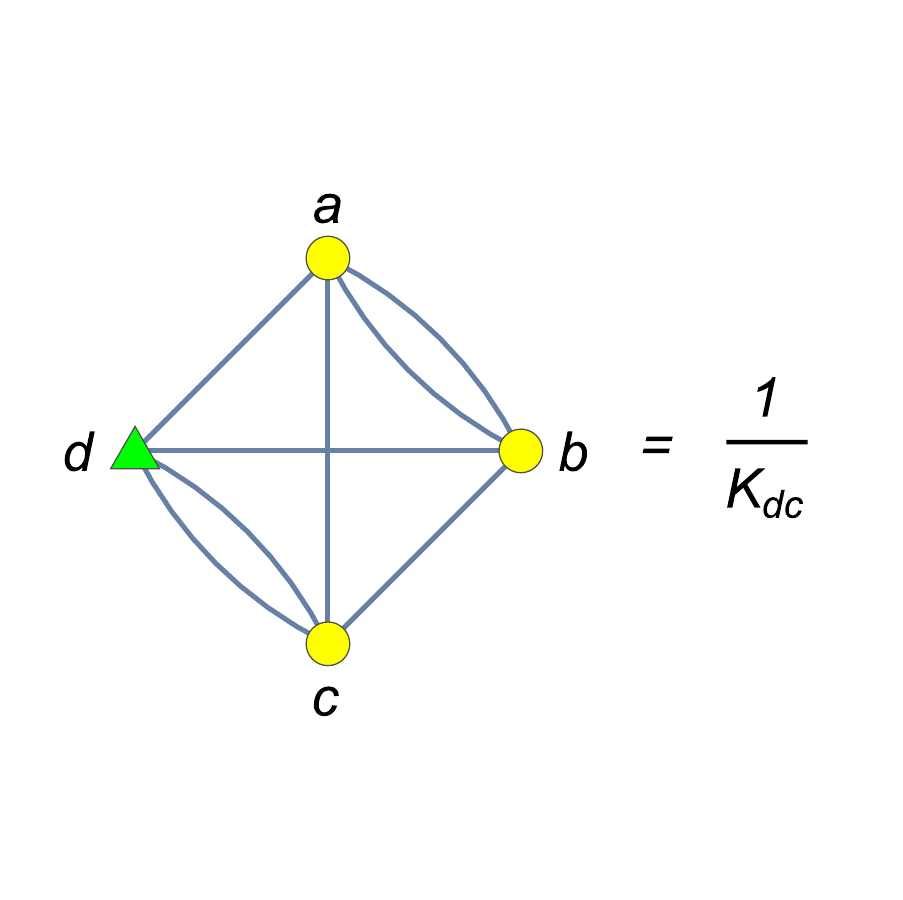}\,\,\,.
\begin{center}
({\bf Fig.6.8})\,{\small {\rm Building Blocks (I)  and (II). \,}}
\end{center}
\end{center}

In order to compute the  the $(II)$ and $(III)$ building blocks, one can note that
on the support of the $E_d$ scattering equation (before performing the residue theorem, section   \ref{residuetheorem})
\begin{equation}
E_d=k_{ad}\,\tau_{d:a}+k_{bd}\,\tau_{d:b}+k_{cd}\,\tau_{d:c}=0\,\,\Rightarrow\,\, -1=\frac{k_{ad}}{k_{cd}}\left(\frac{\tau_{d:a}\tau_{b:c}}{\tau_{b:a}\tau_{d:c}}\right)\,\, .
\end{equation}
So, the $(III)$ and $(IV)$ graphs in  {\bf Fig.6.7} become very simples
\begin{center}
\includegraphics[scale=0.4]{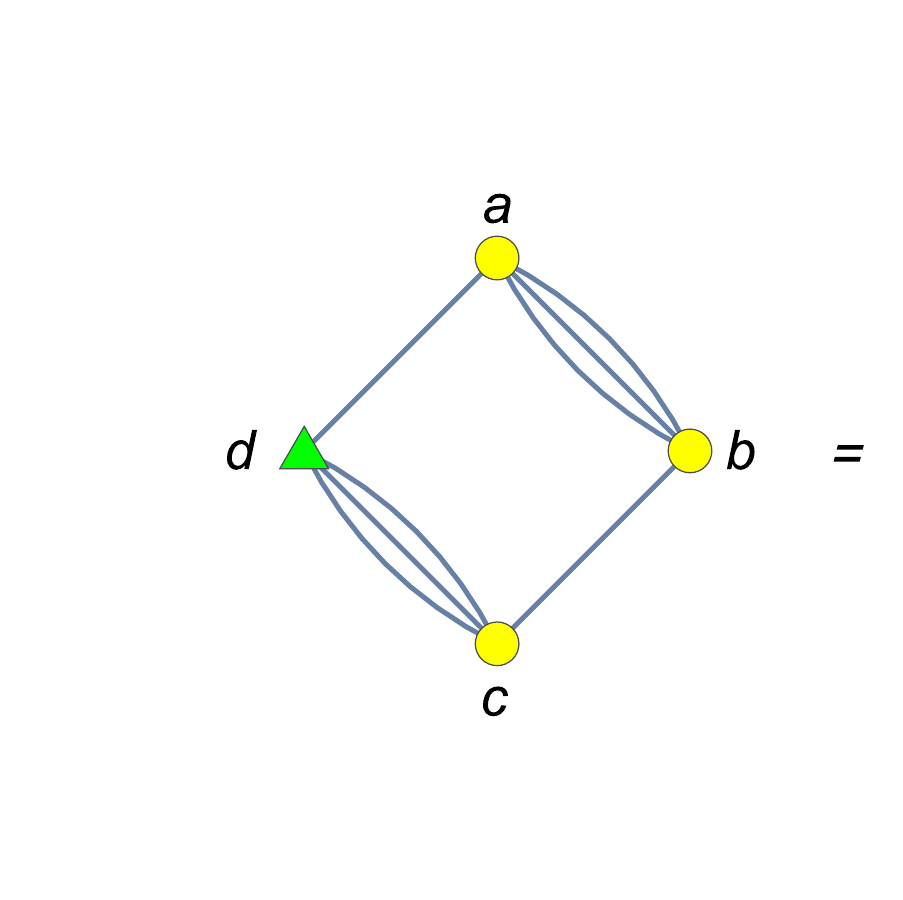}
\includegraphics[scale=0.4]{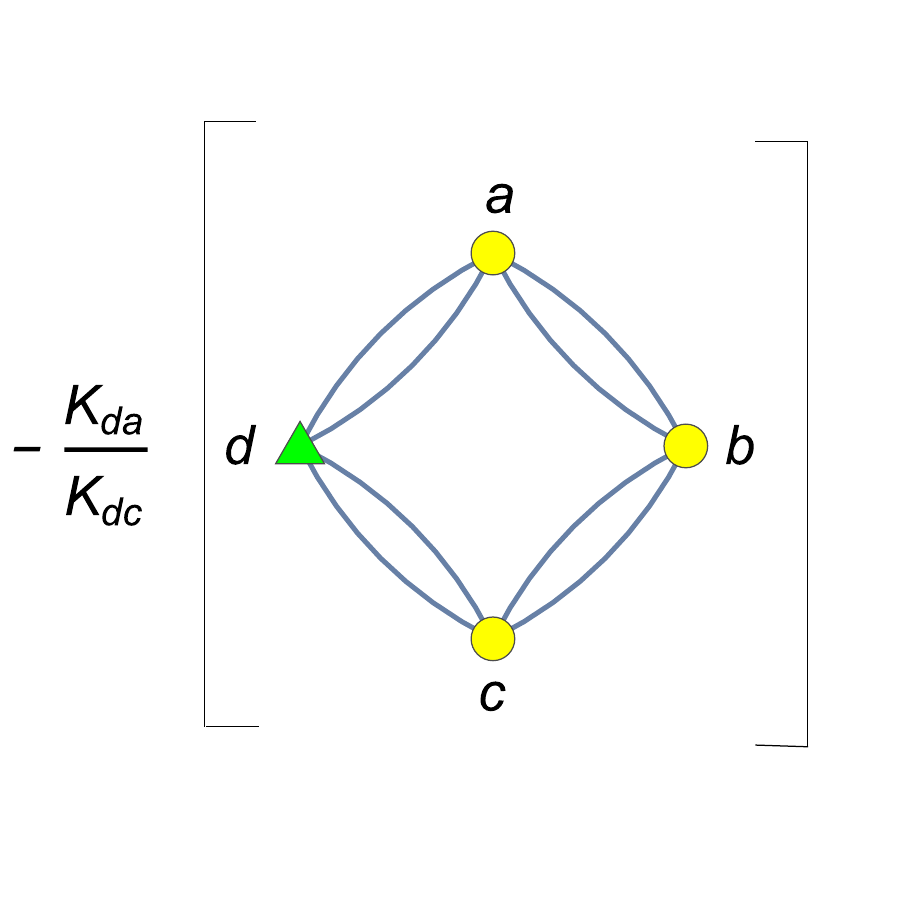}
\includegraphics[scale=0.4]{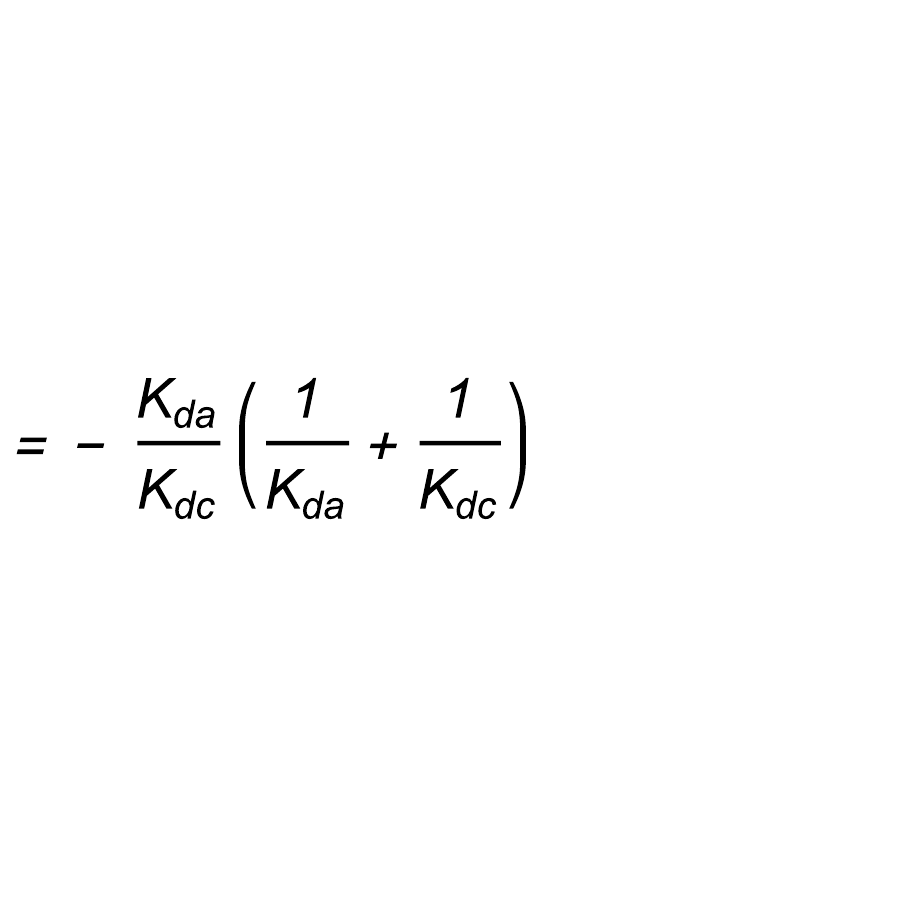},\\
\includegraphics[scale=0.4]{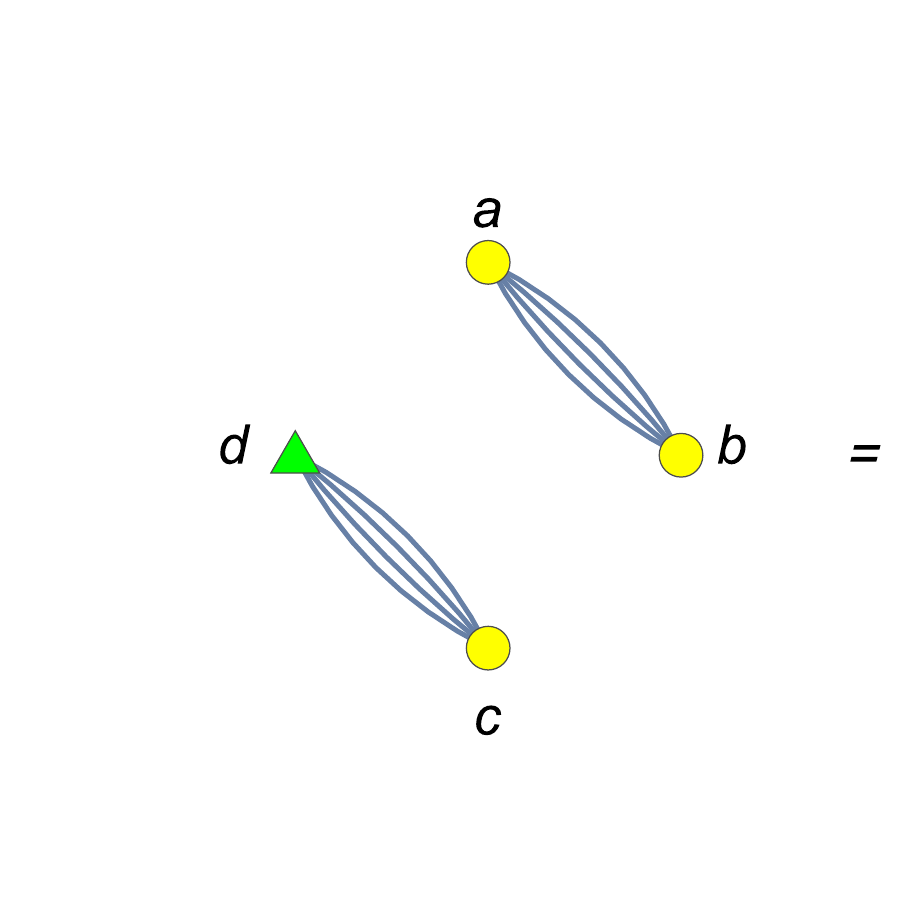}
\includegraphics[scale=0.4]{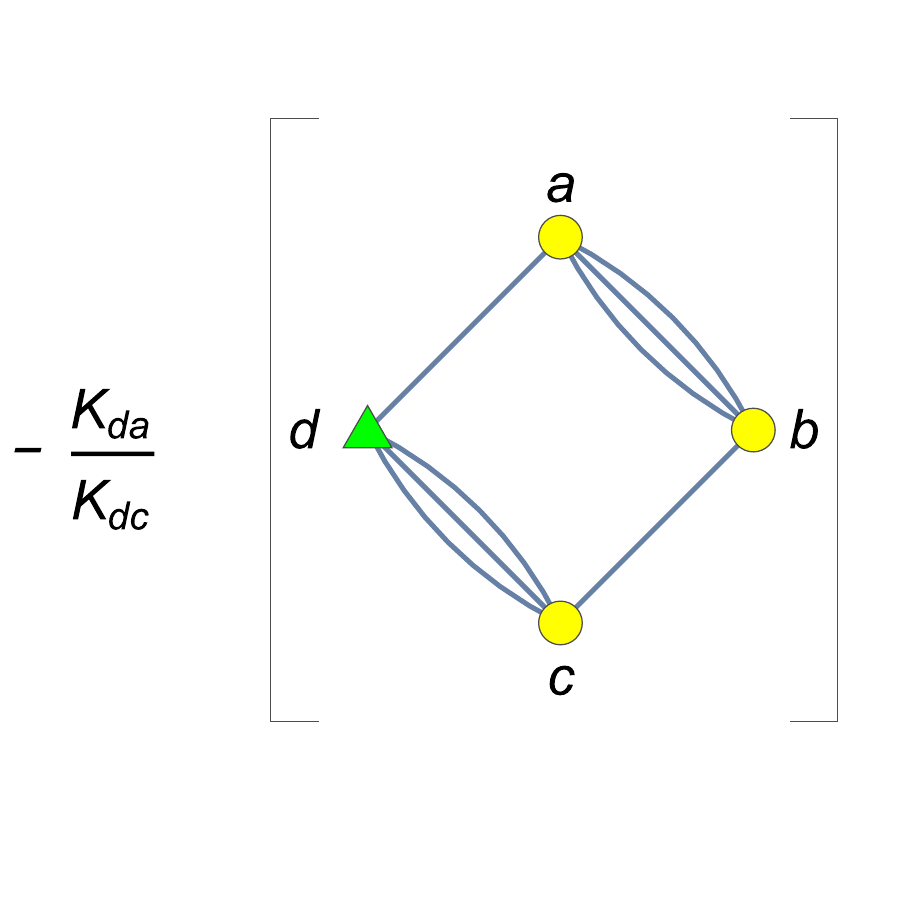}
\includegraphics[scale=0.4]{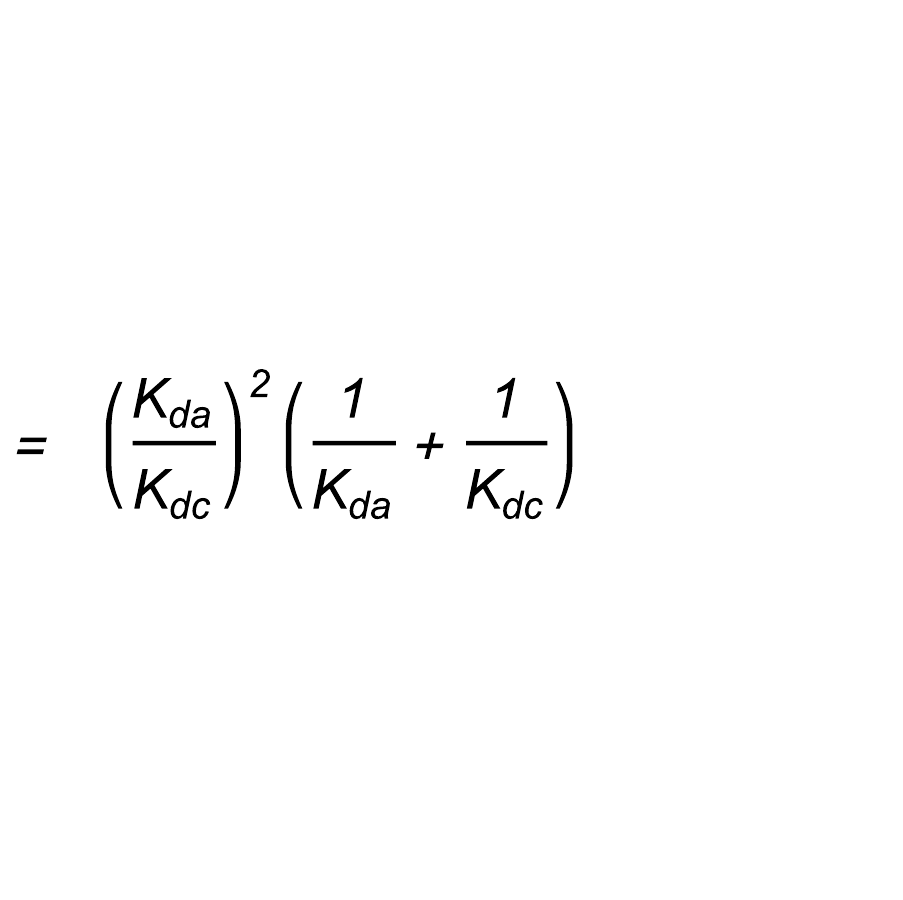}.
\begin{center}
({\bf Fig.6.9})\,{\small {\rm  Building Blocks (III) and (IV).\,}}
\end{center}
\end{center}

Finally, the $(V)$  building block  in  {\bf Fig.6.7} can not be computed using the $\L-$algorithm, because this graph  has a singular configuration. So, we use the algorithm given in \cite{humbertoF} (the general KLT algorithm).

\subsubsection{General KLT algorithm and computation of the $(V)$ building block}\label{fivebb}

In order to apply the general KLT algorithm \cite{humbertoF} on the $(V)$ building block, one must first note that this  building block  has the following decomposition  (in two 2-regular graphs)
\begin{center}
\includegraphics[scale=0.4]{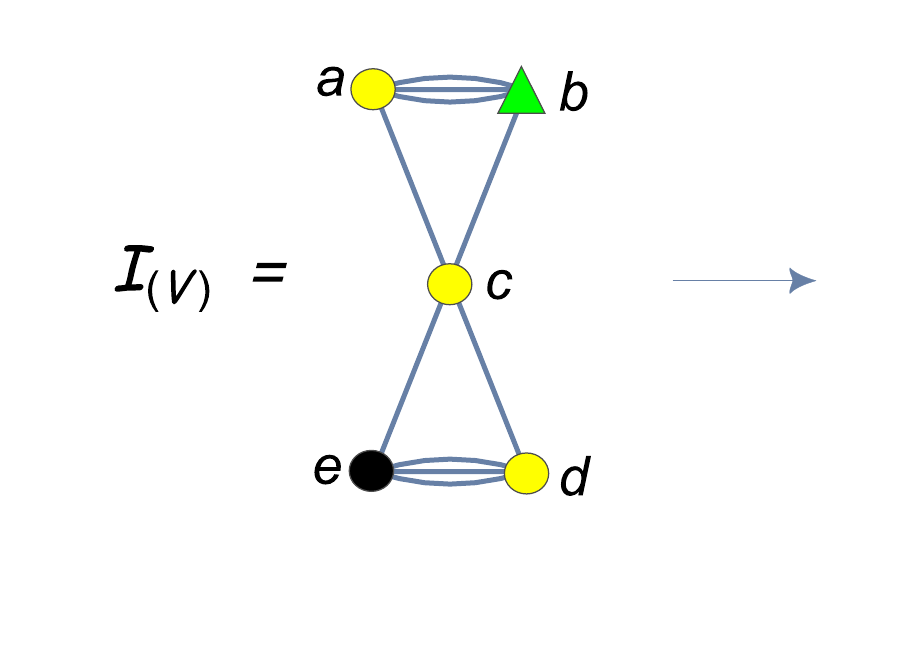}
\includegraphics[scale=0.4]{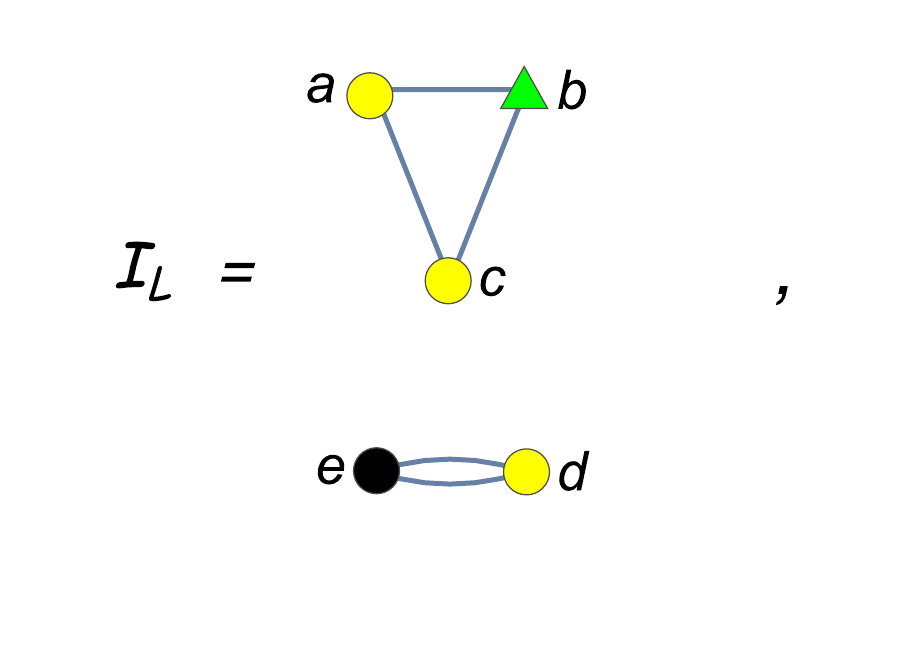}
\includegraphics[scale=0.4]{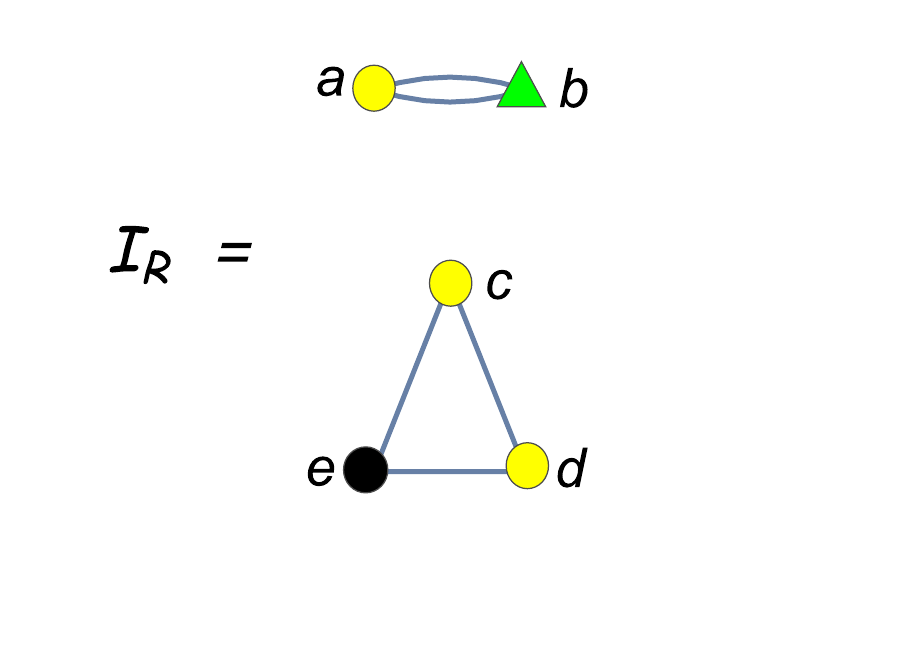}.
\begin{center}
({\bf Fig.6.10})\,{\small {\rm Decomposition in two 2-regular graphs \,}}
\end{center}
\end{center}
The second step is to  find a left and right (Parke-Taylor) base {\it compatible}\footnote{A Parke-Taylor factor is said to be compatible with a 2-regular graph if the union of the two graphs, which is a 4-regular graph, admits a Hamiltonian decomposition, i.e., it is the union of two Parke-Taylor factors.} with the the ${\cal I_L}$ and ${\cal I_R}$ graphs. Choosing the left and right base  as
\begin{center}
\includegraphics[scale=0.4]{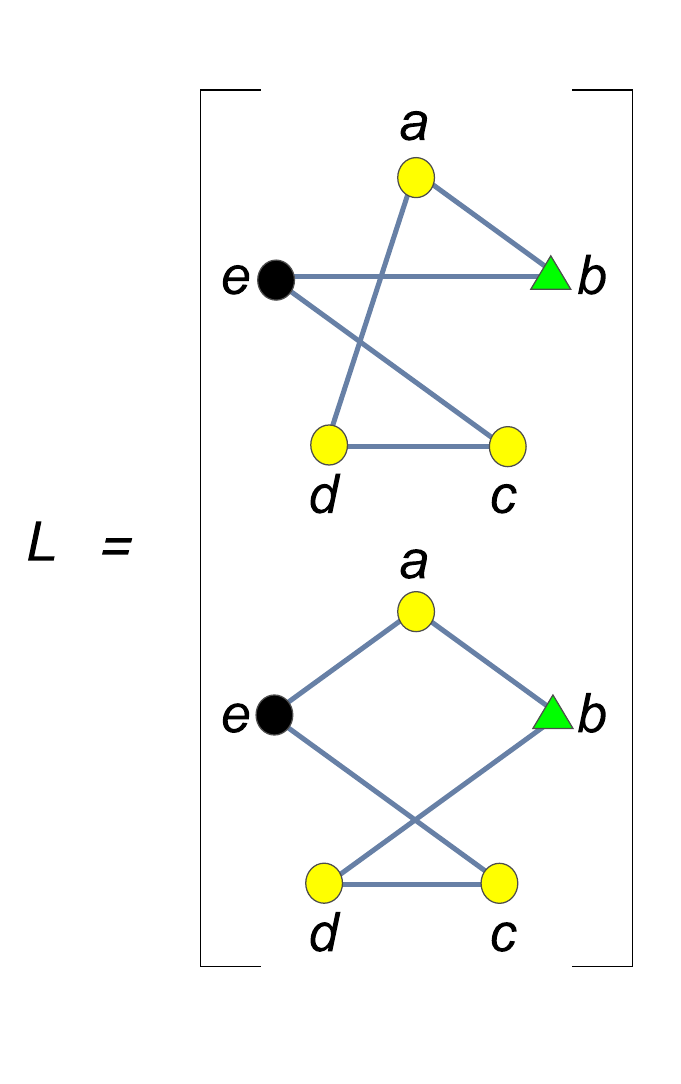}\qquad , \qquad
\includegraphics[scale=0.4]{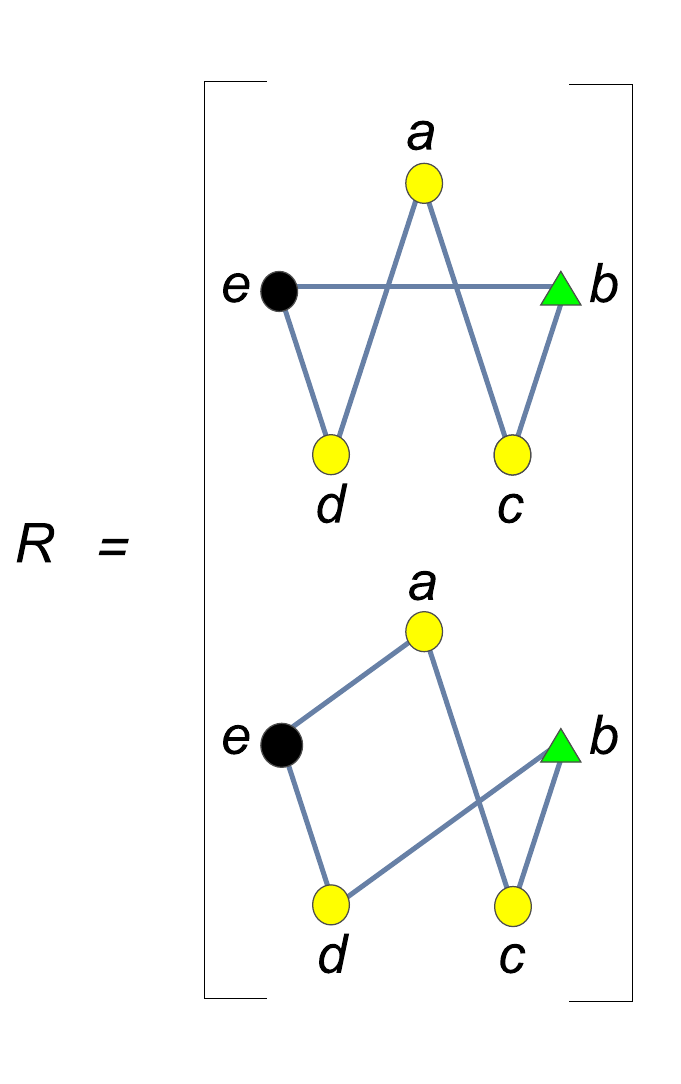}\,\,,
\begin{center}
({\bf Fig.6.11})\,{\small {\rm  Left and right base. \,}}
\end{center}
\end{center}
and following the general KLT algorithm  \cite{humbertoF}, it is straightforward to read the $(V)$ building block as
\begin{center}
\includegraphics[scale=0.4]{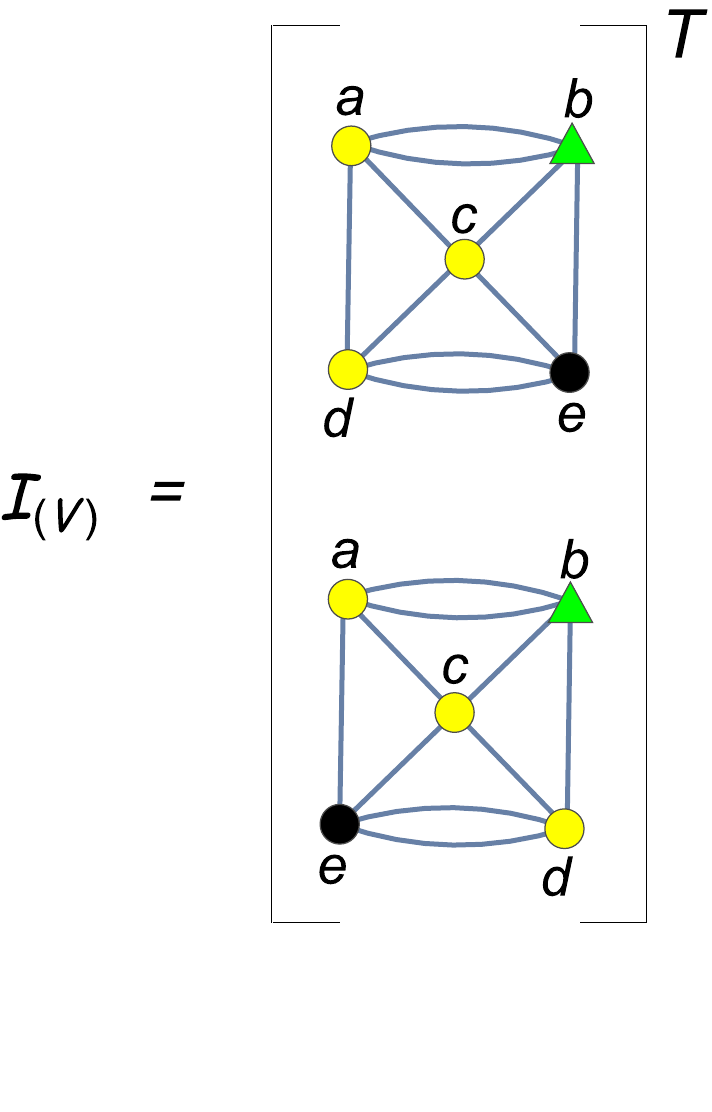}
\includegraphics[scale=0.4]{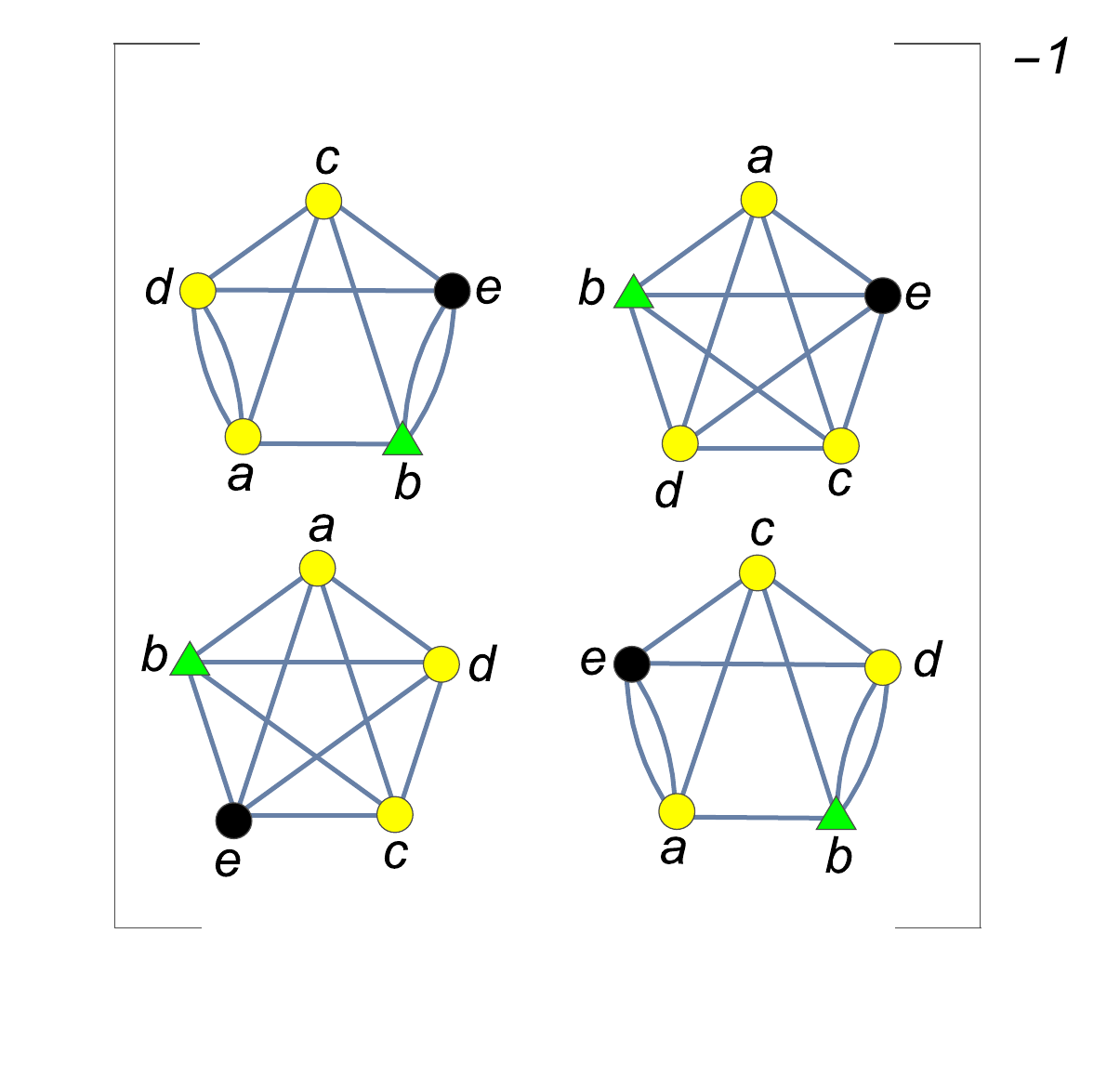}
\includegraphics[scale=0.4]{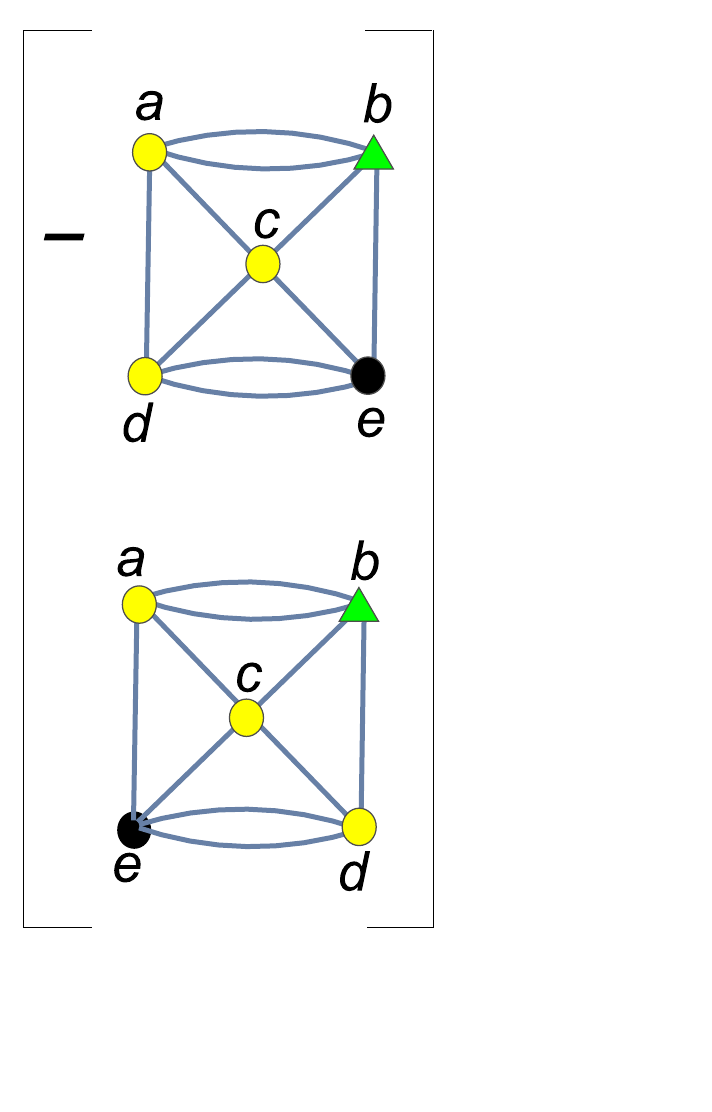},
\begin{center}
({\bf Fig.6.12})\,{\small {\rm Graph  representation of the ${\cal I}_{(V)} = (L\,{\cal I_L} )^{\rm T} (m^{L|R})^{-1} (R\,{\cal I_R} )$ computation. \,}}
\end{center}
\end{center}
where the relative sign was explained  in \cite{Cachazo:2013iea} .  

So as to be consistent  with the initial gauge fixing we must keep it\footnote{This fact is very important, because when the $\L-$algorithm is iterated then massive particles arise  and the gauge fixing must be kept step by step to obtain the right answer.  }, i.e. the color of the vertices.

Although  in \cite{Cachazo:2013iea} was given an algorithm to computed the diagrams found in  {\bf Fig.6.12},  we apply the $\L-$algorithm  since it works when one of the particles is off-shell.

Let us consider the second component of the first vector given in  {\bf Fig.6.12}
\begin{center}
\includegraphics[scale=0.4]{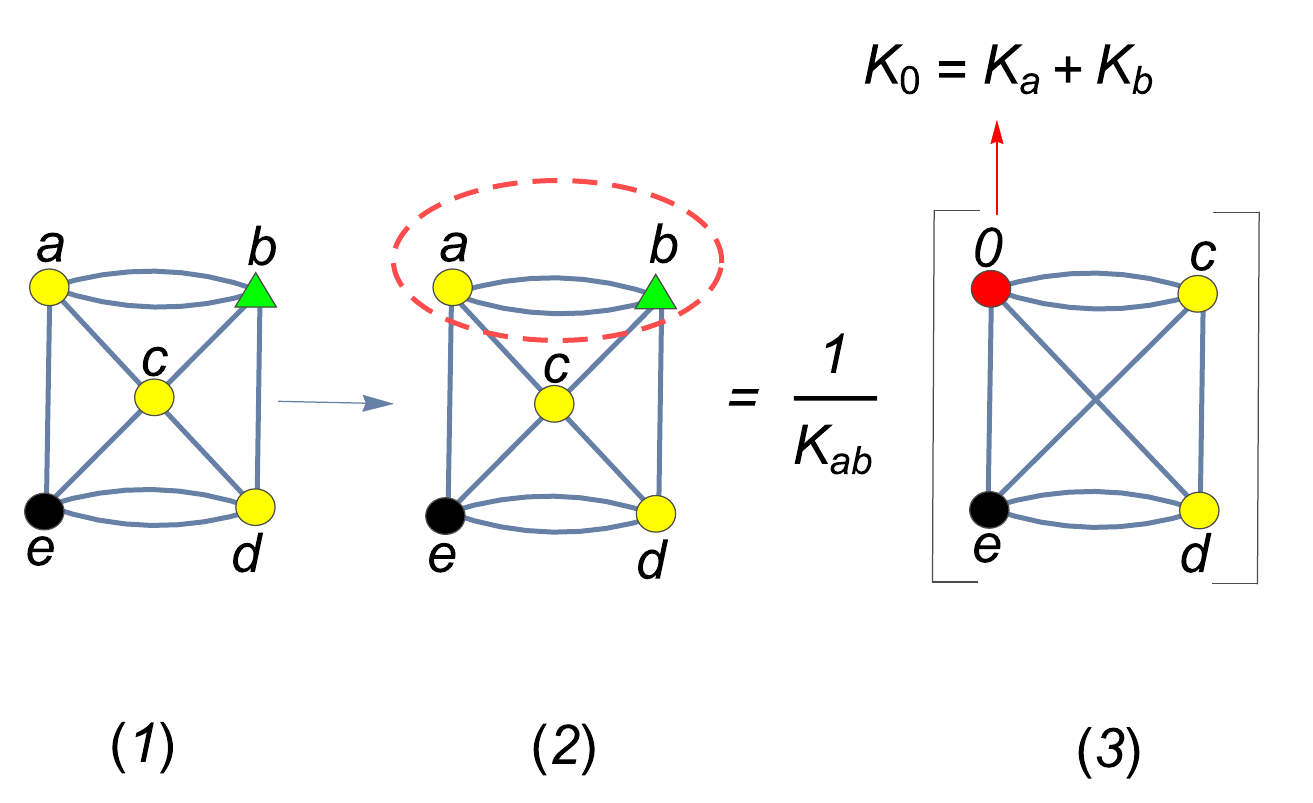}
\includegraphics[scale=0.4]{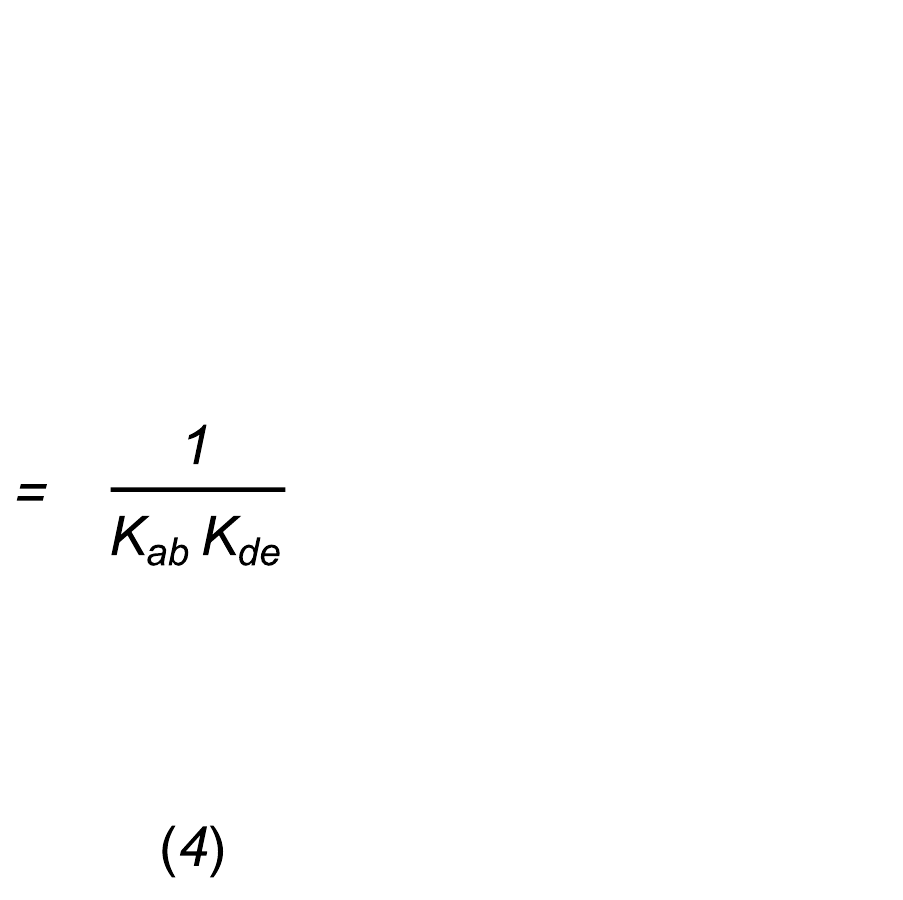}.
\begin{center}
({\bf Fig.6.13})\,{\small {\rm $\L-$algorithm. (1) Integrand. (2) Allowable configurations non zero. (3) Computing the $\L$ integral. (4) Final answer. \,}}
\end{center}
\end{center}
In {\bf Fig.6.13}  we describe step by step the $\L-$algorithm for a particular diagram in {\bf Fig.6.12}:\\
\begin{itemize}
\item {\bf (1)}  We draw the graph to be computed, including the gauge fixing (colored vertices).

\item {\bf (2)} We find the all non-zero allowable configurations, which is only one. 

\item {\bf (3)} We compute the $\L$ integral around $\L=0$. 

\begin{itemize}
\item {\bf (i)}  The scattering equation $1/E_b$ becomes the propagator $1/k_{ab}$. 
\item {\bf (ii)}  The subdiagram obtained on the upper-sheet is a 4-regular graph at three point, which is trivial,, i.e. 1. On the other hand, the 4-regular subdigram obtains on the lower-sheet is a 4-point graph, which is the $(II)$ building block given in {\bf Fig.6.7}.
\item {\bf (iii)} The new massive  particle in the graph on the lower-sheet has momentum $k_0=k_a+k_b$. 
\end{itemize}

\item {\bf (4)} we used the $(II)$ building block  in  {\bf Fig.6.7}  to obtain the final answer.
\end{itemize}

Following the same simple procedure one can compute all graphs  in  {\bf Fig.6.12}, for example, the $m^{L|R}_{22}$ and $m^{L|R}_{12}$ matrix components, respectively,  are given by
\begin{center}
\quad\quad\qquad\qquad\qquad
\includegraphics[scale=0.4]{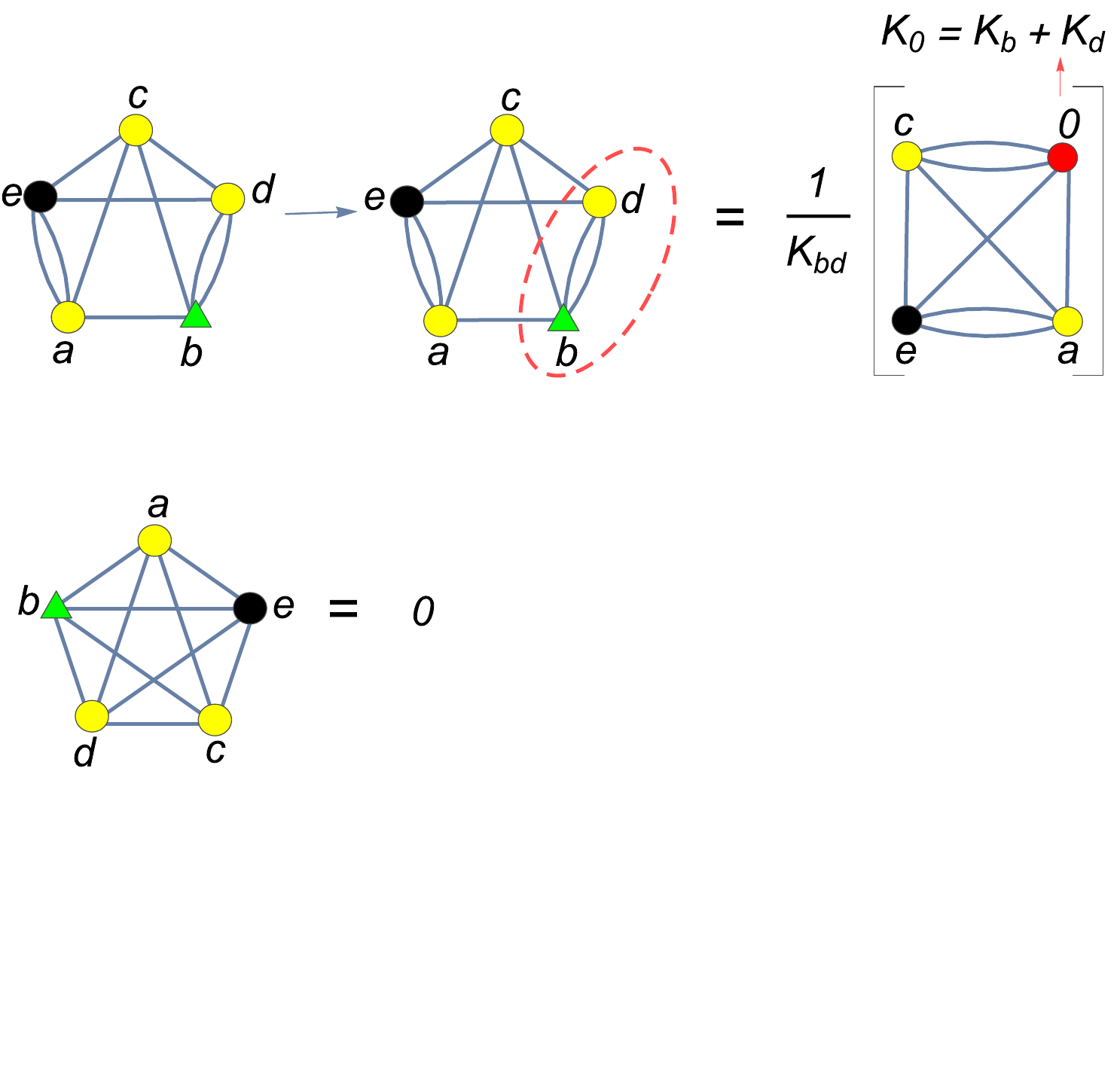}
\includegraphics[scale=0.4]{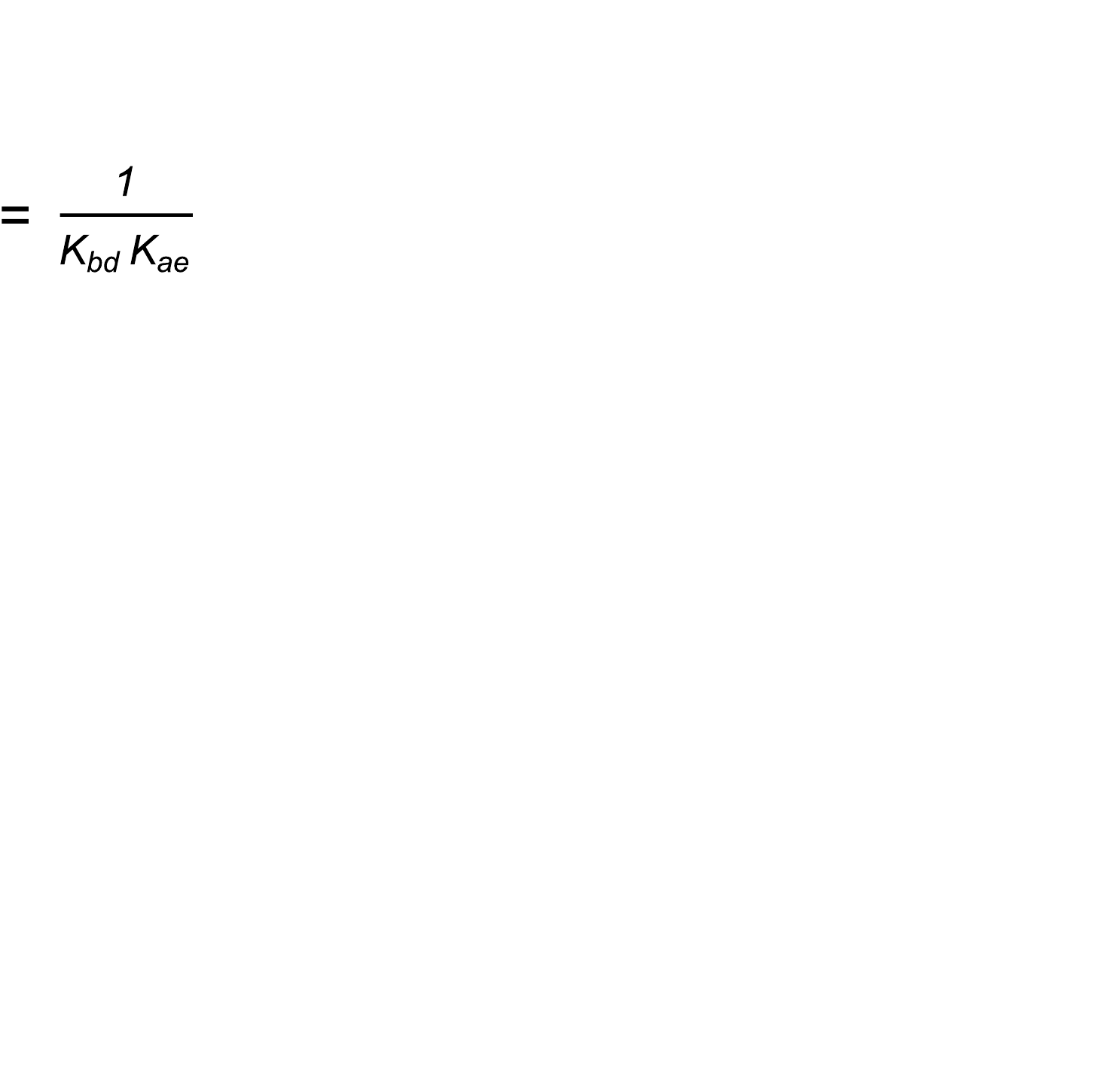} 
\begin{center}
({\bf Fig.6.14})\,{\small {\rm Computation of the  $m^{L|R}_{22}$ and $m^{L|R}_{21}$ matrix components, respectively. \,}}
\end{center}
\end{center}
Therefore, the ${\cal I}_{(V)}$ building block is 
\begin{align}\label{buidingfive}
{\cal I}_{(V)}&=\left(\frac{1}{k_{ab}k_{de}}\right)^{2}
\left(
\begin{matrix}
1\\
1
\end{matrix}
\right)^{\rm T}
\left(
\begin{matrix}
\frac{1}{k_{bce} k_{be}} & 0\\
0 & \frac{1}{k_{ae} k_{bd}}
\end{matrix}
\right)^{-1}
\left(
\begin{matrix}
-1\\
\quad 1
\end{matrix}
\right)\nonumber\\
&=\left(\frac{1}{k_{ab}k_{de}}\right)^{2}
( k_{ae} k_{bd}-k_{bce} k_{be} )\,\, .
\end{align}

\section{Examples}\label{examples}

Although in the previous section we have already applied the $\L-$algorithm, the idea here is to give some non-trivial examples  in order to show the power of this new algorithm.

This section is divided as follows, the first example show us  how to use the $\L$ algorithm, which will be  applied  over a six point  highly non trivial diagram. The idea of the second one is to mix the $\L$ algorithm with the KLT general algorithm \cite{humbertoF}, where we will compute a six point diagram which cannot be performed just with the $\L$ algorithm.  Finally, the last one is given in order to illustrate the using of all building blocks, with this in mind we choose a non trivial  eight point diagram.

\subsection{Six-Point}\label{sixpoint}
 
Let us consider the following two non-trivial six-point examples
\begin{center}
\quad
\includegraphics[scale=0.38]{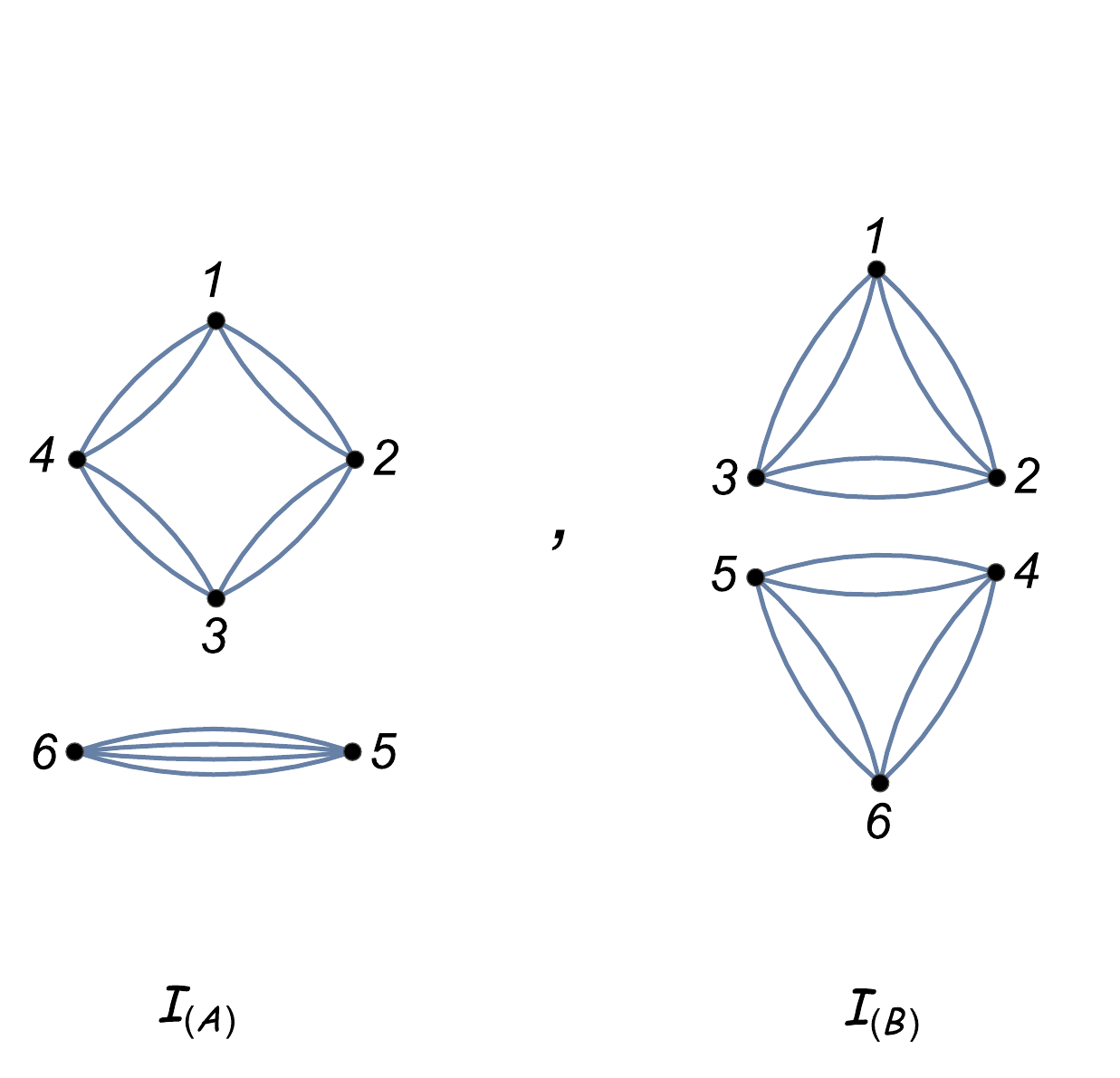}\,\, .
\begin{center}
({\bf Fig.7.1})\,{\small {\rm Six-point  examples. \,}}
\end{center}
\end{center}

The first one, ${\cal I_{(A)}}$, will be computed  just using the $\L-$algorithm. For the second one, ${\cal I_{(B)}}$, the $\L-$algorithm is not enough.  We will combine  the  $\L$ and the general KLT algorithm \cite{humbertoF}  to compute it.

\subsubsection{$\mathcal{I_{(A)}}$-Computation}\label{IAcomputation}

In order to avoid singular configurations we choose the following  gauge fixing
\begin{center}
\quad
\includegraphics[scale=0.37]{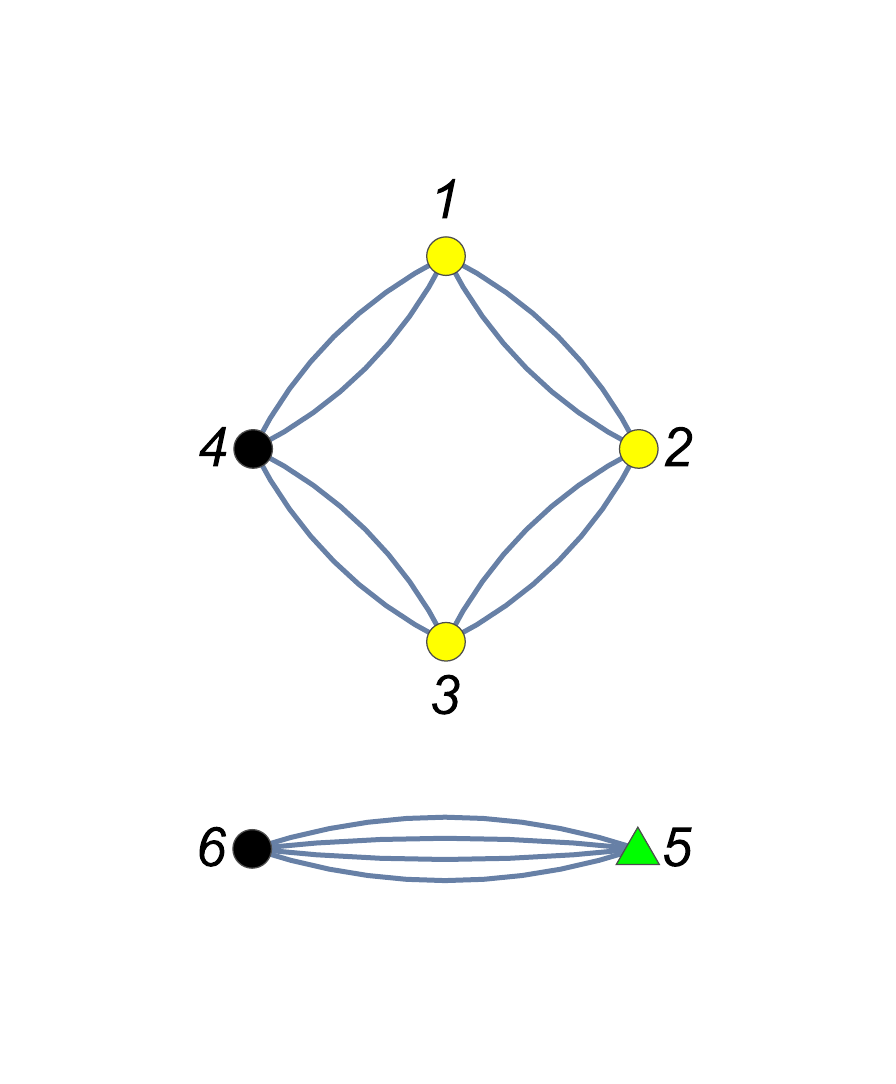}\,\, .
\begin{center}
({\bf Fig.7.2})\,{\small {\rm Gauge Fixing. \,}}
\end{center}
\end{center}

This is straightforward to see that there are two kind of allowable configurations. The first one is given by  the diagrams
\begin{center}
\includegraphics[scale=0.38]{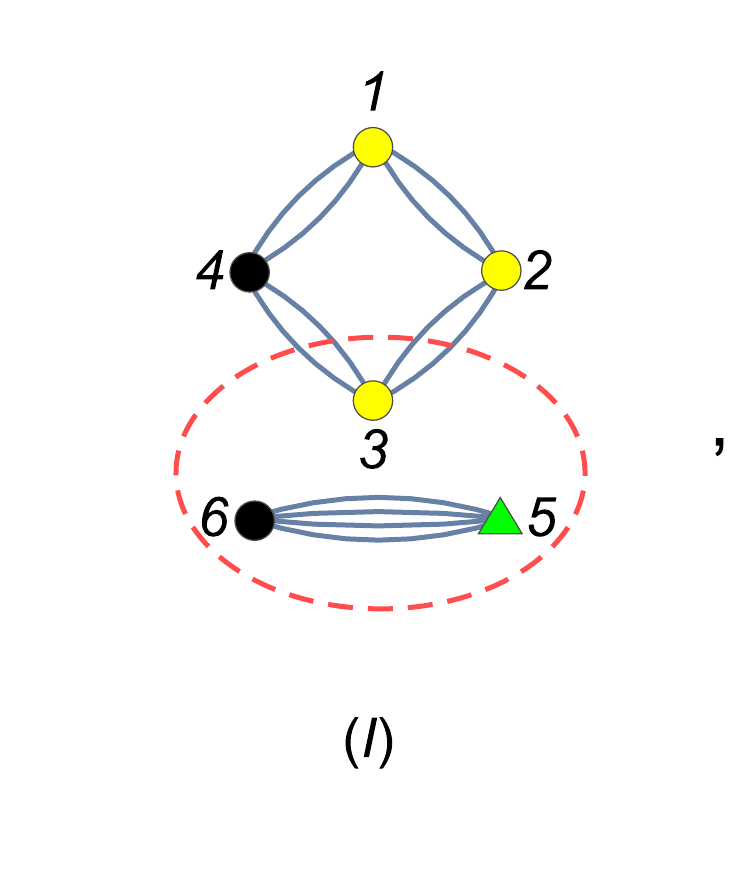}
\includegraphics[scale=0.38]{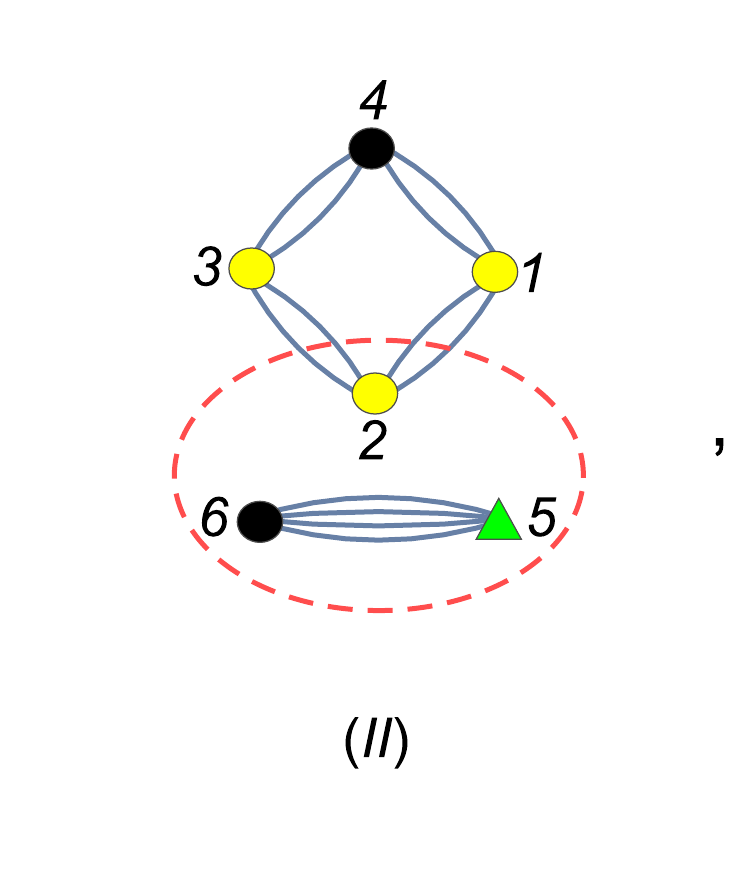}
\includegraphics[scale=0.38]{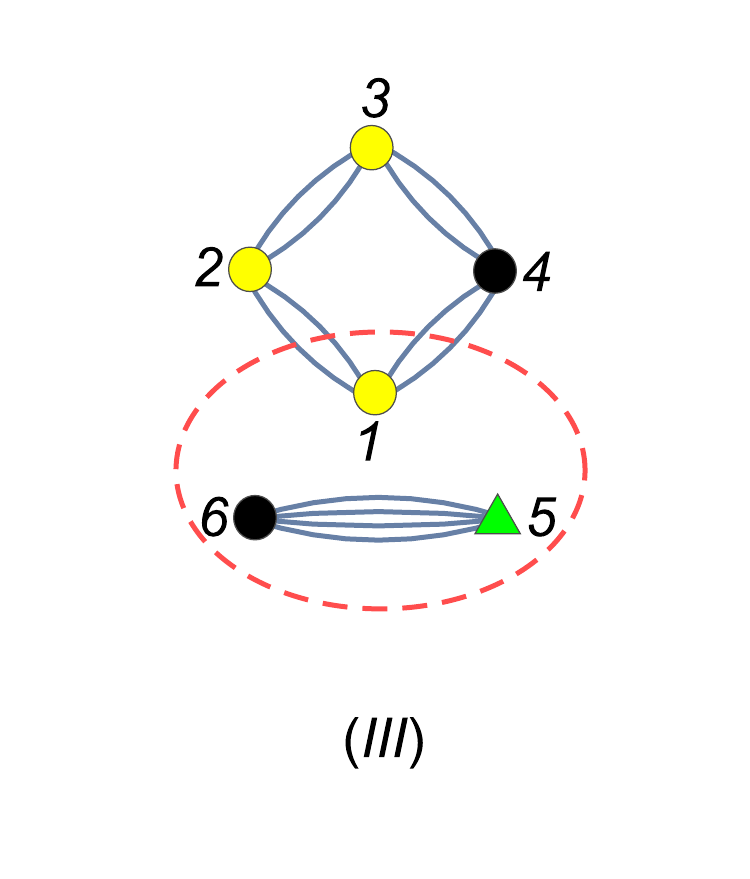}\,\, ,
\begin{center}
({\bf Fig.7.3})\,{\small {\rm Allowable configurations of type 1. \,}}
\end{center}
\end{center}
and the second one by
\begin{center}
\includegraphics[scale=0.38]{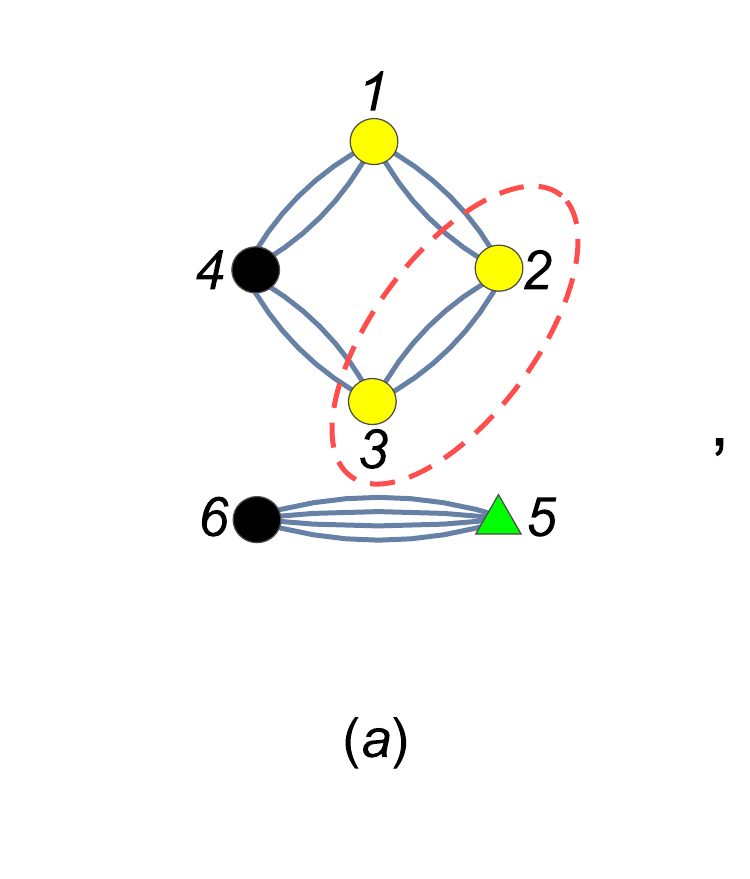}
\includegraphics[scale=0.38]{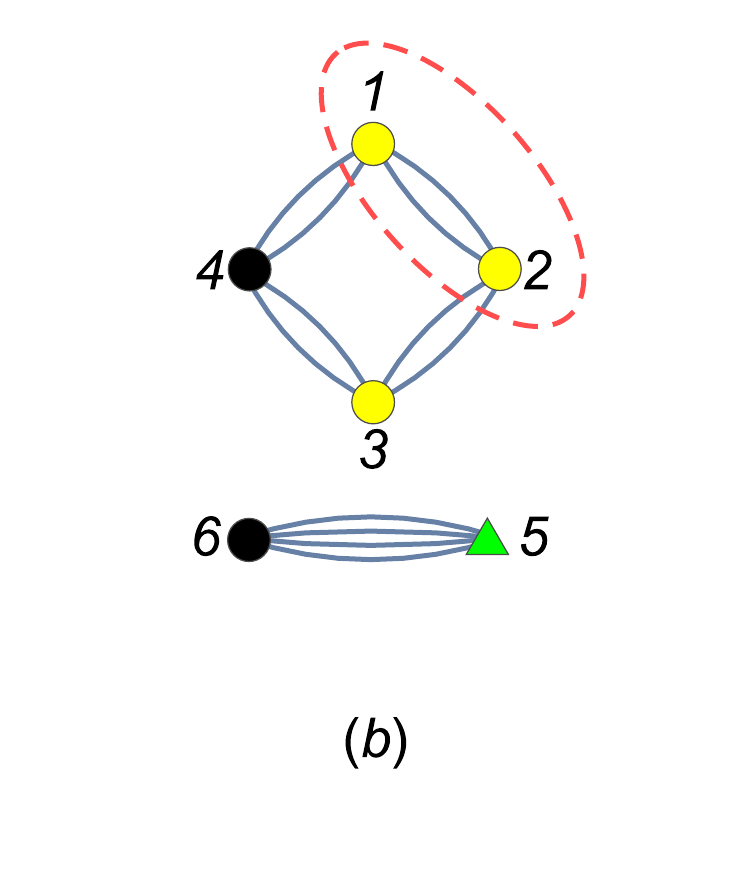}\,\, .
\begin{center}
({\bf Fig.7.4})\,{\small {\rm Allowable configurations of type 2. \,}}
\end{center}
\end{center}

Since the elements of each type are totally analogues then we only compute one of each set.  

Let us begin  by computing the $(I)$ configuration  in  {\bf Fig.7.3}. Applying the $\L-$algorithm, the $E_5$ scattering equation becomes the $1/k_{356}$ propagator and the diagram breaks into two graphs (upper and lower sheet) in the original CHY approach 
\begin{center}
\includegraphics[scale=0.42]{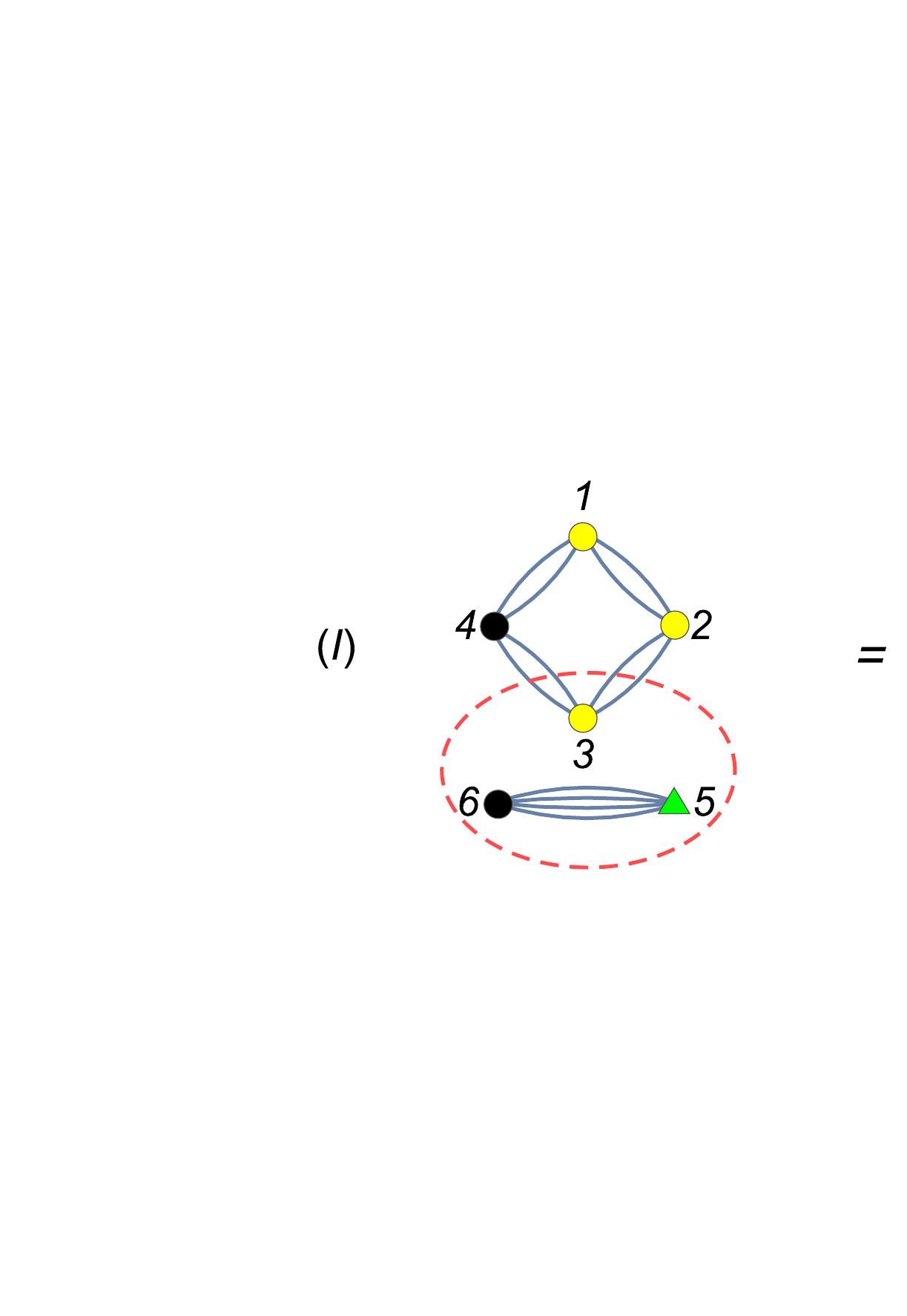}
\includegraphics[scale=0.42]{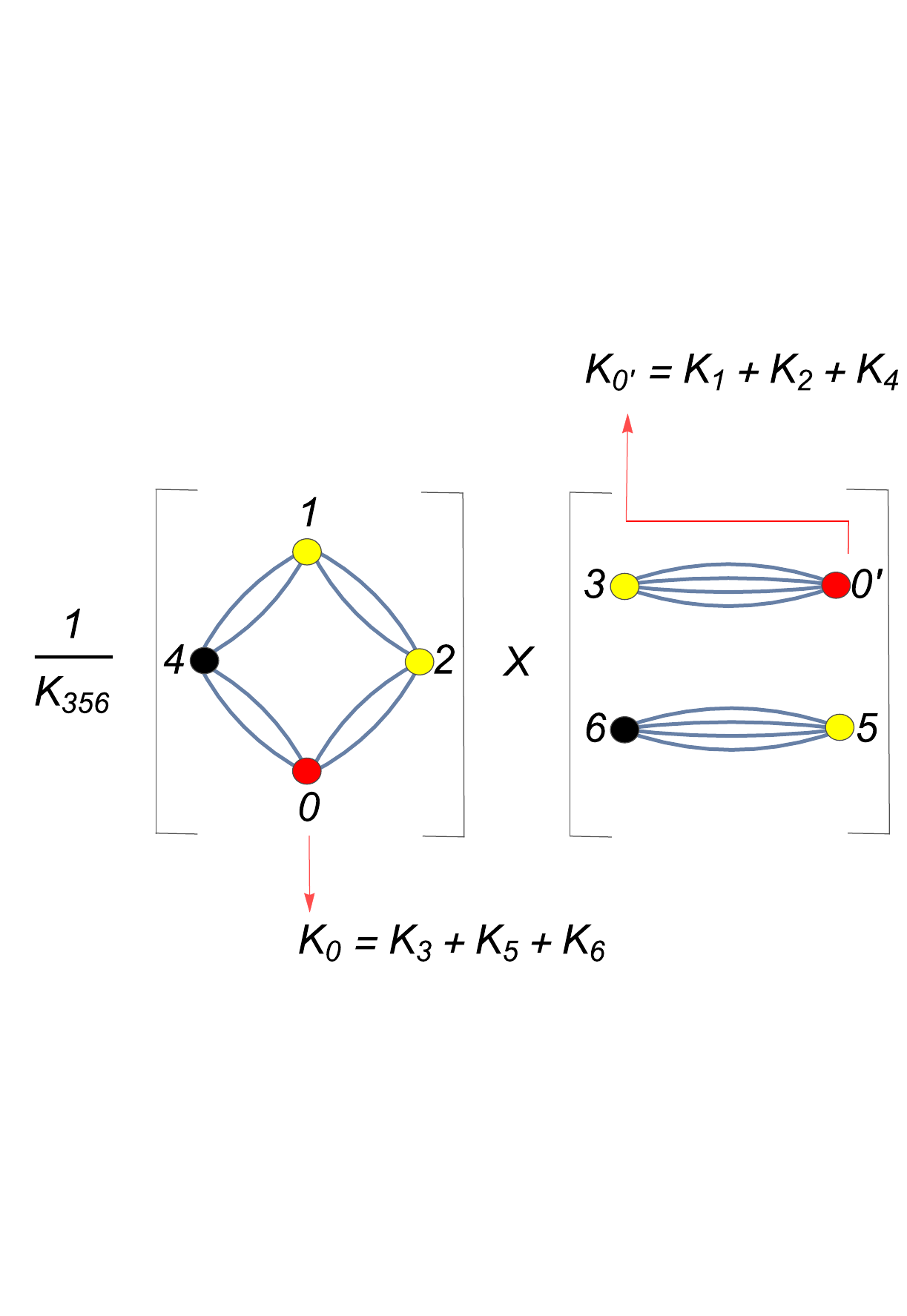}\,\, \qquad\qquad\qquad .
\begin{center}
({\bf Fig.7.5})\,{\small {\rm Computing the $(I)$ diagram. \,}}
\end{center}
\end{center}
Using the building blocks given in section \ref{bblocks} (see  {\bf Fig.6.8} and {\bf Fig.6.9}),  we are able to find the final  answer  for the $(I)$ configuration in  {\bf Fig.7.3}. Thus,  following the same procedure for the  $(II)$ and $(III)$ configurations one obtains 
\begin{center}
\includegraphics[scale=0.65]{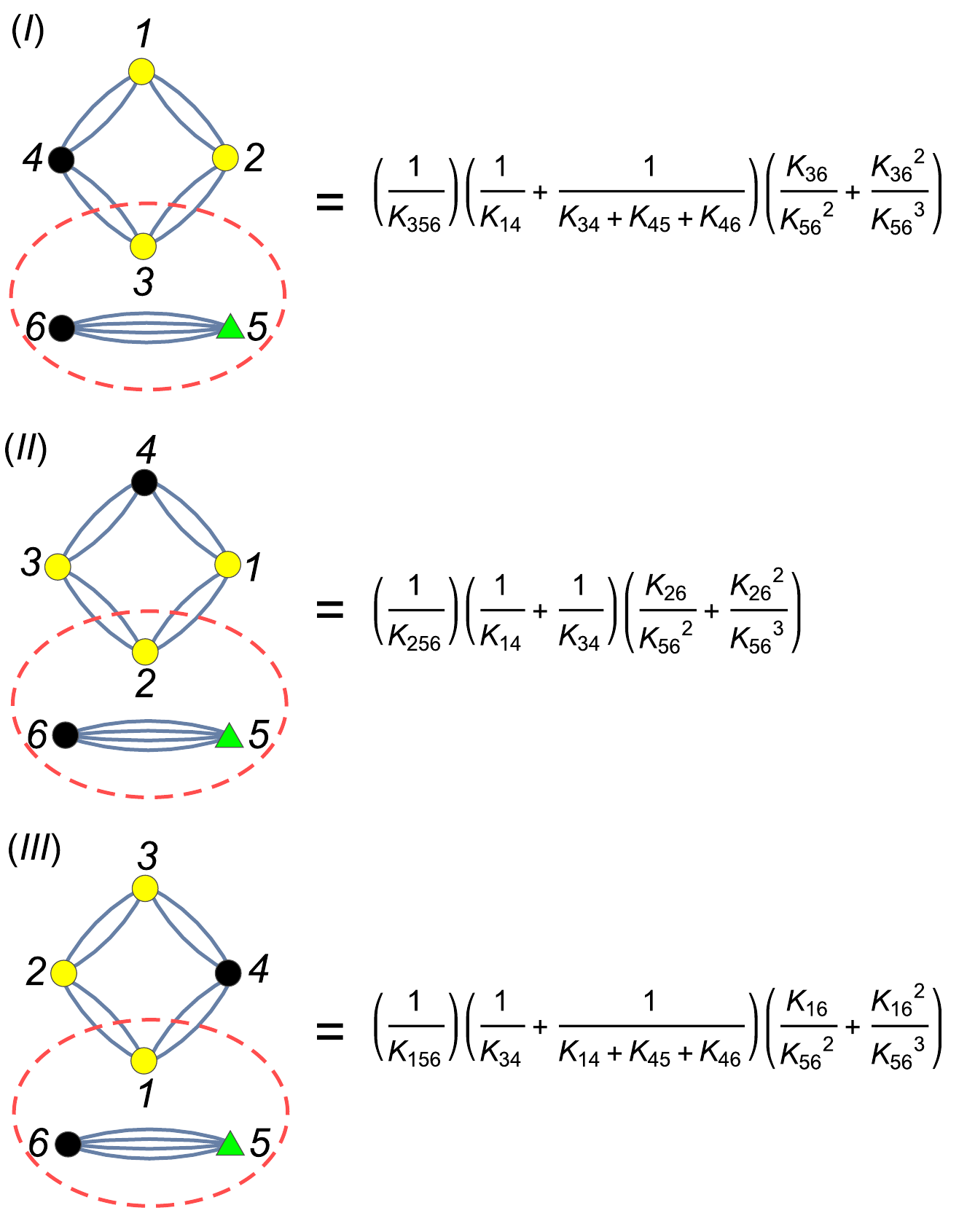}
\begin{center}
({\bf Fig.7.6})\,{\small {\rm  Results of the  $(I),(II)$ and $(III)$ configurations. \,}}
\end{center}
\end{center}

We must now compute the  (a) and (b) configurations  in  {\bf Fig.7.4}. Let us start with  the (a) configuration. From the $\L-$algorithm one has
\begin{center}
\includegraphics[scale=0.45]{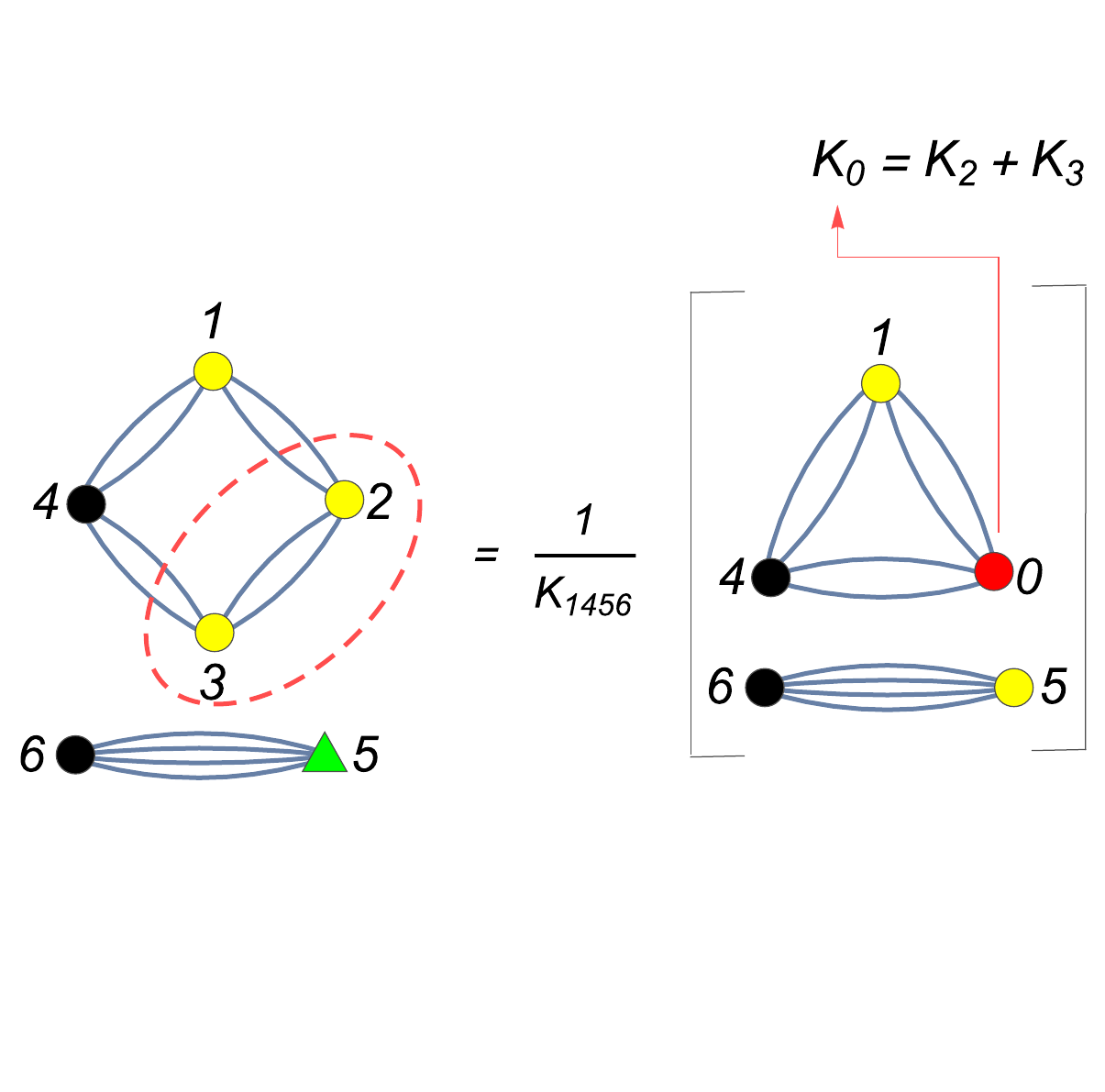}\,\,\, .
\begin{center}
({\bf Fig.7.7})\,{\small {\rm Computing the (a) configuration in {\bf Fig.7.4}. \,}}
\end{center}
\end{center} 
So as to apply  the $\L-$algorithm  on the resulting 5-point graph, we  must fix the scale symmetry (S.S) . We gauge the  $\s_4$ puncture 
\begin{center}
\includegraphics[scale=0.55]{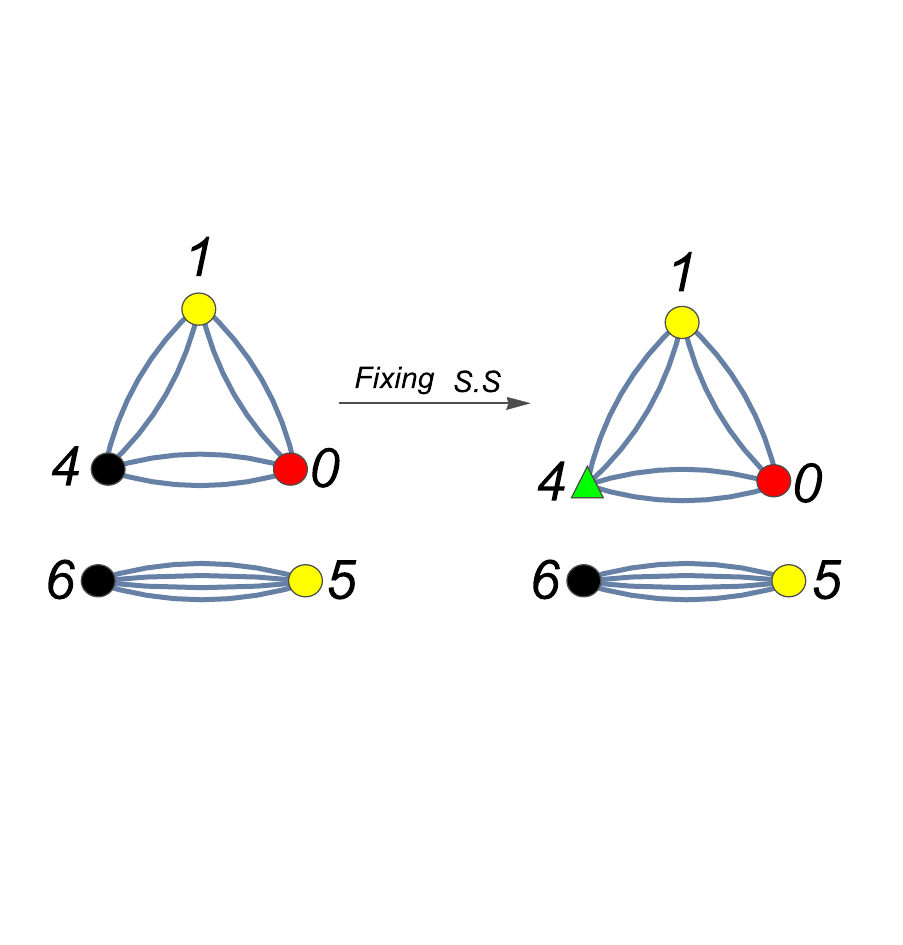}\,\,\, .
\begin{center}
({\bf Fig.7.8})\,{\small {\rm Gauging the Scale Symmetry (Iterative process)\,.}}
\end{center}
\end{center} 
It is simple to see that the non-zero allowable configurations in {\bf Fig.7.8} are given by
\begin{center}
\includegraphics[scale=0.48]{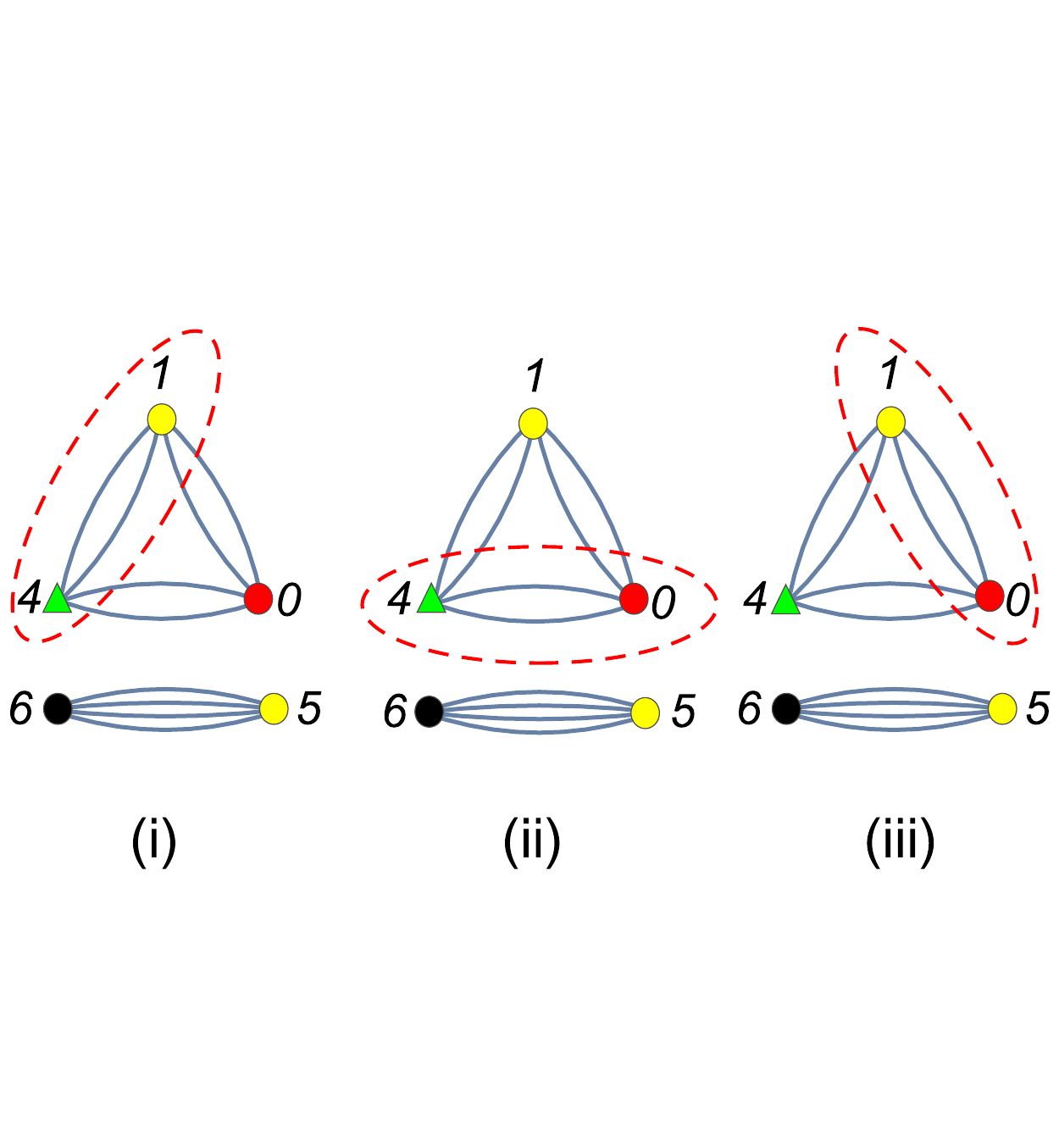}\,\,\, .
\begin{center}
({\bf Fig.7.9})\,{\small {\rm Allowable configurations 5-point graph (Iterative process)\,.}}
\end{center}
\end{center} 
These three configurations are straightforward to compute applying the $\L-$algorithm 
\begin{center}
\includegraphics[scale=0.45]{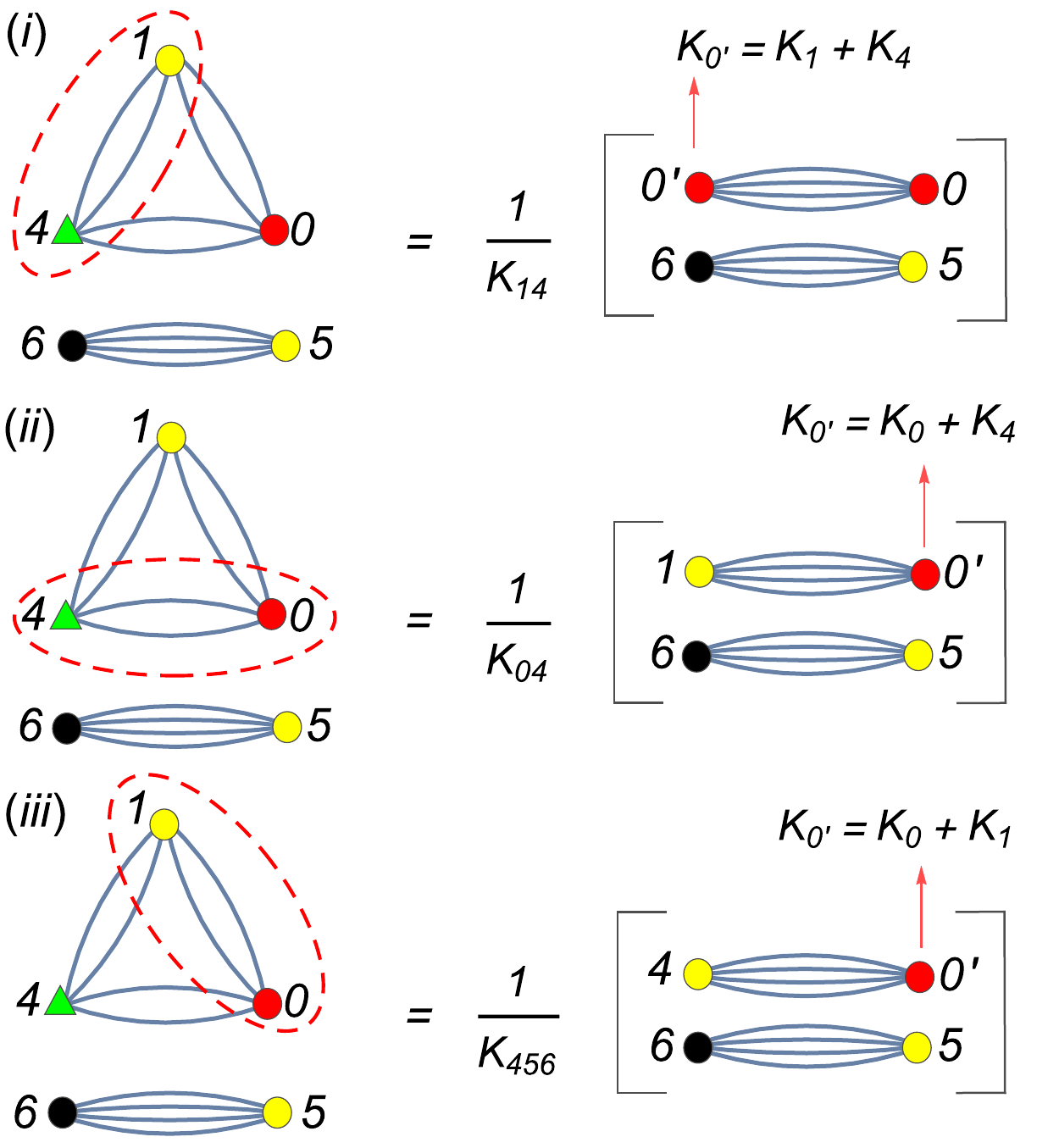}\,\,\, ,
\begin{center}
({\bf Fig.7.10})\,{\small {\rm Allowable configurations five-point graph (Iterative process)\,.}}
\end{center}
\end{center} 
where it is useful to  remember that $k_0=k_2+k_3$, {\bf Fig.7.7}. From the building blocks of the section \ref{bblocks}, {\bf Fig.6.9}, we obtain the final answer for the (a) configuration in  {\bf Fig.7.4}
\begin{center}
\includegraphics[scale=0.67]{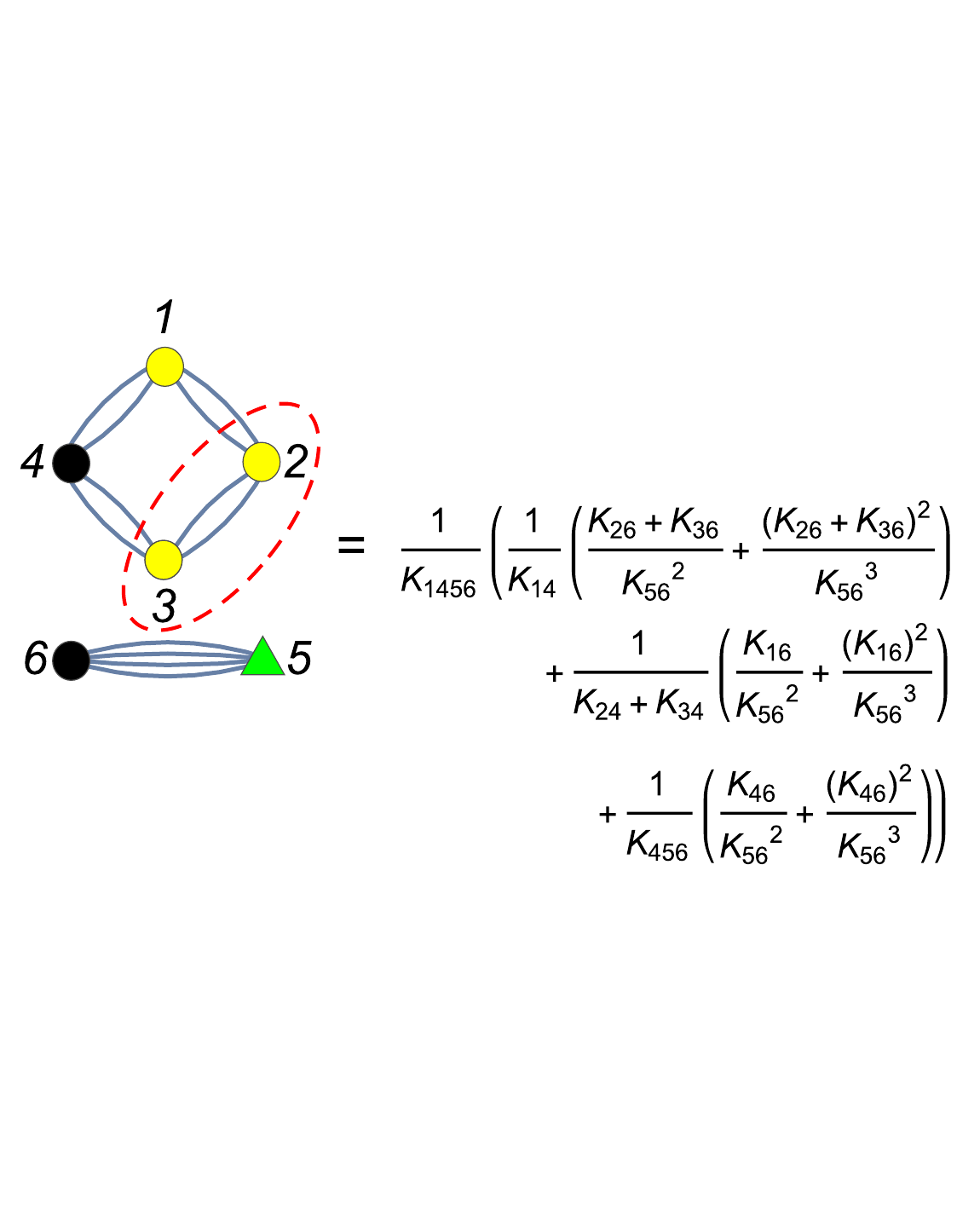}\,\,\, .
\begin{center}
({\bf Fig.7.11})\,{\small {\rm Result (a) configuration.}}
\end{center}
\end{center} 
Performing the same procedure for the (b) configuration one obtains 
\begin{center}
\includegraphics[scale=0.67]{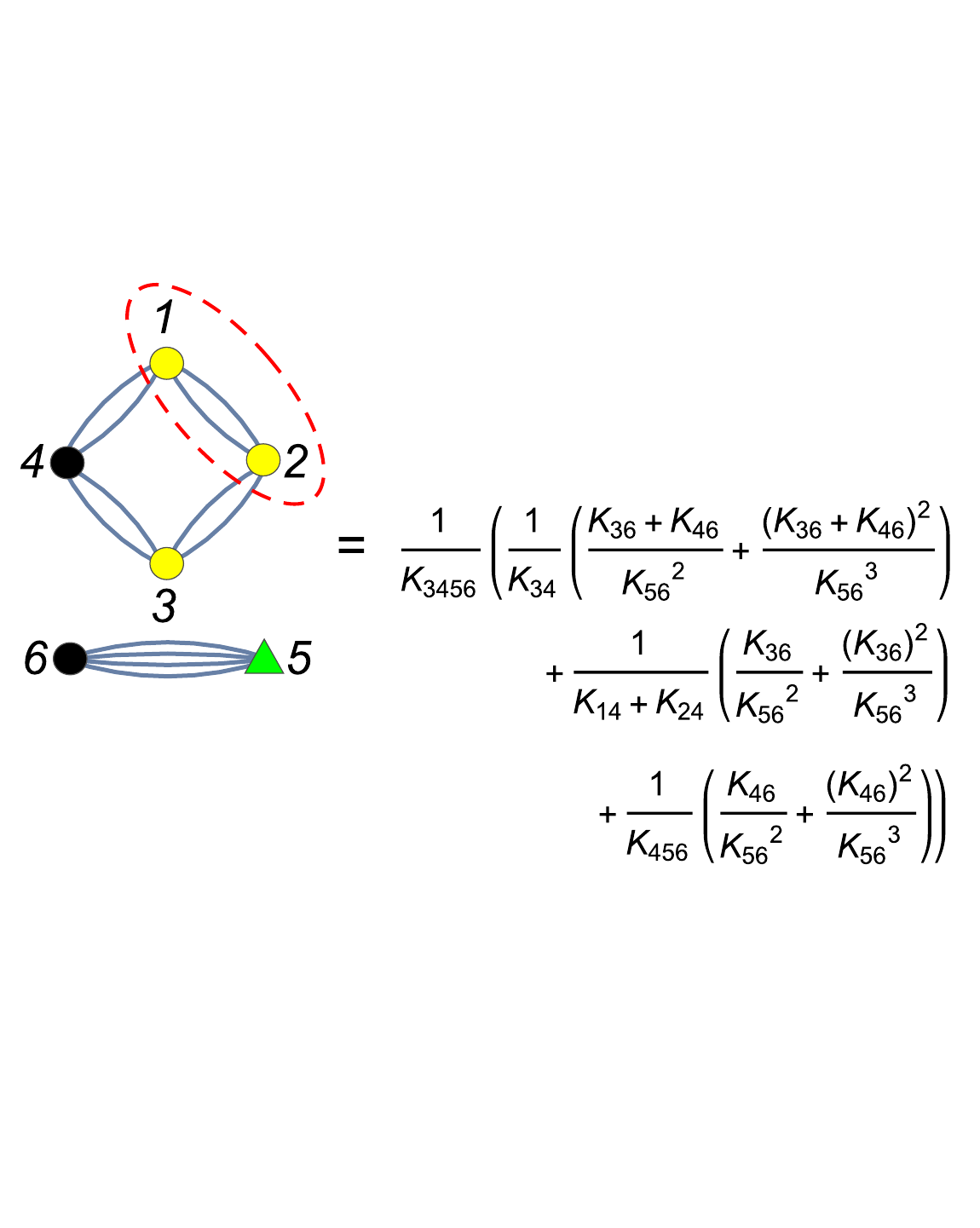}\,\,\, .
\begin{center}
({\bf Fig.7.12})\,{\small {\rm Result (b) configuration.}}
\end{center}
\end{center}

Therefore, summing over all allowable configurations we obtain the total answer for the ${\cal I_{(A)}}$ graph, which is given by the non trivial expression 
\begin{align}\label{rsixpoint}
{\cal I_{(A)}}&=(I) +(II)+(III)+({\rm a}) + ({\rm b})\nonumber\\
 &=  \frac{ B[14:34+45+46] }{k_{356}}\frac{(k_{36})^2 B[56:36]}{(k_{56})^2} 
 + 
\frac{ B[14:34] }{k_{256}}
\frac{(k_{26})^2 B[56:26] }{ (k_{56} )^2 }
\nonumber\\
& + \frac{B[34:14+45+46]}{k_{156}}\,
\frac{ (k_{16})^2 B[56:16] }{ (k_{56})^2 }
\\
&+  \frac{1}{k_{1456}}\left[
\frac{ (k_{26}+k_{36})^2 B[56:26+36] }{k_{14} (k_{56})^2 }
+
\frac{ (k_{16})^2 B[56:16] }{ (k_{24}+k_{34}) (k_{56})^2  }
+\frac{ (k_{46})^2 B[56:46] }{k_{456}  (k_{56})^2   }
\right]\nonumber\\
&+ \frac{1}{k_{3456}}\left[\frac{ (k_{36}+k_{46})^2 B[56:36+46]}{   k_{34} (k_{56})^2   }  
+
\frac{  (k_{36})^2 B[56:36] }{  (k_{14}+k_{24}) (k_{56})^2  }
+
\frac{ (k_{46})^2 B[56:46]   }{k_{456}  (k_{56})^2   }
\right]\nonumber,
\end{align}
where we have defined
\begin{equation}
B[A+ B+\cdots I : C + B+\cdots J]:=\frac{1}{k_{A}+ k_{B}+\cdots k_I}+
\frac{1}{k_{C}+ k_{D}+\cdots k_J},
\end{equation}
and the labels $A,B,C,D,I$ and $J $ mean a index set, for example $k_A:=k_{a_1\cdots a_m}$.

The \eqref{rsixpoint} result was checked numerically.

\subsubsection{$\mathcal{I_{(B)}}$-Computation (General KLT and $\L$ Algorithms)}\label{IBcomputation}

In  section \ref{bblocks} we have combined the general KLT algorithm \cite{humbertoF} and the $\L-$algorithm, respectively,  in order to compute the  $(V)$  building block, however,  in this section our idea is the opposite.  First, we apply the $\L-$algorithm as far as it is possible. From this method we will obtain subdiagrams with less vertices than the original one.  Second, we perform the general KLT algorithm on these subdiagrams and finally we will be able to use the $\L-$algorithm, again, to  compute the diagrams into the vectors and matrix, such as it was done in section \ref{fivebb}.

Let us remember that in the general KLT algorithm \cite{humbertoF} one must  find a base, left $({\cal L})$ and right $({\cal R})$,  such that all graphs  have a Hamiltonian decomposition, i.e. the integrands are product of two Parke-Taylor factors.  One of its main drawback is to compute the inverse of the Gram matrix given by the product among the left and right base, $m^{\cal L|R}$.  For example, in six-point it is necessary to invert a $6\times 6$ matrix.

However, since our idea is first to apply the $\L-$algorithm then this drawback is softened.

Let us consider the ${\cal I_{(B)}}$ example  in  {\bf Fig.7.1}. In order to avoid singular allowable configurations we  set the following gauge fixing 
\begin{center}
\includegraphics[scale=0.4]{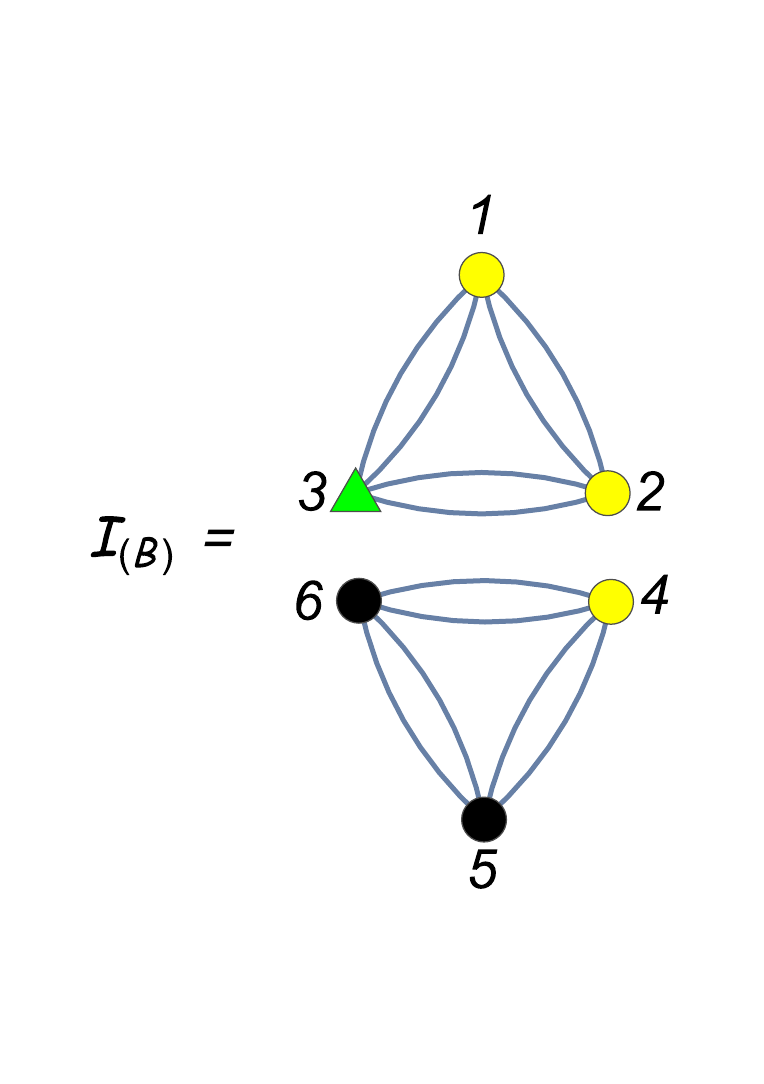}\,\,\, .
\begin{center}
({\bf Fig.7.13})\,{\small {\rm Gauge Fixing\,.}}
\end{center}
\end{center} 
There are only three non-zero allowable configurations
\begin{center}
\includegraphics[scale=0.4]{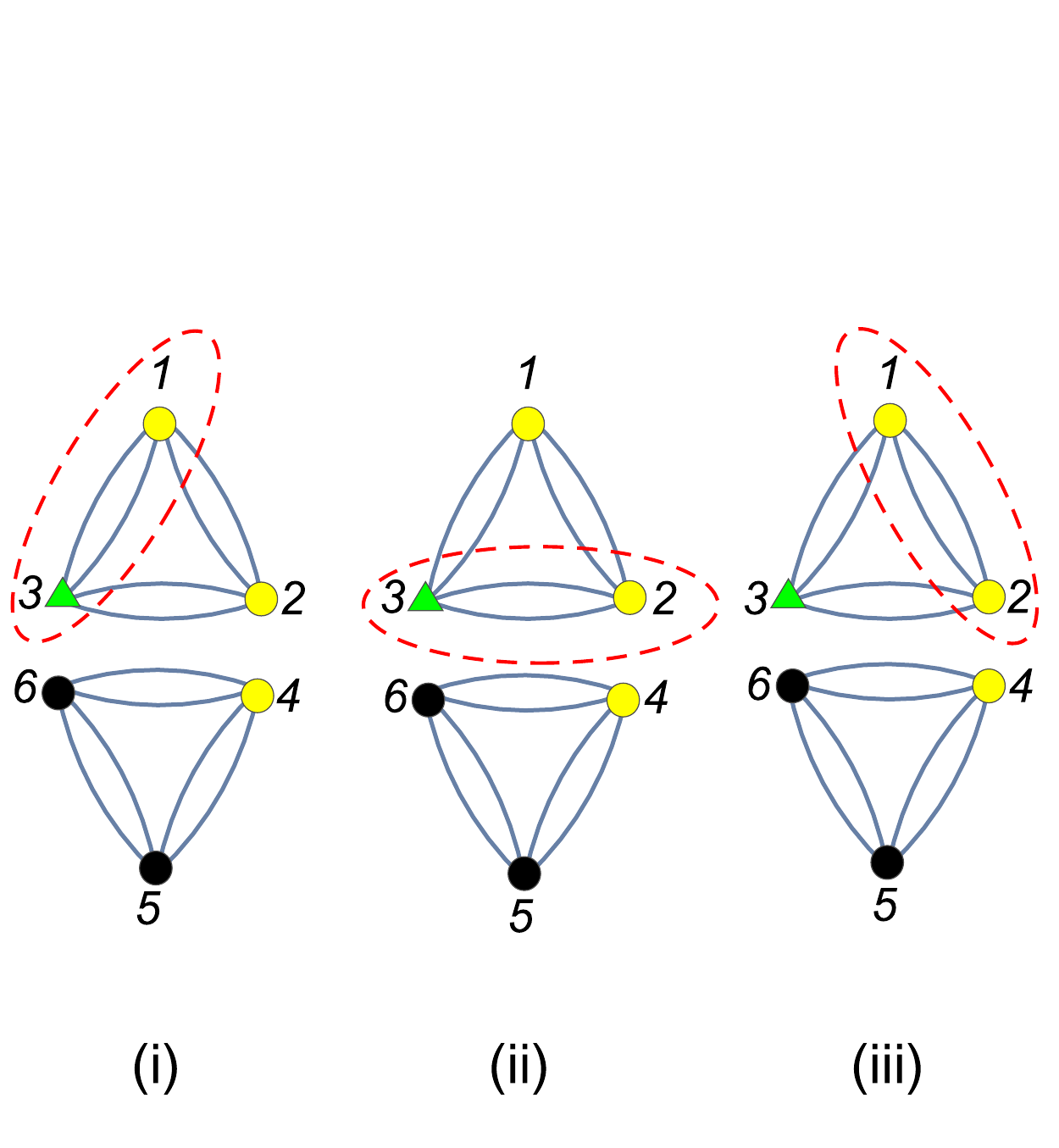}\,\,\, .
\begin{center}
({\bf Fig.7.14})\,{\small {\rm Non-zero configurations\,.}}
\end{center}
\end{center}
Since these three configurations are the same up to relabel the (1,2,3) vertices then it is enough just to compute one of them,  for example  we choose the first one,  (i) configuration. Following the techniques presented in the section \ref{IAcomputation} one obtains
\begin{center}
\includegraphics[scale=0.45]{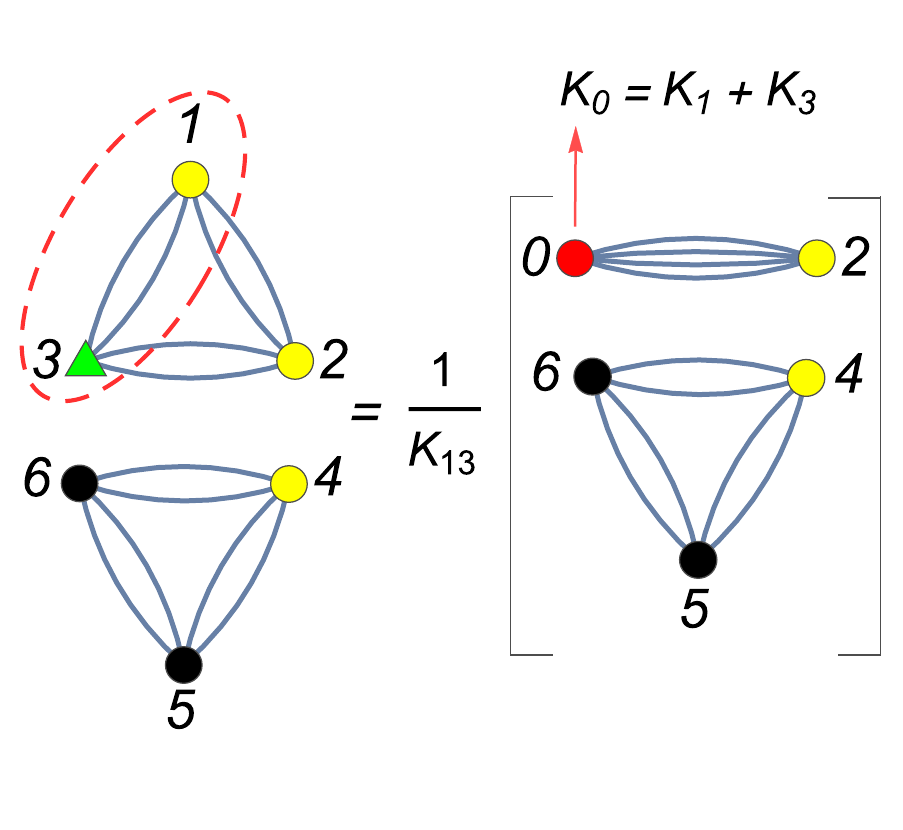}\,\,\, .
\begin{center}
({\bf Fig.7.15})\,{\small {\rm Computing the (i) configuration.}}
\end{center}
\end{center} 
The 5 point graph on the right hand side can not  be computed using the $\L-$algorithm presented in section \ref{Lalgorithm}, therefore we use the general KLT-algorithm \cite{humbertoF}.  

Following the same procedure used to compute the $(V)$ building block  in {\bf Fig.6.7} (general KLT algorithm), we must break the 5 point graph (4-regular graph)  into two 2-regular graphs  (Left and Right)    
\begin{center}
\includegraphics[scale=0.45]{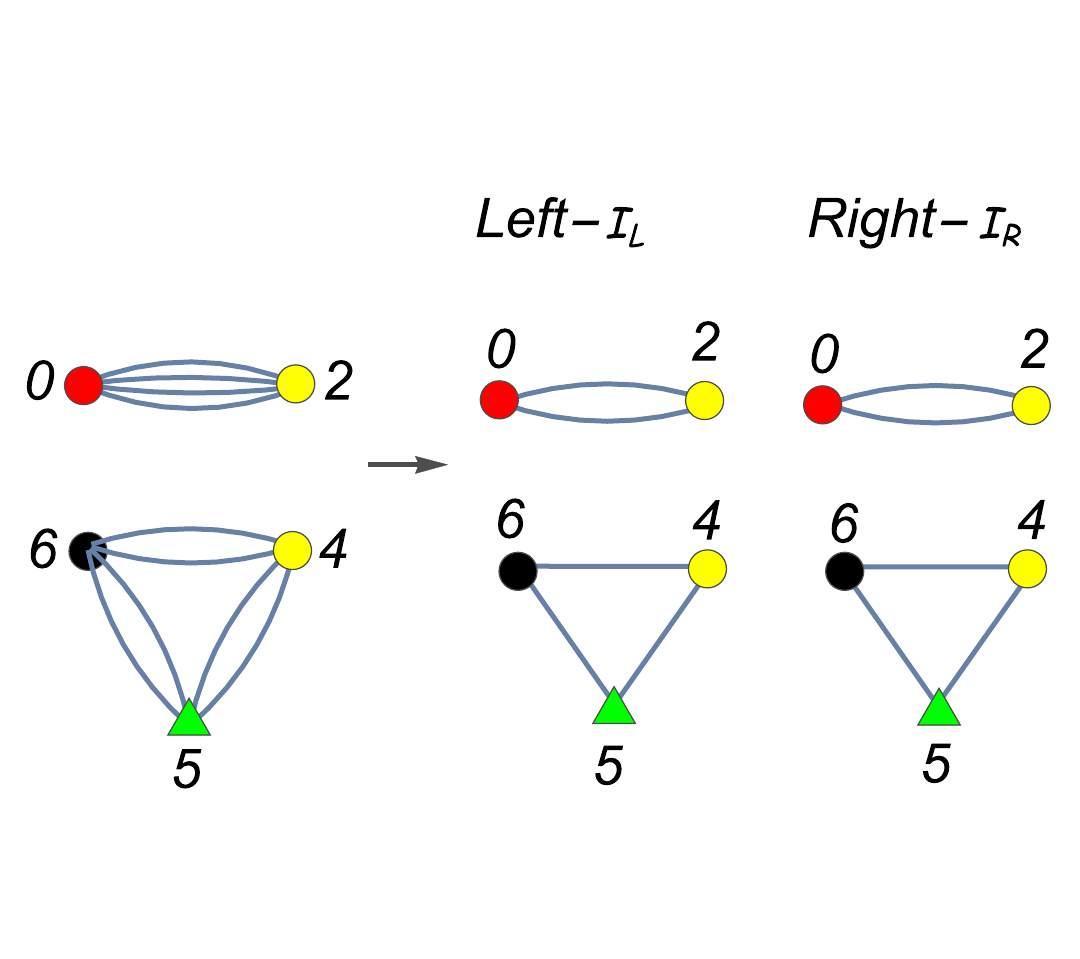}\,\,\, ,
\begin{center}
({\bf Fig.7.16})\,{\small {\rm Splitting the 5 point 4-regular graph in two 2-regular graphs (Left and Right)\,. The ``5" vertex  has been fixed by the scale symmetry.}}
\end{center}
\end{center} 
where we have fixed the vertex number 5 using the scale symmetry.  We choose the  left and right base as\footnote{Unlike of the general KLT algorithm presented in \cite{humbertoF},  we must keep the initial gauge fixing. This is important because the $\L-$algorithm has generated a massive particle (red vertex). }
\begin{center}
\includegraphics[scale=0.4]{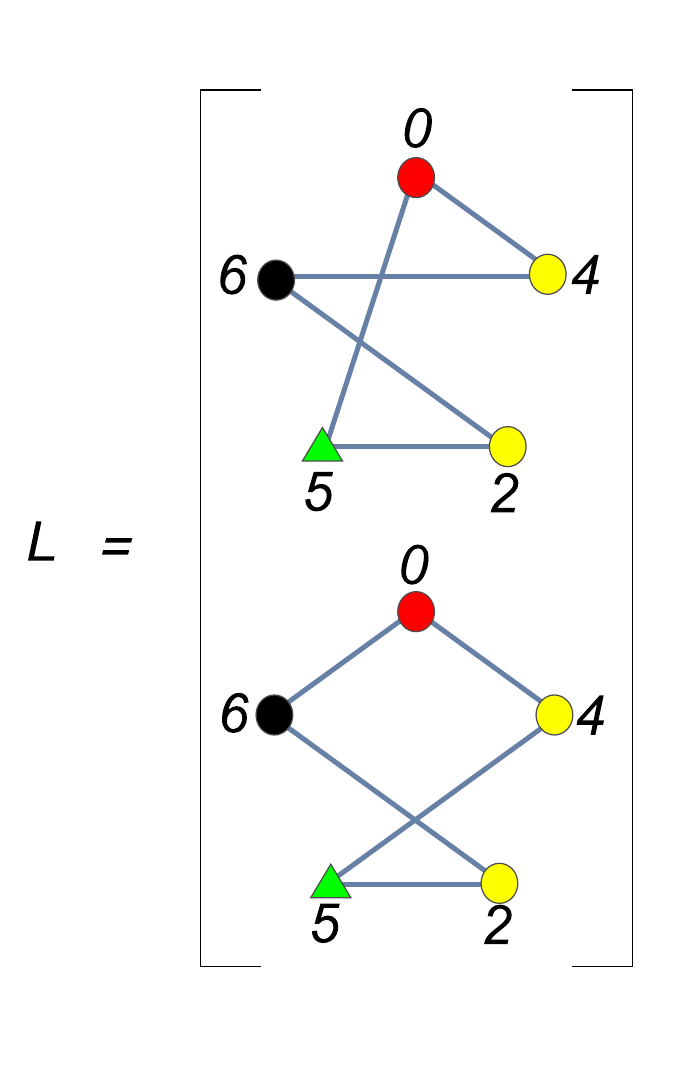}\quad , \quad
\includegraphics[scale=0.4]{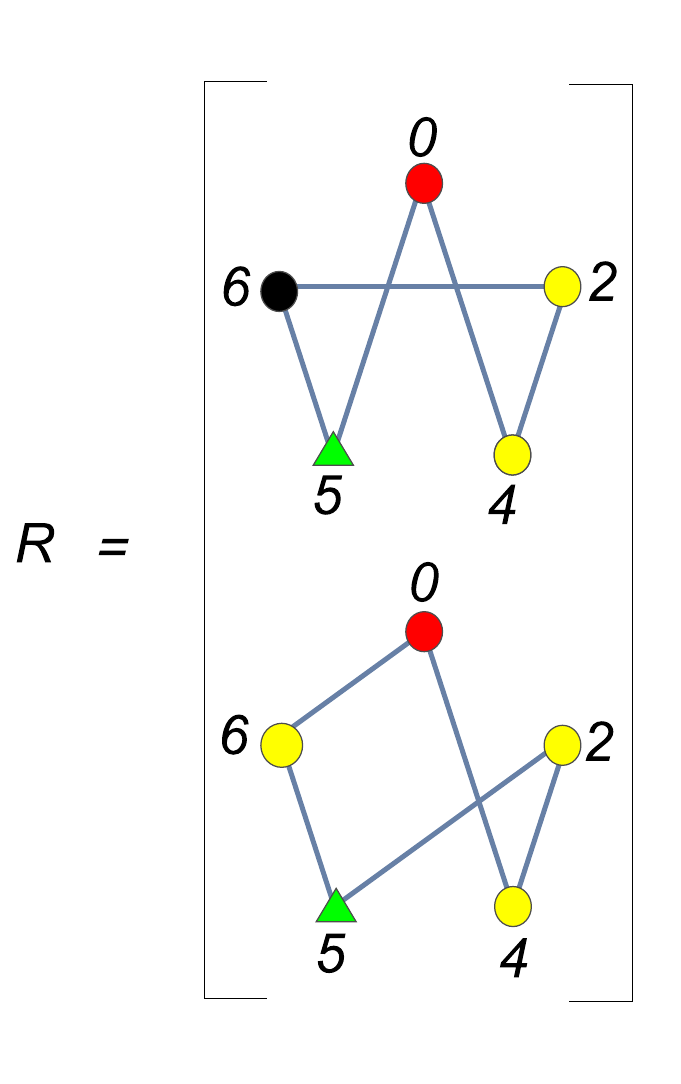}
\,\,\, ,
\begin{center}
({\bf Fig.7.17})\,{\small {\rm Left and Right base. }}
\end{center}
\end{center} 
such that  the diagrams in the  $(L {\cal I_{L}}) $ and $(R {\cal I_{R}}) $ vectors have a Hamiltonian decomposition \cite{graph1,graph2}. Thus, we can write the 5-point diagram in {\bf Fig.7.16} as the matrix product  $(L{\cal I_L}) (m^{L|R})^{-1} (R{\cal I_R})$, diagrammatically  one has
\begin{center}
\includegraphics[scale=0.4]{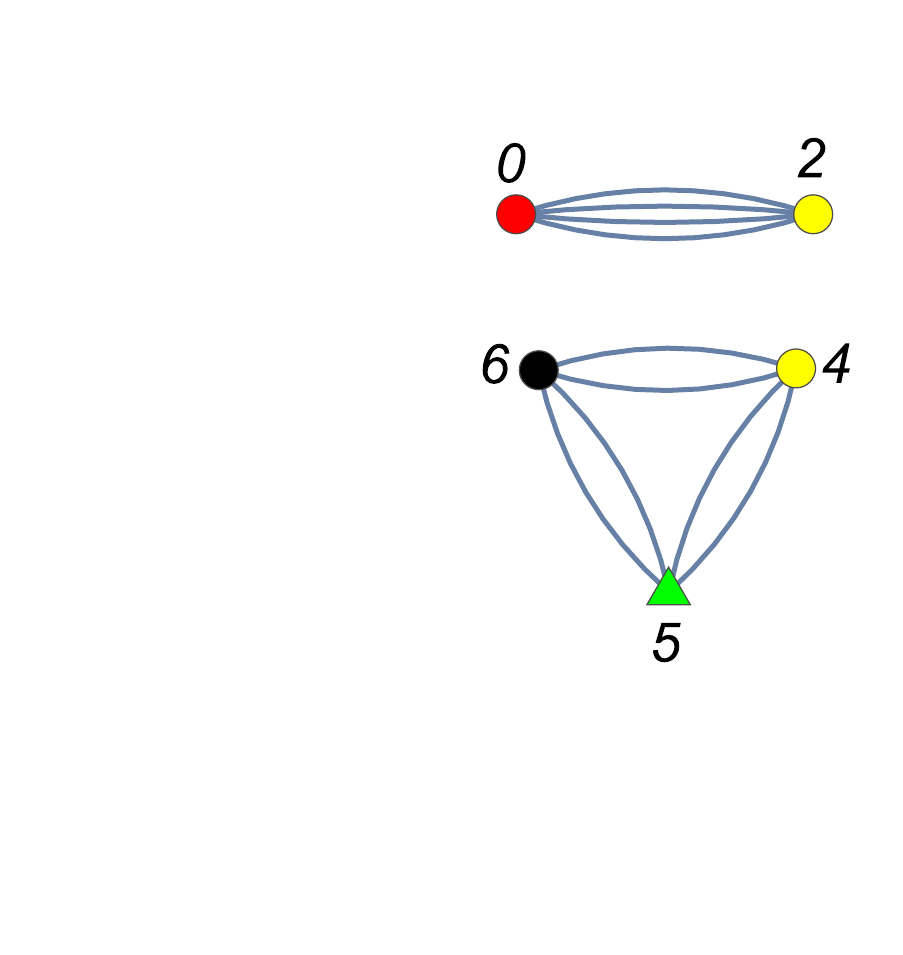}
\includegraphics[scale=0.4]{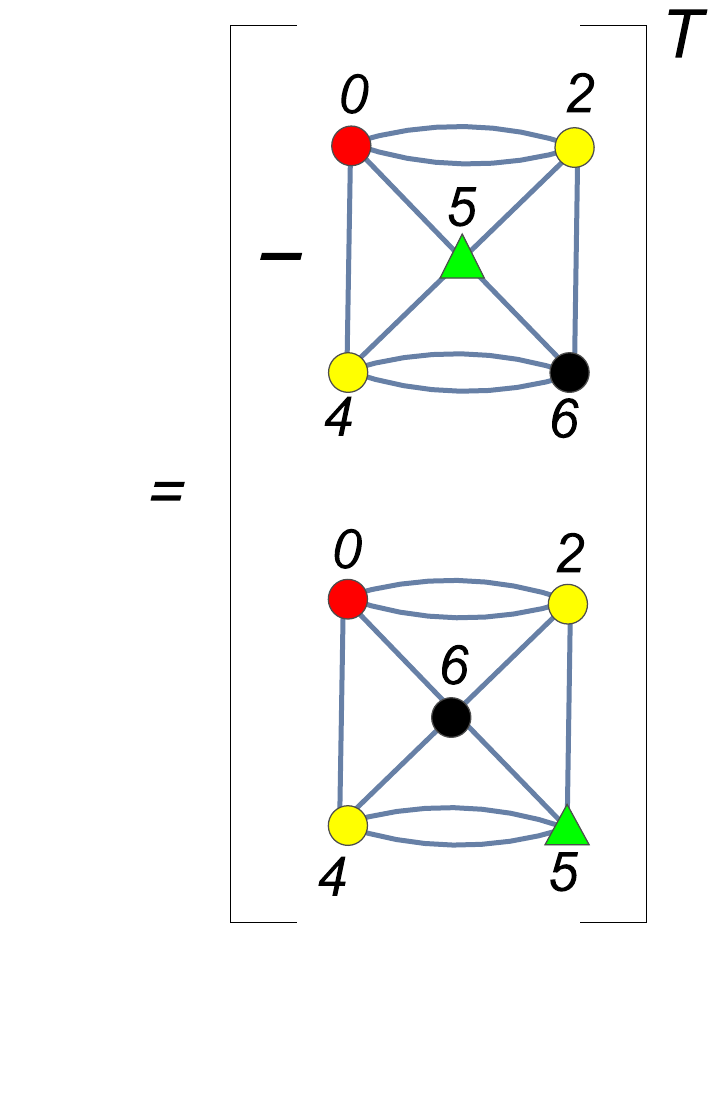}
\includegraphics[scale=0.4]{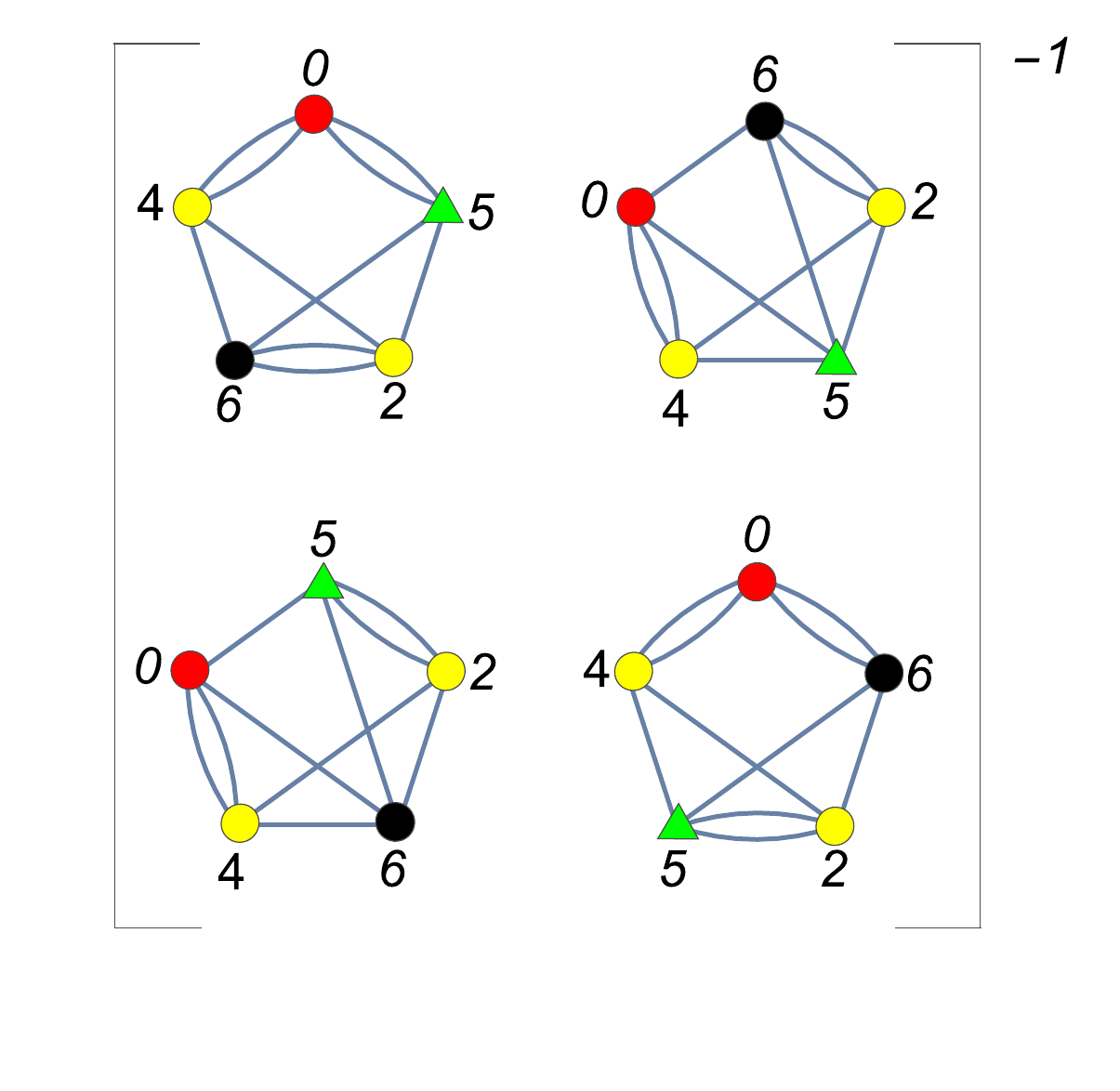}
\includegraphics[scale=0.4]{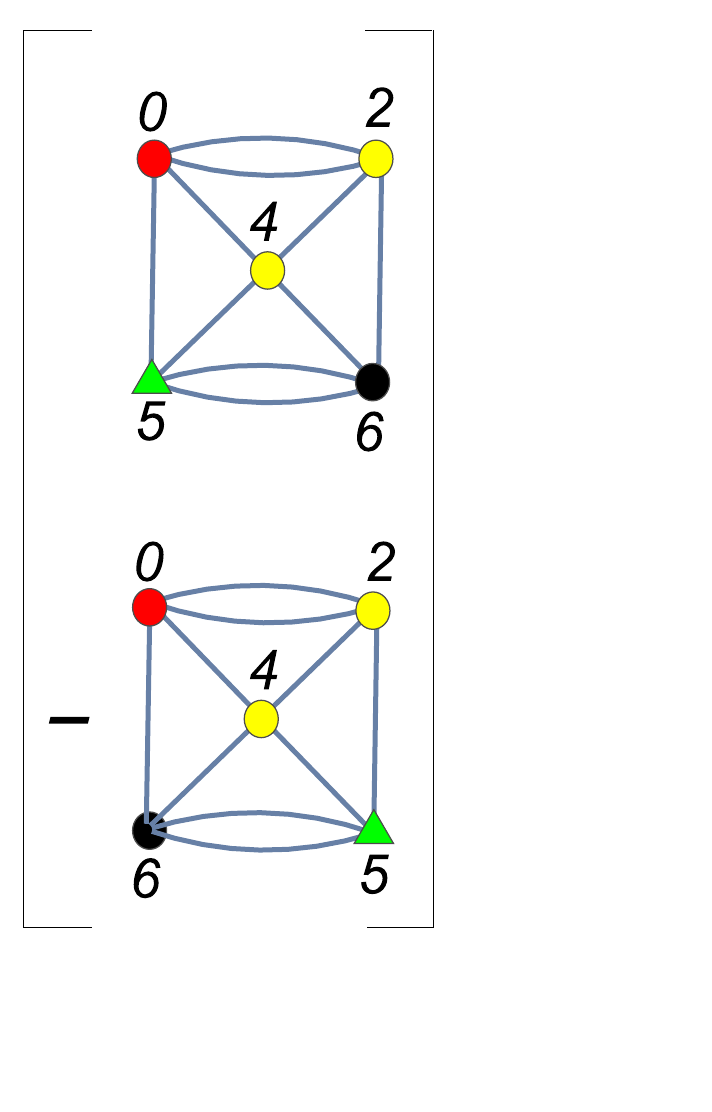}
\,\,\, .
\begin{center}
({\bf Fig.7.18})\,{\small {\rm General KLT algorithm. }}
\end{center}
\end{center} 
Using the $\L-$algorithm  we compute each diagram in {\bf Fig.7.18}, and the result is
\begin{center}
\includegraphics[scale=0.67]{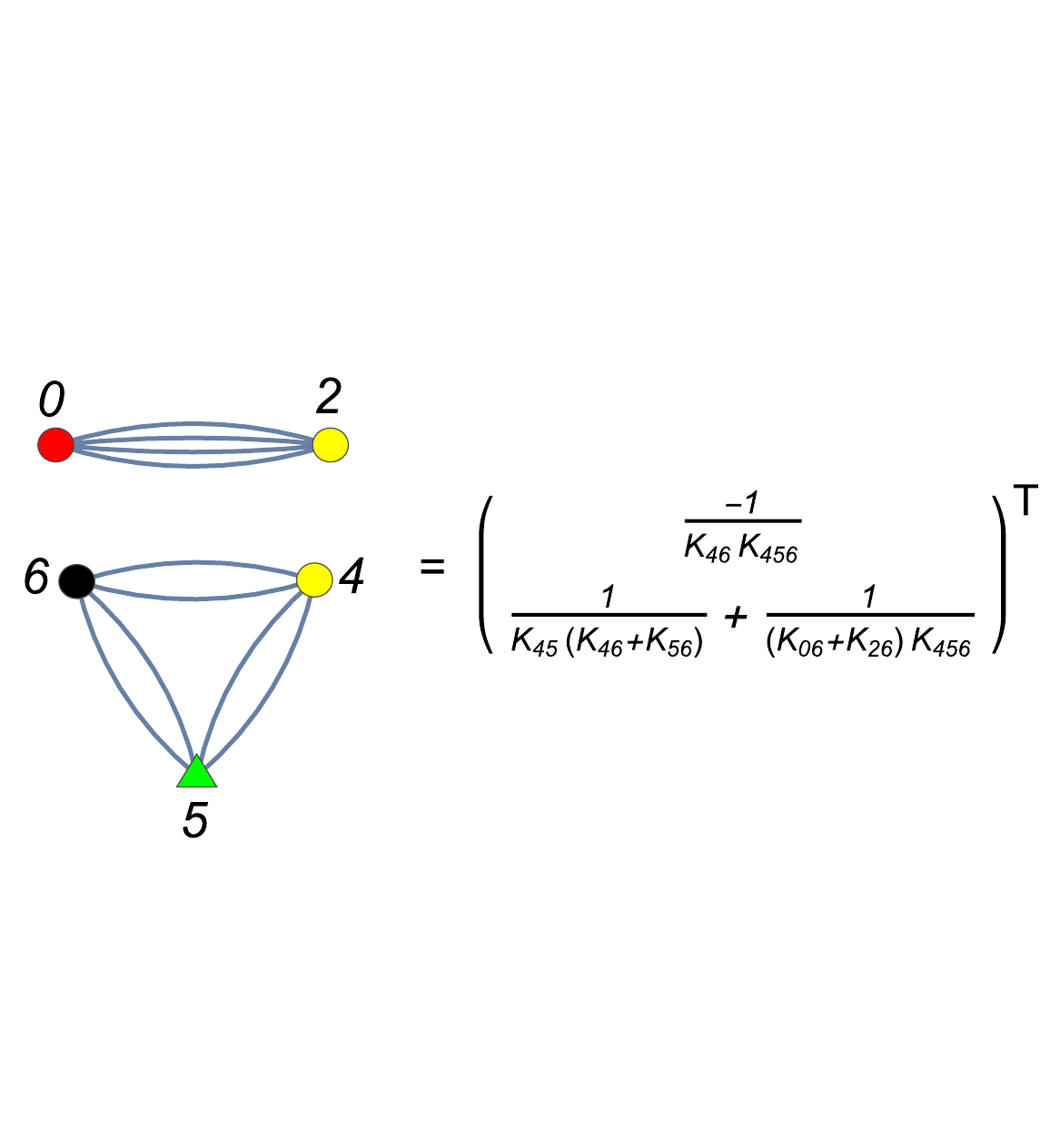}
\includegraphics[scale=0.68]{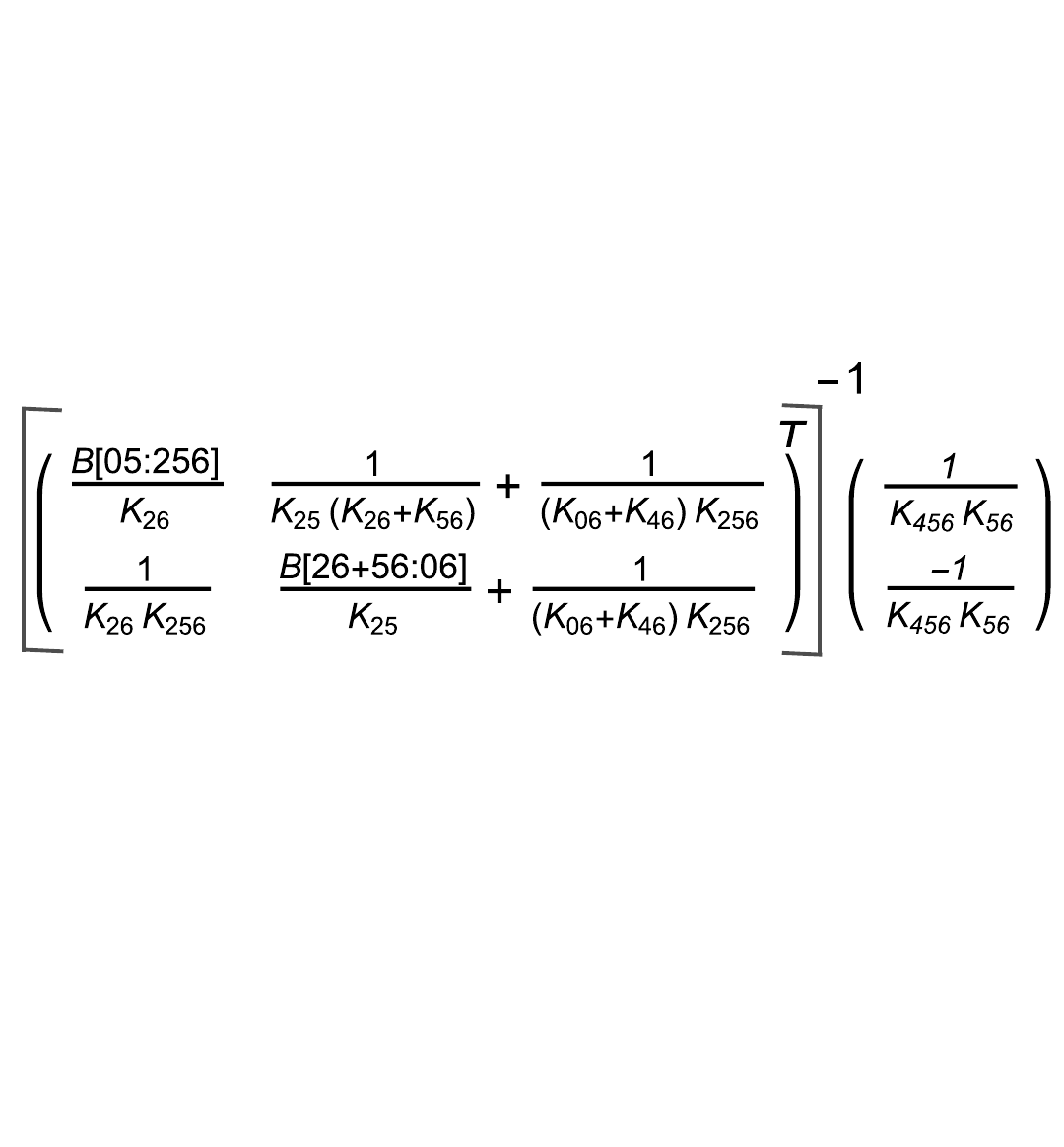}
\begin{center}
({\bf Fig.7.19})\,{\small {\rm Result from the general KLT algorithm.  }}
\end{center}
\end{center} 

Relabeling the (1,2,3) indices one can write the final answer as 
\begin{align}
{\cal I_{(B)}}=&\left(
\begin{matrix}
\frac{-1}{k_{456} k_{46}} \\
\frac{1}{k_{456} (k_{16}+k_{26}+k_{36}  ) }  +  \frac{1}{k_{45} (k_{46} + k_{56})}
\end{matrix}
\right)^{\rm T}  \\
&
\left[\frac{1}{k_{13}}
\left(
\begin{matrix}
\frac{B[15+35:256]}{k_{26}} 
& 
\frac{1}{k_{26}k_{256}}\\
\frac{1}{k_{25}(k_{26}+k_{56})} + \frac{1}{k_{256}(k_{16}+k_{36}+k_{46})} 
& 
\frac{1}{k_{256}(k_{16}+k_{36}+k_{46})}+\frac{B[16+36:26+56]}{k_{25}} 
\end{matrix}
\right)^{-1}\right. \nonumber\\
&
+
\frac{1}{k_{23}}
\left(
\begin{matrix}
\frac{B[25+35:156]}{k_{16}} 
& 
\frac{1}{k_{16}k_{156}}\\
\frac{1}{k_{15}(k_{16}+k_{56})} + \frac{1}{k_{156}(k_{26}+k_{36}+k_{46})} 
& 
\frac{1}{k_{156}(k_{26}+k_{36}+k_{46})}+\frac{B[26+36:16+56]}{k_{15}} 
\end{matrix}
\right)^{-1}\nonumber \\
&
+
\left.\frac{1}{k_{12}}
\left(
\begin{matrix}
\frac{B[25+15:156]}{k_{36}} 
& 
\frac{1}{k_{36}k_{356}}\\
\frac{1}{k_{35}(k_{36}+k_{56})} + \frac{1}{k_{356}(k_{26}+k_{16}+k_{46})} 
& 
\frac{1}{k_{356}(k_{26}+k_{16}+k_{46})}+\frac{B[26+16:36+56]}{k_{35}} 
\end{matrix}
\right)^{-1}
\right]
\left(
\begin{matrix}
\frac{1}{k_{456} k_{56}} \\
\frac{-1}{k_{456} k_{56}}
\end{matrix}
\right) \nonumber ,
\end{align}
which was checked numerically.
 
\subsection{Eight-Point}\label{Epoint} 

In this section we consider a non-trivial 8 point graph, which has the following left and right decomposition
\begin{center}
\includegraphics[scale=0.5]{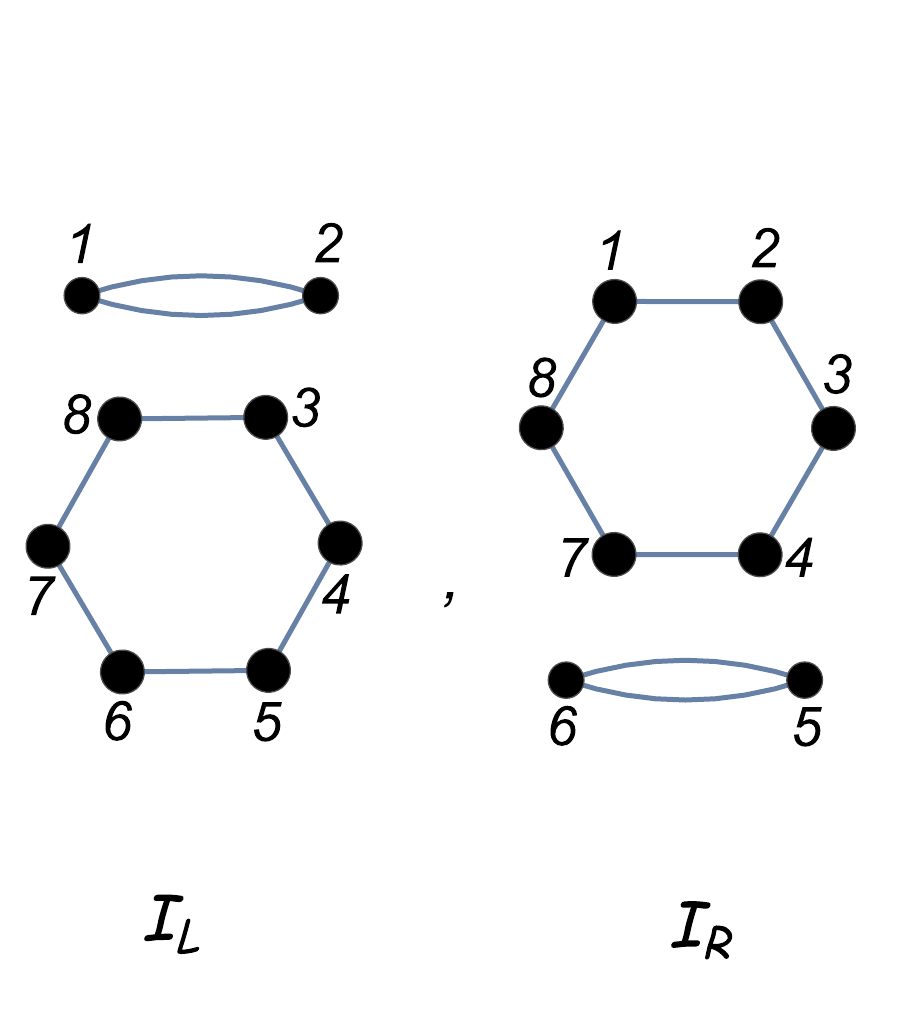}
\,\,\, .
\begin{center}
({\bf Fig.7.20})\,{\small {\rm Left and Right base. }}
\end{center}
\end{center} 
In addition to continue testing the power of the algorithm, this kind of graph was chosen in order to use the $(V)$ building block ({\bf Fig.6.7}).

First we fix a gauge such that there is no a singular configuration, for example we choose the gauge fixing
\begin{center}
\includegraphics[scale=0.5]{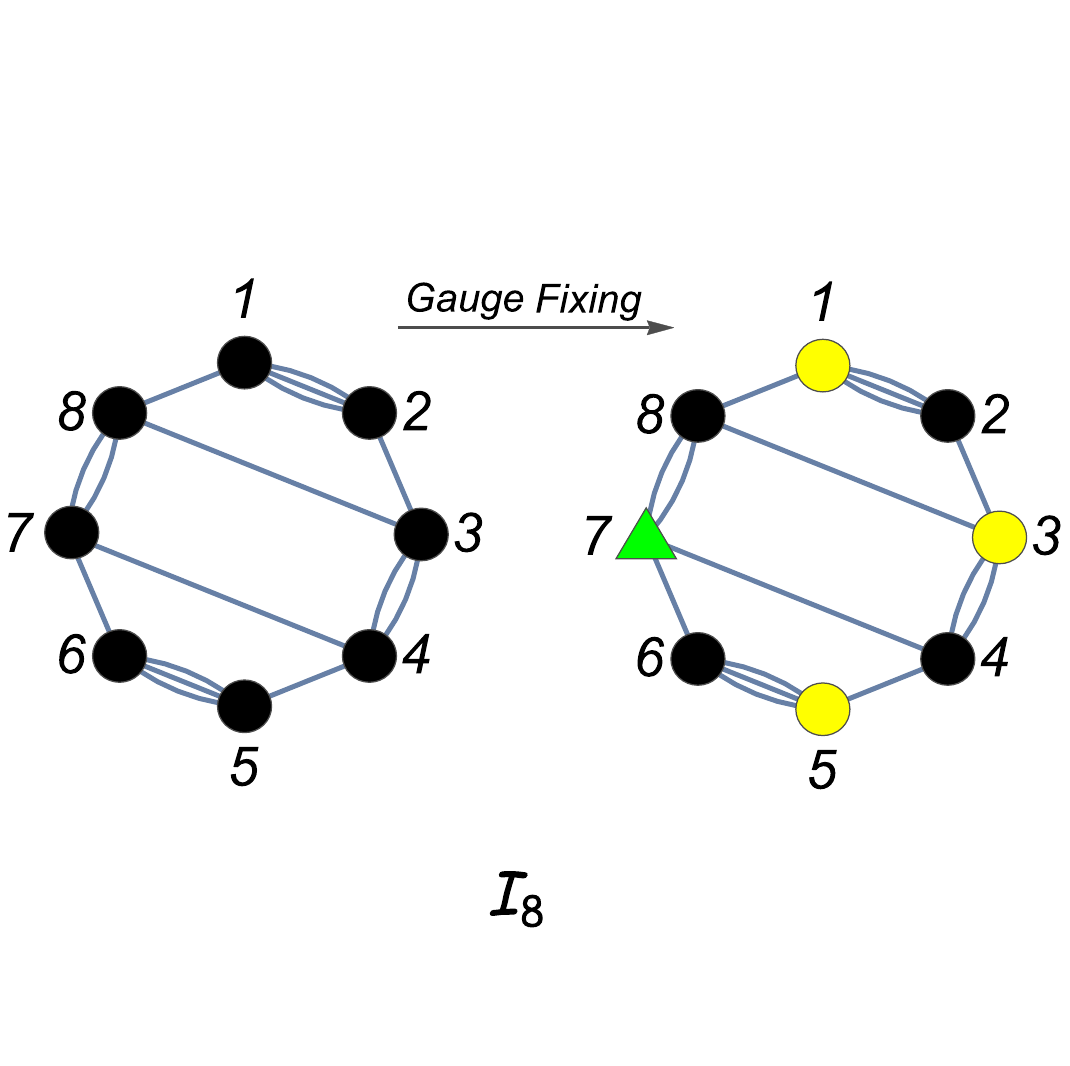}
\,\,\, .
\begin{center}
({\bf Fig.7.21})\,{\small {\rm Gauge Fixing. }}
\end{center}
\end{center} 
Using this gauge we find  three types of non-zero allowable configurations  
\begin{center}
\includegraphics[scale=0.5]{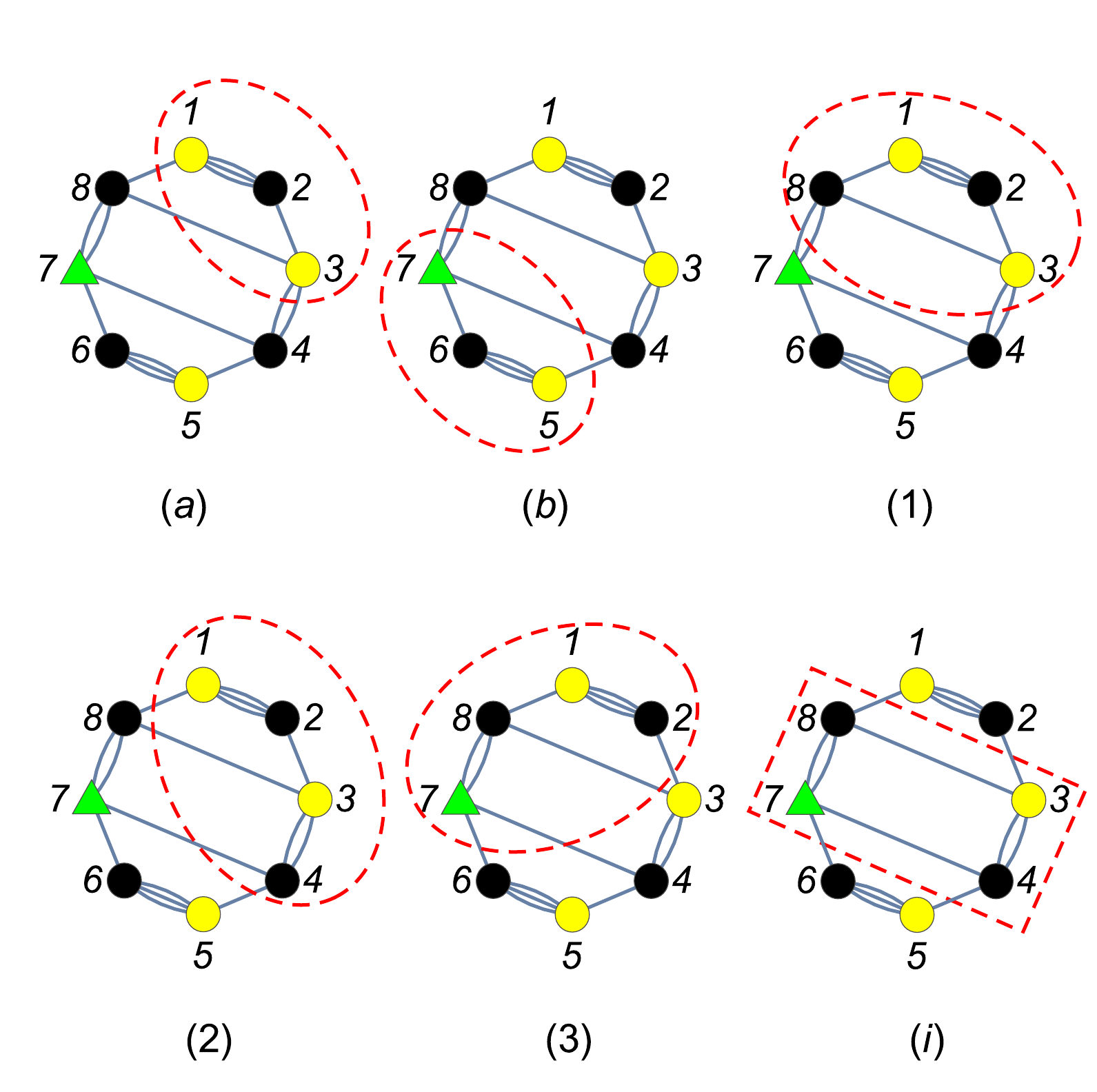}
\,\,\, .
\begin{center}
({\bf Fig.7.22})\,{\small {\rm Non-zero allowable configurations. Type (I): (a) and (b) configurations.\\ Type (II): (1), (2) and  (3) configurations. Type(III): (i) configuration. }}
\end{center}
\end{center}
Applying the $\L-$algorithm one obtains that from the Type (I),  $\{({\rm a}),({\rm b})\}$,  arises a  6-point subdiagrams with the graph 
\begin{center}
\includegraphics[scale=0.5]{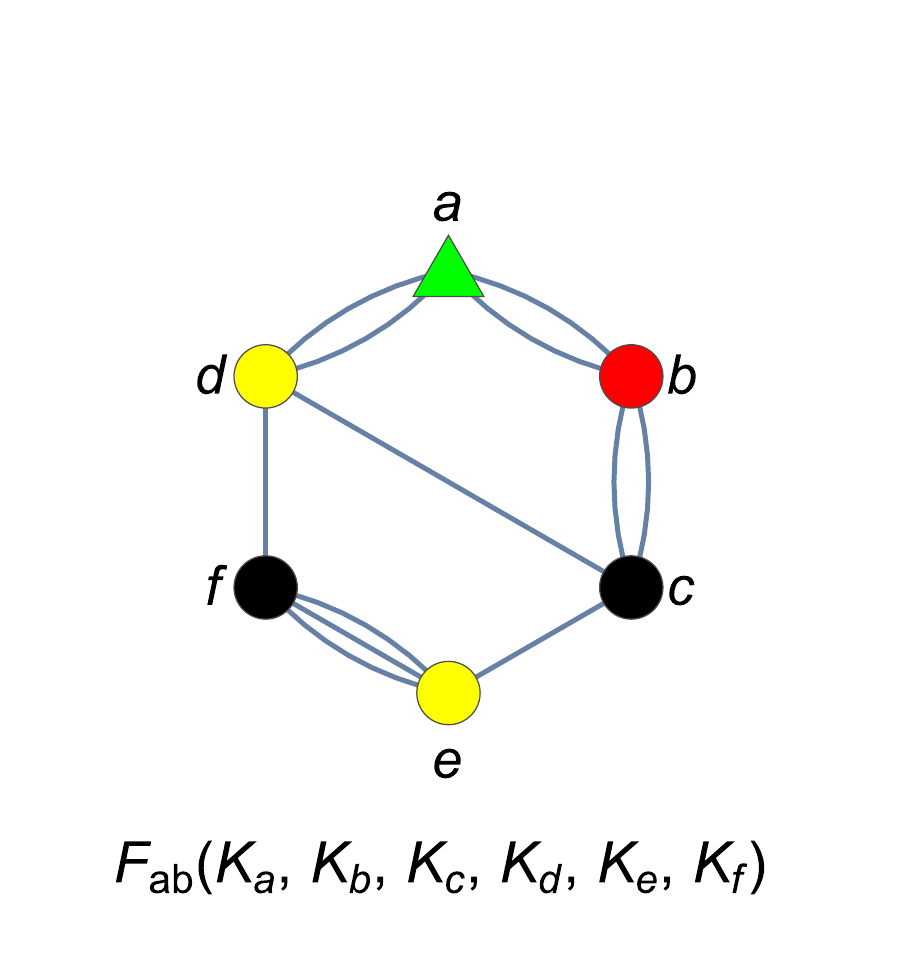}
\,\,\, ,
\begin{center}
({\bf Fig.7.23})\,{\small {\rm Six point graph from the Type (I) configuration. }}
\end{center}
\end{center}  
where we have fixed the ``a"-puncture (by scale symmetry)  in order to avoid singular configurations. This graph is very similar to one given in {\bf Fig.7.2} and its computation is totally analog. Using the $\L-$algorithm the result for this graph is  the function
\begin{align}
&F_{ab}(k_a,k_b,k_c,k_d,k_e,k_f)= -\left( \frac{k_{df} B[ef:df] }{k_{ab} (k_{ac}+k_{bc})k_{ef}}  +
\frac{( k_{af} + k_{bf} + k_{df} ) B[ef:af+bf+df] }{k_{ab} k_{cef} k_{ef}} 
\right) \nonumber \\
& \qquad\qquad\qquad\quad -
\left( \frac{ ( k_{af} + k_{df}) B[ef:af+df] }{k_{ad} k_{bc} k_{ef}}  +
\frac{( k_{af} + k_{bf} + k_{df} ) B[ef:af+bf+df] }{ k_{ad} k_{cef} k_{ef}} 
\right)\nonumber\\
& \qquad\qquad\qquad\quad -
\frac{ k_{df} B[bc:cd+ce+cf ] B[ef : df]  }{k_{abc} k{ef} }\,\,.
\end{align} 
Therefore, the (a) and (b) configurations  can be written as
\begin{align}
({\rm a}) &= -\frac{k_{23} B[12:23]}{k_{123} k_{12}} F_{ab}(k_8,k_1+k_2+k_3 , k_4, k_7 , k_5, k_6),\\
({\rm b}) &= -\frac{k_{57} B[56:57]}{k_{567} k_{56}} F_{ab}(k_4,k_5+k_6+k_7 , k_8, k_3 , k_1, k_2).
\end{align}

From the type (II) configurations, $\{ (1),(2),(3)\}$,  one obtains the following 5-point subdiagrams after applying the $\L-$algorithm
\begin{center}
\includegraphics[scale=0.65]{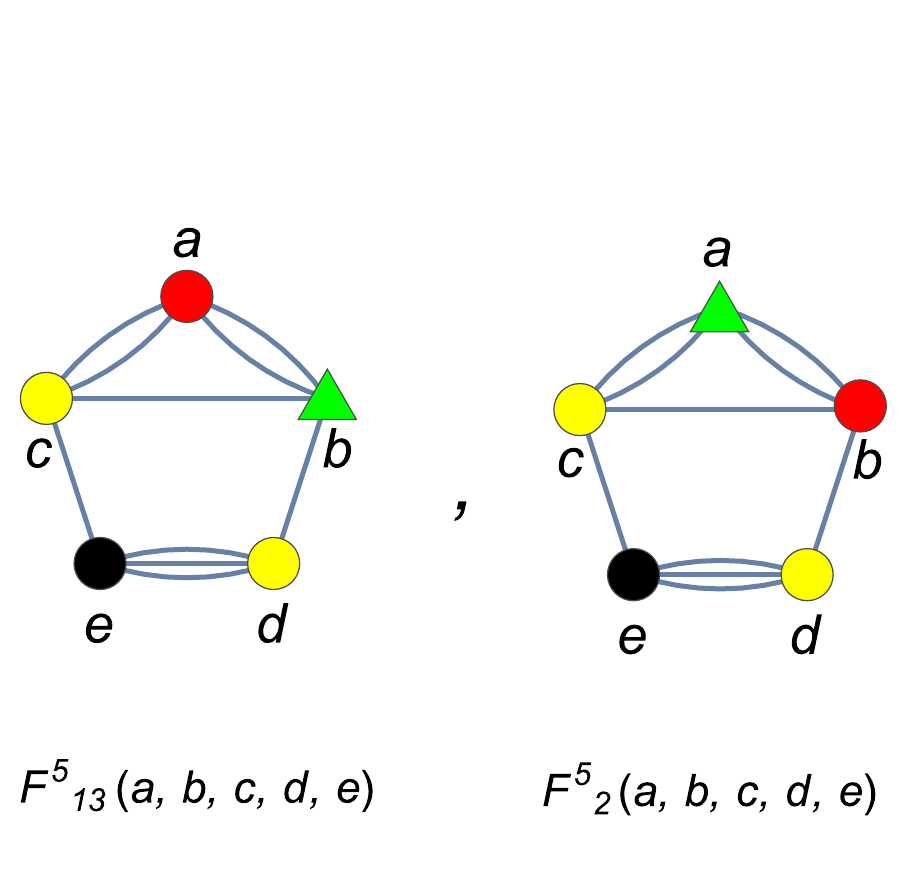}
\,\,\, .
\begin{center}
({\bf Fig.7.24})\,{\small {\rm Five point graphs from the Type(II) configurations. }}
\end{center}
\end{center}  
The first one, $F_{13}^5(k_a,k_b,k_c,k_d,k_e)$, is a subdiagram obtained from the $(1)$ and $(3)$ configurations and the second one,  $F_{2}^5(k_a,k_b,k_c,k_d,k_e)$, is obtained from the $(2)$ configuration. These two subdiagrams are very similar to one obtained in {\bf Fig.7.8} and their computations are very simple using the $\L-$algorithm. The results for these two graphs are
\begin{align}
F_{13}^5(k_a,k_b,k_c,k_d,k_e) &= -\frac{ k_{ce} B[de:ce]  }{k_{ab} k_{de}   }
 -\frac{  ( k_{ae} +  k_{ce}  ) B[de: ae+ ce]  }{k_{bde} k_{de}   }, \\
 F_{2}^5(k_a,k_b,k_c,k_d,k_e) &= -\frac{ k_{ce} B[de:ce]  }{k_{ab} k_{de}   }
 -\frac{  ( k_{ae} +  k_{ce}  ) B[de: ae+ ce]  }{k_{ac} k_{de}   } .
\end{align}
Note that the two answers are totally different,  this is because $k_{ac}\neq k_{bde}$ since there is a massive particle. We can now write the results for the Type (II) configurations, $\{(1), (2), (3)\}$, as
\begin{align}
(1) &= \frac{F_{13}^5(k_4+k_5+k_6+k_7 , k_8,k_3,k_1,k_2) F_{13}^5(k_1+k_2+k_3+k_8 , k_4,k_7,k_5,k_6)}{k_{4567}} , \\
(2) &= \frac{F_{2}^5(k_4,k_5+k_6+k_7 + k_8,k_3,k_1,k_2) F_{2}^5(k_8,k_1+k_2+k_3+k_4 ,k_7,k_5,k_6)}{k_{5678}}  ,  \\
(3) &= \frac{F_{13}^5(k_7,k_8,k_3+k_4 + k_5+k_6,k_1,k_2) F_{13}^5(k_3,k_4,k_1+k_2 + k_7+k_8,k_5,k_6)}{k_{4567}}   .
\end{align}

Finally, from the Type (III) configuration, $\{(i)\}$, one obtains the subdiagrams  
\begin{center}
\includegraphics[scale=0.55]{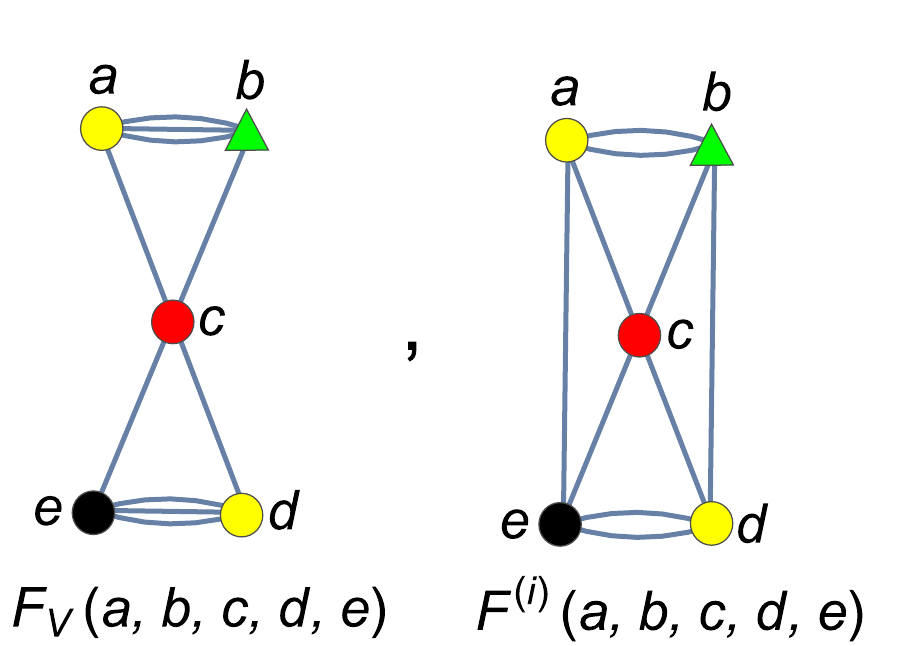}
\,\,\, .
\begin{center}
({\bf Fig.7.25})\,{\small {\rm Five point graphs from Type(III) configuration. }}
\end{center}
\end{center}  
The first one, $F_V(k_a,k_b,k_c,k_d,k_e)$, is the $(V)$ building block computed in section \ref{fivebb} and it is given by
\begin{equation}
F_V(k_a,k_b,k_c,k_d,k_e) = \left( \frac{1}{k_{ab} k_{de} }  \right)^2 (k_{ae} k_{bd} - k_{bce} k_{be}).
\end{equation}
The second one was also computed in section \ref{fivebb} using the $\L-$algorithm  and its result is very simple
\begin{equation}
F^{(i)}(k_a,k_b,k_c,k_d,k_e) =  \frac{1}{k_{ab} k_{de} }  .
\end{equation}
Thus,  the $(i)$ configuration can be read as 
\begin{equation}
(i) = \frac{ F_V (k_1, k_2 , k_3+k_4+k_7+k_8 , k_5 , k_6)\,\, F^{(i)} (k_7 , k_8 , k_1+k_2 + k_5+k_6,k_3,k_4 )  }{k_{3478}}
\end{equation}

The full answer is the sum over all configurations given in {\bf Fig.7.22}, i.e.
\begin{align}
{\cal I}_8= & ({\rm a}) + ({\rm b}) + (1) + (2) + (3) +(i) \\
= &-\frac{k_{23} B[12:23]}{k_{123} k_{12}} F_{ab}(k_8,k_1+k_2+k_3 , k_4, k_7 , k_5, k_6)\nonumber\\
& -\frac{k_{57} B[56:57]}{k_{567} k_{56}} F_{ab}(k_4,k_5+k_6+k_7 , k_8, k_3 , k_1, k_2)\nonumber \\ 
&+ \frac{F_{13}^5(k_4+k_5+k_6+k_7 , k_8,k_3,k_1,k_2) F_{13}^5(k_1+k_2+k_3+k_8 , k_4,k_7,k_5,k_6)}{k_{4567}} \nonumber\\
& + \frac{F_{2}^5(k_4,k_5+k_6+k_7 + k_8,k_3,k_1,k_2) F_{2}^5(k_8,k_1+k_2+k_3+k_4 ,k_7,k_5,k_6)}{k_{5678}} \nonumber \\
& + \frac{F_{13}^5(k_7,k_8,k_3+k_4 + k_5+k_6,k_1,k_2) F_{13}^5(k_3,k_4,k_1+k_2 + k_7+k_8,k_5,k_6)}{k_{4567}}  \nonumber\\
&+  \frac{ F_V (k_1, k_2 , k_3+k_4+k_7+k_8 , k_5 , k_6)\,\, F^{(i)} (k_7 , k_8 , k_1+k_2 + k_5+k_6,k_3,k_4 )  }{k_{3478}}  \nonumber  .
\end{align}
This result was checked numerically. 

\section{The Baadsgaard, Bohr, Bourjaily and Damgaard Rules (BBBD) Vs The $\L$-Algorithm}\label{jacobvslambda}
  
In \cite{jacobrules}, Baadsgaard {\bf et al}, formulated some rules in order to compute the same kind of integrals or diagrams that we have studied so far. Nevertheless, although their rules are a certain sum over all possible factorization limits, similar to the $\L$-algorithm, these two algorithms present important differences.  For example, the $\L$-algorithm depends of the gauge fixing, such as it has been explained and shown in  section \ref{mainsection}. This particular characteristic  is in fact  a powerful tool,  for instance, using the BBBD rules, which are  independent of the choice of  gauge,  it is not possible to compute directly integrands such as ones given by the  diagrams in {\bf Fig.  7.2, 7.8, 7.21, 7.23 or 7.24}. The reason is because there are four or three edges connecting two vertices. Nevertheless,  as it has already been shown,  these kind of diagrams can be easily computed using the $\L$-algorithm.



At the same way as in \cite{jacobrules},  the $\L$-algorithm can also be directly used to integrands with non trivial numerator.  For example, let us consider the same diagram as in \cite{jacobrules} 
\begin{center}
\includegraphics[scale=0.53]{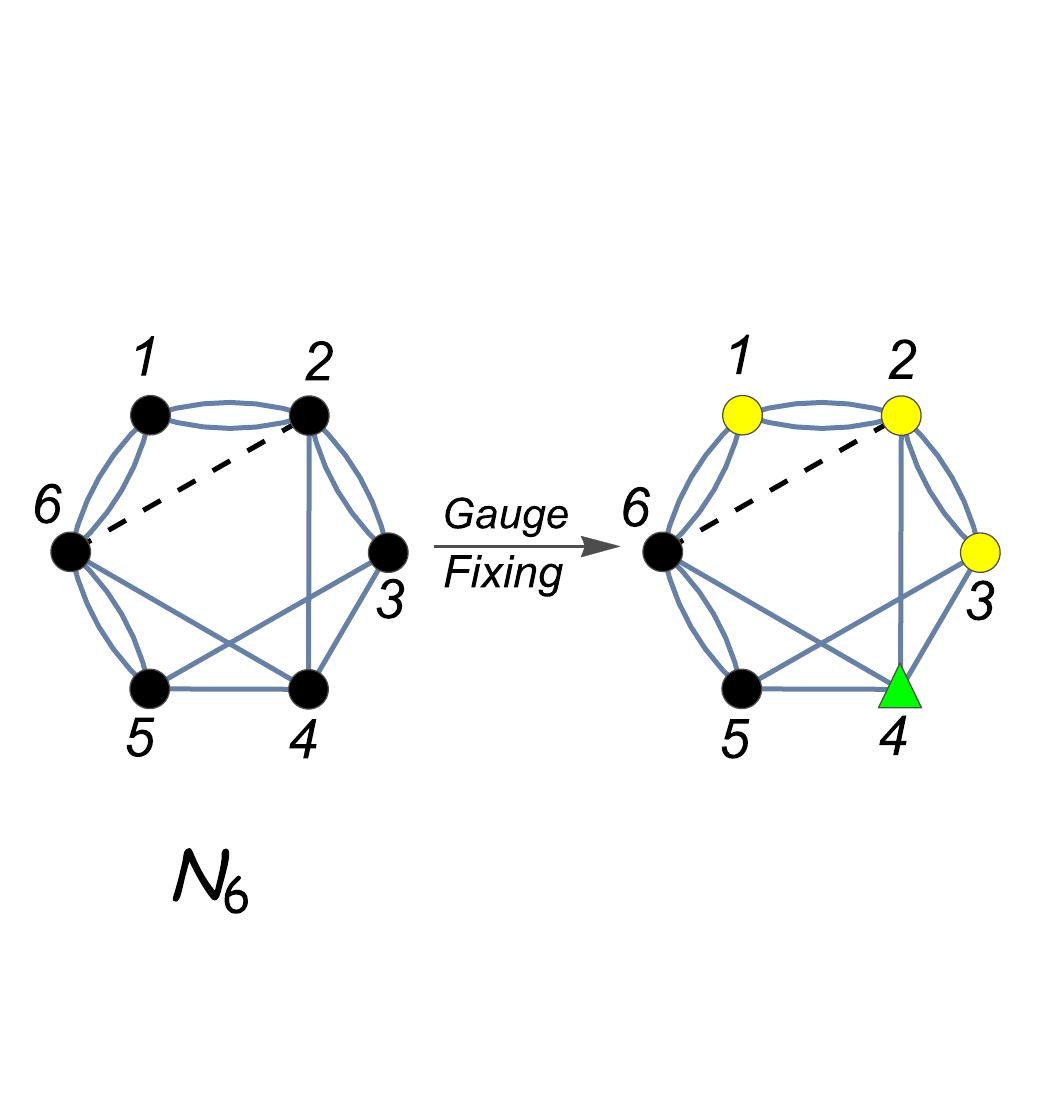}\,\, ,
\begin{center}
({\bf Fig.8.1})\,{\small {\rm Gauge fixing for the ${\cal N}_6$ numerator diagram.}}
\end{center}
\end{center}  
where the black dotted line counts as negative solid line (antiline), i.e.  it carries negative weight.  Clearly this diagram corresponds to the integrand   
\begin{equation}
H^{{\cal N}_6} =\frac{ (\tau_{1:2}\tau_{2:3}\tau_{3:4}\tau_{4:5}\tau_{5:6}\tau_{6:1}) (\tau_{1:2}\tau_{2:3}\tau_{3:5}\tau_{5:6}\tau_{6:1})(\tau_{2:4}\tau_{4:6}\tau_{6:2})}{(\tau_{2:6}\tau_{6:2})},
\end{equation}
or using the CHY variables one can write it as
\begin{equation}
H^{{\cal N}_6}_{{\rm CHY}}=\frac{(z_2-z_6)}{ (z_1-z_2)^2 (z_1-z_6)^2  (z_2-z_3)^2  (z_5-z_6)^2 (z_2-z_4)  (z_3-z_4) (z_3-z_5) (z_4-z_5) (z_4-z_6) }\nonumber  .
\end{equation}
Note that the lines and the antilines connecting the same two vertices cancel each other\footnote{This fact means that a numerator cancels with one denominator.}. Moreover,
due to the presence of a non trivial  numerator, the
denominator has more factors than otherwise, this is  so as to retain the
$SL(2,\mathbb{C})$ invariance.

Obviously, the ${\cal N}_6$ diagram in {\bf Fig.8.1} is not a 4-regular graph, but the subtraction between the  number of lines and antilines must always be four (on each vertex)  in order to keep the $SL(2,\mathbb{C})$ symmetry.

To compute the ${\cal N}_6$ diagram we are obliged to extend the $\L$-Theorem  
\begin{itemize}
\item $\L-${\bf Theorem (Extension)}\qquad\qquad\qquad\\
Let $C$ be an allowable configuration, then the integrand ${\cal I}=|ijk|\Delta_{FP}(ijk,d) \, H(\s)$ 
on the $C$ configuration    has the $\Lambda-$behavior 
\begin{equation}
\mathcal {I}\,\Big|_{\L\rightarrow 0}^C\, \sim\, \L^{(L-A)-4}\,+\,{\cal O}(\L^{(L-A)-3}),
\end{equation}
around $\L= 0$, where  $L$ is the number of lines and $A$ is the number of antilines which are intersected by the red line.
\end{itemize}

Using  the $\L-${\bf Theorem\,(Extension)} it is simple to see there are only two non zero allowable configurations
\begin{center}
\includegraphics[scale=0.5]{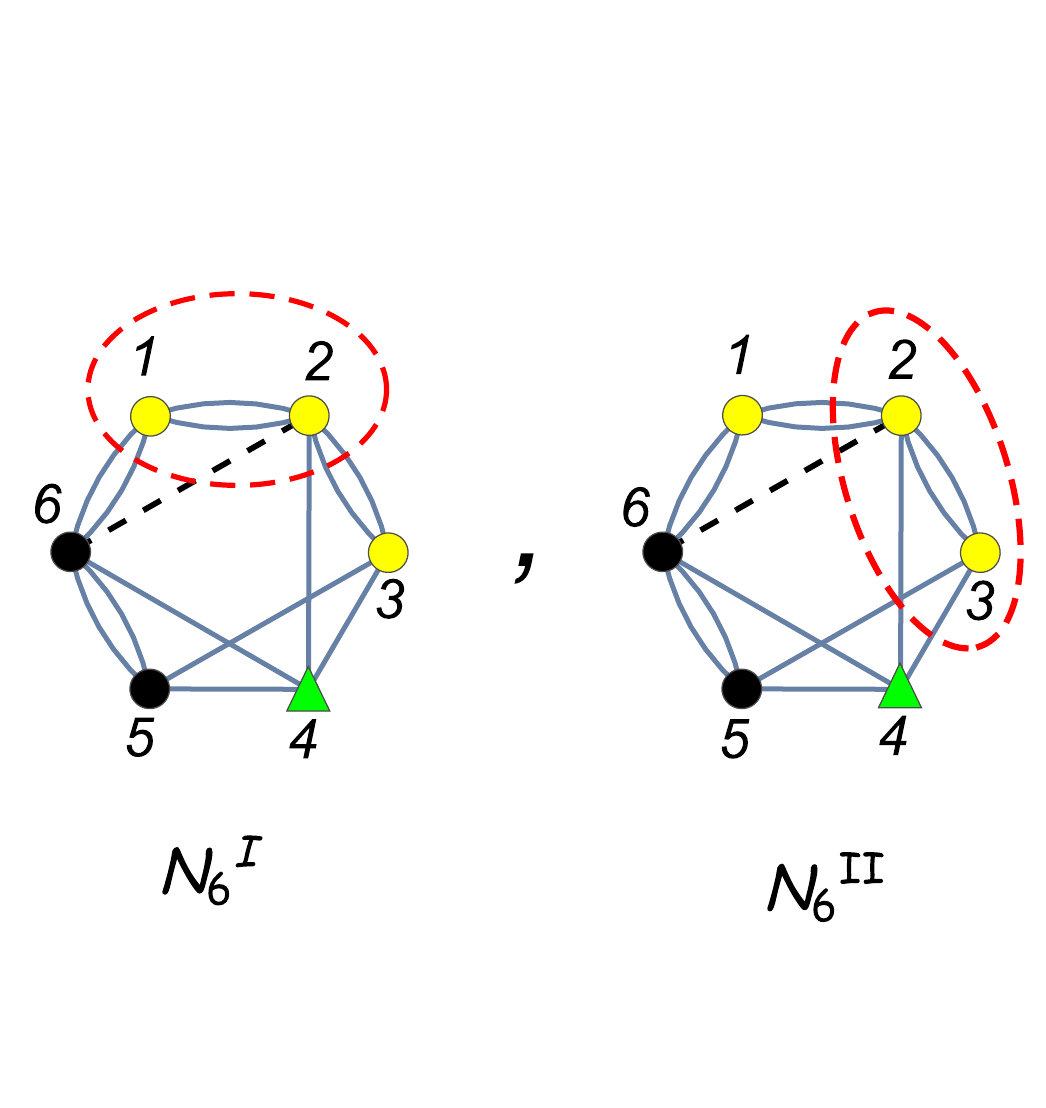}\,\, .
\begin{center}
({\bf Fig.8.2})\,{\small {\rm Non zero configurations for the ${\cal N}_6$ diagram. }}
\end{center}
\end{center}  
These two configurations are easily calculated from the rules in section \ref{Lalgorithm}, so the first one configuration reads 
\begin{equation}
{\cal N}_6^I=\frac{1}{k_{3456}\, k_{456}\, k_{56}}  \,\,\, ,
\end{equation}
and the second one as
\begin{equation}
{\cal N}_6^{II}= \frac{1}{k_{1456}} \left[ \frac{1}{k_{456}\, k_{56}} + \frac{1}{k_{156}}\left(\frac{1}{k_{56}}+\frac{1}{k_{16}} \right) \right].
\end{equation}
Therefore,  the final result  can be written 
\begin{eqnarray}
{\cal N}_6 &= & {\cal N}_6^I +  {\cal N}_6^{II}  \\
&=&\frac{1}{k_{3456}\, k_{456}\, k_{56}}  + \frac{1}{k_{1456}} \left[ \frac{1}{k_{456}\, k_{56}} + \frac{1}{k_{156}}\left(\frac{1}{k_{56}}+\frac{1}{k_{16}} \right) \right],
\end{eqnarray}
which is the same answer found in \cite{jacobrules}. 

This show how powerful is the $\L-$algorithm,  which can be applied to solve highly non trivial integrands.


\section{Discussions}\label{discussion}

In this paper we gave a new representation for the CHY integrals.  We call this new representation as the $\L-$prescription.  The $\L-$prescription is supported on an algebraic curve of degree two, which is  embedded in $\mathbb{C}P^2$, i.e. this  is a sphere.   This curve can be thought as a Riemann surface with two sheets connected by a branch cut.  

The new scattering equations ({\bf the $\L-$scattering equations}) must contain  information about the branch where the particles (punctures) are localized. For example,  the $\L$ scattering equations are given by the expression 
$$
E_a:=\sum_{b\neq a }^n k_a\cdot k_b \,\,\tau_{a:b}, ~~{\rm where} ~ \tau_{a:b}:=\frac{1}{2\,y_a}\left(\frac{y_a+y_b+\s_{ab}}{\s_{ab}}\right), ~~{\rm and} ~ y_a^2=\s_a^2-\L^2,
$$
with $a=1,\ldots n$. When $y_a=\sqrt{\s_a^2-\L^2}$ one says  that the particle (puncture)  is on the upper sheet and when 
$y_a=-\sqrt{\s_a^2-\L^2}$ then one says that the particle is on the lower sheet. Note that the quadratic curves, $y_a$, have an additional parameter, $\L$,  which controls the opening of the branch cut.  When this parameter is promoted as a variable then a new symmetry arise (scale symmetry), which can be used to fix one more particle (puncture).  

In section \ref{residuetheorem} we performed the global  residue theorem over this new variable, $\L$. After integrating $\L$  one  obtains that the $\L$ prescription must be evaluated at the point $\L=0$ ($\L=\infty$), i.e. at the limit when the branch cut collapses in a line. So, the initial integral is broken into two new smaller integrals, which are now written as in the original CHY approach. In addition, these two new integrals are multiplied by  a propagator, which is associated to the collapsed branch cut, it is kind a factorization limit. This is an iterative process, i.e. it can be applied over each one of these two new integrals. All this procedure is encoded into what we call the {\bf $\L-$algorithm}.

The $\L-$algorithm  allow us to expand a given integral in terms of fundamental building blocks, given in {\bf Fig.6.7}. 

Unlike to the other algorithms, the $\L-$algorithm depends  totally  of the gauge fixing. Although this does not look like to be a good thing, in fact it is.   For example, diagrams such as ones given in {\bf Fig.7.1}, which are very complicated using other type of algorithms, they are easily computed from the $\L$ algorithm, obviously, after choosing a good gauge. 

The $\L$  algorithm is a powerful, simple and beautiful tool because it is a pictorial algorithm.  Nevertheless, this mechanism has some limitations, i.e. there are some CHY integrals which can not be performed just using this algorithm.  This is due  we do not know the behavior of the {\bf singular allowable configurations}, which is the reason why one must choose a good gauge.  It will be very interesting to know how to extend the $\L$ algorithm to singular allowable configurations. 

We know that the $\L$ algorithm can be used on a big spectrum of CHY integrals, the main idea is to choose a gauge such that the all allowable configurations will not be singular. In particular, we know diagrams on  which this fact always happens. These diagrams are given by  all possible combinations of the following two  2-regular graphs    
\begin{center}
\includegraphics[scale=0.5]{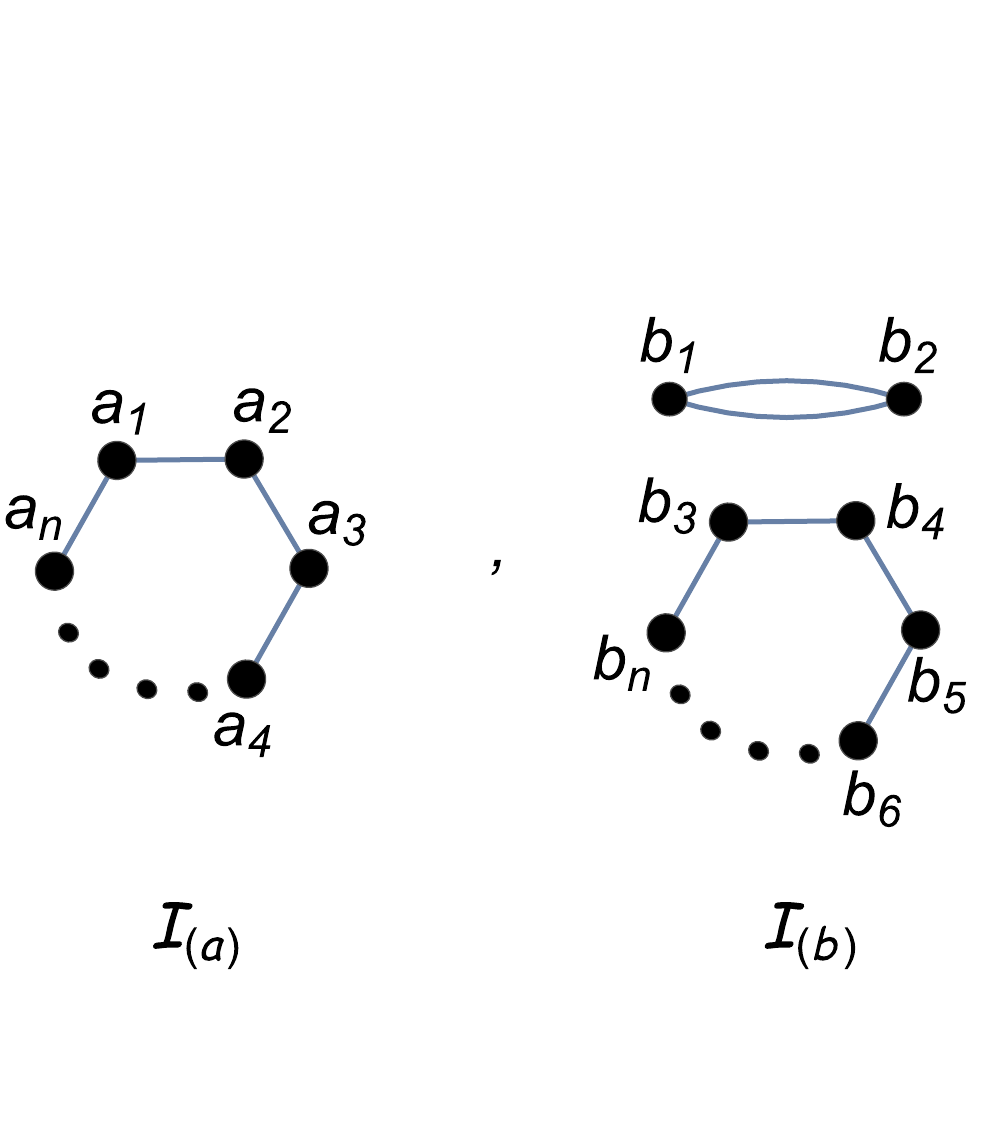}\,\, .
\begin{center}
({\bf Fig.9.1})\,{\small {\rm Two 2-regular graphs. ${\cal I}_a$ is a Parker-Taylor graph. ${\cal I}_b$ is a bubble with a regular polygon  graph. }}
\end{center}
\end{center}  
The ${\cal I}_{a}$ graph is clearly a Parker-Taylor factor, therefore, the diagram given by the integrand $H(\s)={\cal I}_a ~{\cal I}_a$ is just the $m(\a|\b)$ kernel, which is very simple to compute. The other two options given by the integrands $H(\s)={\cal I}_a ~{\cal I}_b$  and $H(\s)={\cal I}_b ~{\cal I}_b$, which are non trivial diagrams, they can be easily computed using the $\L$ algorithm. 

The $\L$ algorithm has two more advantages. As we saw, some massive particles arise in the process, so this algorithm supports off shell particles.  The other one is that this algorithm could be used on integrands with non trivial numerators, such as one given in {\bf Fig.8.1}.  These two characteristics are very important in order to compute diagrams at loop level, for example, the diagram given by 
\begin{center}
\includegraphics[scale=0.5]{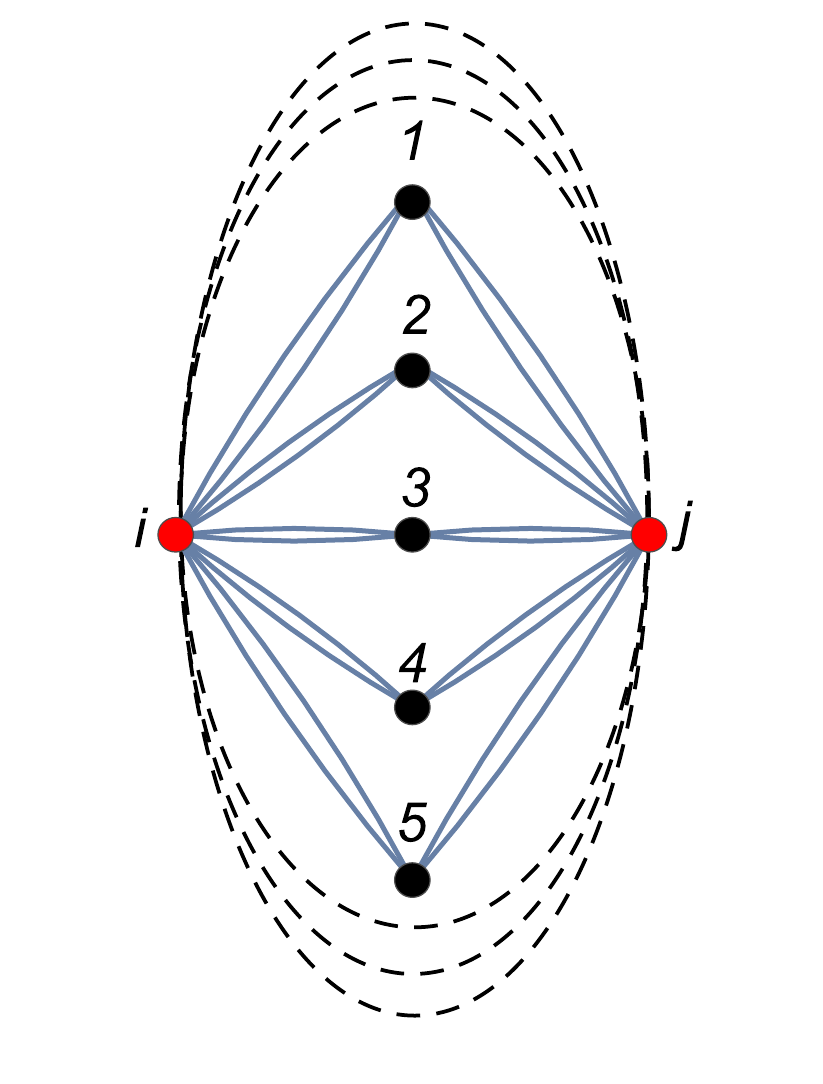}\,\, ,
\begin{center}
({\bf Fig.9.2})\,{\small {\rm 5-gon CHY diagram representation. }}
\end{center}
\end{center}  
which appears at 1-loop computation of the 5-gon, it can easily be computed using the $\L-$algorithm \cite{humbertoC}.

Finally, note that the integrand in the  $\L$ prescription is basically obtained from the original CHY approach just changing  the $\frac{1}{z_{ab}}$ form by  the $\tau_{a:b}$ form.  However, although $\frac{1}{z_{ab}}$ is an antisymmetric form, i.e $\frac{1}{z_{ab}}=-\frac{1}{z_{ba}}$,  the $\tau_{a:b}$ form is not, $\tau_{a:b}\neq -\tau_{a:b}$.  So, the antisymmetric matrix, $\Psi_{\a\b}$, which was defined in \cite{Cachazo:2013hca}, it is not any more antisymmetric when $(z_{ab})^{-1}$ is replaced by $\tau_{ab}$. Therefore, the Pfaffian of $\Psi_{\a\b}(\tau_{a:b})$ is not well defined. Naively, in order to give an  interpretation for the Yang-Mills theory from the $\L$ prescription one can replace the Pfaffian of $\Psi_{\a\b}(z_{a:b})$ by $\sqrt{{\rm det}\,(\Psi_{\a\b}(\tau_{a:b}))}$, but we leave this for future research.

\acknowledgments
The author would like to thank  F. Cachazo for his initial collaboration in this paper.
The author thanks to F. Cachazo, C. Cardona and C. Kalousios for carefully reading the draft,  for comment and  useful discussion. 
The author would like to thank the hospitality of Perimeter Institute,  Universidade de S\~ao Paulo (USP)
and Universidad Santigo de Cali,  where this work was developed. The author thanks to the string theory group of the USP, where this work was presented.
HG is supported by CNPq  grant 403178/2014-2.

\appendix

\section{$\L$-Theorem}\label{appendix}
In this appendix we prove the $\L$-theorem, which was given in section \ref{ltheorem}.
\\

\begin{tabular}{| l |}
 \cline{1-1}  
 $\L-${\bf Theorem}\qquad\qquad\qquad\\
Let $C$ be an allowable configuration, then the integrand ${\cal I}=|ijk|\Delta_{FP}(ijk,d) H^D(\s)$ \\
on the $C$ configuration    has the $\Lambda-$behavior \\
\\
\hspace{5cm}$\mathcal {I}\,\Big|_{\L\rightarrow 0}^C\, \sim\, \L^{L-4}\,+\,{\cal O}(\L^{L-3})$\\
\\
around $\L= 0$, where  $L$ is the number of edges which are intersected by the red line.\\
\cline{1-1}
\end{tabular}
\\
\\

{\bf Proof}

Let us consider an allowable configuration, i.e. two fixed punctures on the upper branch and the others two fixed punctures on the lower branch. Without loss of generality,  one  can consider  the  1 and 2 puntures  fixed on the upper branch and the 3 and 4 puntures  fixed on the lower branch. Under this consideration it is straightforward to check that the Faddeev-Popov determinant has a behavior 
\begin{equation}
|1,2,3|\Delta_{FP}(1,2,3|4)={2^5 \,\sigma_1^2 \,\,\s_2^2\,\sigma_3^3 \,\s_4 \,(\sigma_1 - \sigma_2)^2(\sigma_3 - \sigma_4)   \over \Lambda^4} - {1\over \sigma_4\,\, \Lambda^2}+{\cal O}(\L^0).
\end{equation}
Before showing  that $H^D(\s)\sim \L^L$, we analyse  the $\tau_{a:b}$ form. When the $\s_a$ and $\s_b$ punctures  are on upper sheet, i.e. $y_a=\sqrt{\s_a^2-\L^2}$ and  $y_b=\sqrt{\s_b^2-\L^2}$,  the $\tau_{a:b}$ form has the behavior 
\begin{equation}\label{Ltau1}
\tau_{a:b}\Big|^{a,b} ={1\over \sigma_{ab}}-{\L^2 \over 4 \,\sigma_a ^2 \,\sigma_b}-
{ (\sigma_a^2 + \sigma_a \,\sigma_b + 
   3\, \sigma_b^2) \Lambda^4   \over  2^4 \,\sigma_a^4 \,\sigma_b^3 } +{\cal O}(\L^6),
\end{equation} 
where $\tau_{a:b}\Big|^{a,b}$ means that $\s_a$ and $\s_b$ are on the upper branch cut. For the others three more configurations, $(a\rightarrow \,{\rm upper}, b\rightarrow\,{\rm  lower})$, $(a\rightarrow \,{\rm lower}, b\rightarrow\, {\rm upper})$ and  $(a\rightarrow \,{\rm lower}, b\rightarrow\, {\rm lower})$,  the $\L$ expansion is read as
\begin{align}\label{Ltau2}
\tau_{a:b}\Big|^{a}_{~~ b} &=
{1\over \sigma_a} + {(\sigma_a + 2 \sigma_b) \Lambda^2 \over 
 2^2\, \sigma_a^3 \sigma_b} + {(\sigma_a^3 + \sigma_a^2 \sigma_b + 
    3 \sigma_a \sigma_b^2 + 6 \sigma_b^3) \Lambda^4 \over 
 2^4 \,\sigma_a^5 \sigma_b^3}
+{\cal O}(\L^6),\nonumber \\
\tau_{a:b}\Big|^{~~b}_a &=
{\Lambda^2 \over 
 2^2 \,\sigma_a^2 \sigma_b}  + {(\sigma_a^2 + \sigma_a \sigma_b + 
    3 \sigma_b^2) \Lambda^4  \over 2^4\, \sigma_a^4 \sigma_b^3}
+{\cal O}(\L^6),\\
\tau_{a:b}\Big|_{a,b} &=
{\sigma_b  \over \sigma_a^2 - \sigma_a \sigma_b} - {(\sigma_a + 
    2 \sigma_b) \Lambda^2  \over 
 2^2 \,\sigma_a^3 \sigma_b} - {(\sigma_a^3 + \sigma_a^2 \sigma_b + 3 \sigma_a] \sigma_b^2 + 
    6 \sigma_b^3) \Lambda^4  \over 2^4 \,\sigma_a^5 \sigma_b^3}
 +{\cal O}(\L^6).\nonumber
\end{align} 

Now, let us remember that the $H^D(\s)$ integrand is given by the products of chains, i.e. the products of factors such as 
\begin{equation}
\left[a_1,\ldots a_k\right]=(\tau_{a_1:a_2}\,\tau_{a_2:a_3} \,\cdots \,\tau_{a_{k-1}:a_k}\,\tau_{a_k:a_1}).
\end{equation}
This implies the number of edges which are intersected  by the red line is a even number, it is because for each  $\tau_{a_i:a_j}\Big|^{a_i}_{~~ a_j} $ term into the chain, it must also have a term such as  $\tau_{a_m:a_n}\Big|_{a_m}^{~~ a_n}$, in order to close it.  So, from  this fact and using the $\L$ expansion given in \eqref{Ltau1} and \eqref{Ltau2}, it is straightforward to conclude that
\begin{equation}
H^D(\s)\sim \L^L,
\end{equation}
where $L$ is the number of edges which are intersected  by the red line. Thus the $\L-$theorem has been proved $_\blacksquare$

The proof for the $\L-$theorem (Extension), given in section \ref{jacobvslambda}, is completely analogous.

\bibliography{references_L_S}
\bibliographystyle{JHEP}

\end{document}